\documentclass[acmsmall]{acmart}

\setcopyright{rightsretained}
\acmJournal{TOSEM}
\acmYear{2023}
\acmVolume{1}\acmNumber{1}
\acmArticle{1}
\acmMonth{1}
\acmPrice{}
\acmDOI{10.1145/3624742}
 \usepackage{soul}          \usepackage{xcolor}
\usepackage{fontawesome5}  \usepackage[inline]{enumitem} \setlist[itemize]{leftmargin=20pt}
\usepackage{xurl}
\usepackage{subcaption}    

\usepackage{siunitx} 

\usepackage{algorithm} \usepackage{algpseudocode}

\usepackage{tikz}
\usepackage{pgfplots}
\pgfplotsset{compat=1.17}
\usetikzlibrary{arrows.meta, shapes, fit}
\usepgfplotslibrary{fillbetween}
\usetikzlibrary{patterns}
\usepgfplotslibrary{groupplots}
\usepgfplotslibrary{statistics}  

\newcommand{\partparagraph}[1]{\vspace{1mm}\noindent\textbf{#1:}}

\newcommand{\funof}[1]{\left(#1\right)}           \newcommand{\AGain}[0]{\alpha}
\newcommand{\tGain}[0]{\gamma}
\newcommand{\shape}[0]{\phi} \newcommand{\shapefcn}[1]{\phi\funof{#1}}
\newcommand{\DFT}[0]{\mathit{DFT}}
\newcommand{\dof}[0]{\mathfrak{d}_\mathit{f}}
\newcommand{\dnl}[0]{\mathfrak{d}_\mathit{nl}}
\newcommand{\dnlth}[0]{\mathfrak{t}_\mathit{nl}}

\pgfplotsset{colormap={dnl}{rgb255(0)  =( 99,170, 45);
                            rgb255(0.5)=(105,175,210);
                            rgb255(1)  =(119, 41,158)}}

\pgfplotsset{colormap={sat}{rgb255(0)  =( 99,170, 45);
                            rgb255(0.5)=(105,175,210);
                            rgb255(1)  =(119, 41,158)}}

\definecolor{linear}{RGB}{   99,170, 45}
\definecolor{boundary}{RGB}{105,175,210}
\definecolor{nonlinear}{RGB}{119, 41,158}
 
\usetikzlibrary{external}
\begin{document}

\title[Control-Based CPS Stress Testing]{Stress Testing Control Loops in Cyber-Physical Systems}

\author{Claudio Mandrioli}
\authornote{Part of this work was done while the author was a PhD student at Lund University, Sweden.}
\email{claudio.mandrioli@uni.lu}
\orcid{0000-0002-7013-1191}
\affiliation{\institution{University of Luxembourg}
\country{Luxembourg}}
\author{Seung Yeob Shin}
\email{seungyeob.shin@uni.lu}
\orcid{0000-0001-9025-7173}
\affiliation{\institution{University of Luxembourg}
\country{Luxembourg}}
\author{Martina Maggio}
\email{maggio@cs.uni-saarland.de}
\orcid{0000-0002-1143-1127}
\affiliation{\institution{Saarland University}
\country{Germany}
\and
\institution{Lund University}
\country{Sweden}}
\author{Domenico Bianculli}
\email{domenico.bianculli@uni.lu}
\orcid{0000-0002-4854-685X}
\affiliation{\institution{University of Luxembourg}
\country{Luxembourg}}
\author{Lionel Briand}
\email{lionel.briand@uni.lu}
\orcid{0000-0002-1393-1010}
\affiliation{\institution{University of Luxembourg}
\country{Luxembourg}
\and 
\institution{University of Ottawa}
\country{Canada}
} 

\begin{abstract}

Cyber-Physical Systems (CPSs) are often safety-critical and deployed in uncertain environments.
Identifying scenarios where CPSs do not comply with requirements is fundamental but difficult due to the multidisciplinary nature of CPSs.
We investigate the testing of control-based CPSs, where control and software engineers develop the software collaboratively.
Control engineers make design assumptions during system development to leverage control theory and obtain guarantees on CPS behaviour.
In the implemented system, however, such assumptions are not always satisfied, and their falsification can lead to loss of guarantees.
We define stress testing of control-based CPSs as generating tests to falsify such design assumptions.
We highlight different types of assumptions, focusing on the use of linearised physics models.
To generate stress tests falsifying such assumptions, we leverage control theory to qualitatively characterise the input space of a control-based CPS.
We propose a novel test parametrisation for control-based CPSs and use it with the input space characterisation to develop a stress testing approach.
We evaluate our approach on three case study systems, including a drone, a continuous-current motor (in five configurations), and an aircraft.
Our results show the effectiveness of the proposed testing approach in falsifying the design assumptions and highlighting the causes of assumption violations.

\end{abstract}


\maketitle

\section{Introduction}
\label{sec:introduction}
Cyber-Physical Systems (CPSs) are engineering artefacts characterised by the tight coupling of physical and software components~\cite{Lee:2015}.
This tight coupling is created by sensors and actuators that allow the software component to measure and affect the physical part of the system.
For example, a drone uses sensors (such as accelerometer and camera) to estimate its position and actuators (such as propellers) to move.
The desired behaviour is usually that the drone performs stable flight and reaches a desired position.

CPSs are often by nature safety-critical, and they are expected to operate in uncertain environments~\cite{Wu:2017}.
For example, drones and cars operate in environments where people are present and the external conditions are never fully known (e.g., the presence of wind and obstacles for drones, and other vehicles and pedestrians on the road for cars).
In such circumstances, it is of primary importance to identify the scenarios in which the CPS is no longer able to fulfil its requirements.
To identify such scenarios, \emph{stress testing} aims to execute a system under test (SUT) in conditions that are different from the ones expected during the system design~\cite{Naik:2008}.
However, CPS development is known to be multidisciplinary~\cite{Lee:2015}, and the definition of the operating conditions expected during the CPS design depends on the combination of design choices made by different types of engineers (in our case, software engineers and control engineers).
For example, for a drone we want to identify what can limit its capability to avoid a moving obstacle.
The limiting factors can be the hardware design (e.g., the size of the propellers), different software components (e.g., the path planning), or their interaction.

\begin{figure}[t]
    \includegraphics{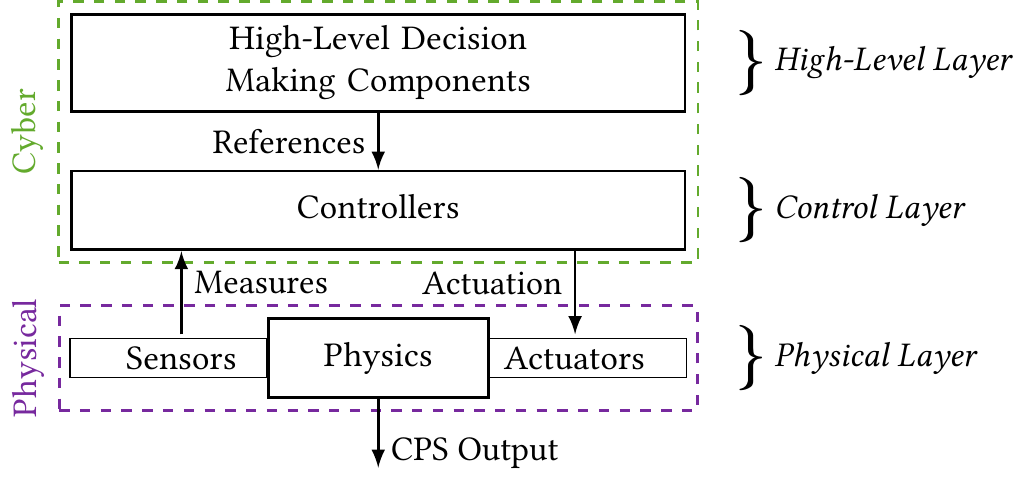}
    \caption{Structure of a CPS, where a cyber component (green dashed box) interacts with a physical component (purple dashed box).
    In the cyber component, we highlight the controllers that implement the control layer and handle the low-level interaction with the physics.
    The control layer receives desired values for given physical quantities (the \emph{references}) and uses sensors and actuators in order to enforce these values in the physical component.
    The measures and actuation signals create a closed-loop interaction between the control layer and the physical layer.
    In fact, CPS requirements are generally defined over quantities that live in the physical part of the system, captured by the \emph{CPS outputs}.
    }
    \label{fig:cps-structure}
\end{figure}

Intuitively, being composed of a physical and software part, CPS development involves both software engineers and engineers with specific knowledge of the physical part of the system (e.g., aerospace engineers for a drone or mechanical engineers for a car).
However, besides these categories of engineers, many CPS applications involve also control engineers~\cite{He:2019}.
We call CPSs that involve control engineers \emph{control-based CPSs}.\footnote{
    CPSs developed with control theory have also been called ``control-CPSs'' in the literature~\cite{He:2019}.
    We believe ``control-based CPSs'' is a more precise characterisation since the word ``control'' can have different meanings depending on the context (e.g., control in the sense of supervision).
}
In such applications, there are multiple layers of decision making, depending on the decision's level of abstraction and time-scale.
Among the different layers, we highlight in Figure~\ref{fig:cps-structure} the role of the \emph{control layer} in the interaction between the cyber component (green dashed box) and the physical component (purple dashed box).
The control layer receives desired values (the \emph{references}) for physical quantities of interest and uses sensors and actuators in real-time to enforce these values in the physical component.
Accordingly, CPS requirements are generally defined over quantities that live in the physical part of the system, captured by the \emph{CPS outputs} in the figure.
For example, for a delivery drone, high-level decision making concerns the definition of the sequences of picking up the object to be delivered, while the control layer receives the desired position, reads the sensors values, and adjusts in real time the propellers commands to reach and maintain the desired position.

In control-based CPSs, the role of control engineers is to design \emph{control algorithms} that are implemented in software as part of the control layer.
When designing control algorithms, control engineers leverage control theory.
To apply such theory, they make \emph{design assumptions} and abstract certain aspects of the design problem.
Such abstractions concern both the physical component as well as the software implementation of the control layer.
For example, for a drone, they simplify how the drone moves in space, they neglect the finite capacity of the motors to generate thrust (also known as actuator saturation) and dismiss other functionalities of the software like flight mode changes.

From a theoretical point of view, control theory provides formal guarantees, under its design assumptions, on the CPS performance (e.g., drone flight speed along a certain trajectory).
In practice, the design assumptions do not necessarily hold for the implemented CPS, and the guarantees might be lost.
We can then distinguish scenarios (e.g., values of the references, environment conditions, system state) in which the design assumptions hold (and therefore also the formal guarantees are valid) from scenarios in which they are \emph{falsified} (and the behaviour of the system is unpredictable).
In general, a CPS is expected to be robust to some level of falsification of a design assumption (e.g., if an assumption is falsified for a limited amount of time) and be still able to provide the guarantees.
Following this observation, we define the \emph{design scope} of a CPS as the set of scenarios in which the design assumptions either hold for the CPS implementation or their falsification does not impact the CPS behaviour.
Conversely, we consider scenarios where the software implementation of the control layer is affected by the falsification of the design assumptions to be outside of the design scope of the CPS.
In other words, the falsification of design assumptions limits the CPS design scope, i.e., it reduces the set of scenarios in which the designed algorithm can provide a priori guarantees.
For example, in the drone, scenarios in which the actuators do not saturate (i.e., they do not reach their maximum or minimum power), or saturate without affecting the flight, are within the design scope.
Conversely, scenarios in which the limited capacity of the motors limits the drone flight capabilities are outside of the design scope.

We argue that, when performing the verification and validation of a control-based CPS, software engineers have to take into account its design scope, which is limited by the falsification of the design assumptions in practice.
In fact, when the CPS is executed within the design scope, the control software can be expected to have predictable behaviour thanks to the a priori guarantees (although other types of faults can of course still appear).
In contrast, when the CPS is outside of the design scope, the control layer can expose unpredictable behaviour.
To give a practical example, a delivery drone can be led out of its design scope by a specific sequence of reference values fed to the control layer, such as those created by unexpected obstacles encountered during its mission.
When this happens, the ability of the drone to perform stable flight is impaired and the delivery is likely compromised.
By identifying the design scope of the CPS, software engineers can decide, for example, if the CPS needs changes in the design, runtime checks, or fall-back solutions to ensure the CPS's stable and reliable operation.

Previous literature on control-based CPS testing focuses the testing process on the identification of test cases that lead the SUT to violate its  requirements~\cite{menghi:2019:oracles}.
Accordingly, several prior works propose various approaches to CPS test case generation based on metrics that quantify the fulfilment of the requirements (e.g., approaches based on search algorithms~\cite{Matinnejad:2017}, or classification trees~\cite{Lamberg:2004}).
In this article, we propose a complementary perspective to the test case generation problem.
Instead of focusing on testing the requirements, we focus on testing to falsify the control design assumptions.
The intuition is that, in the scenarios where such assumptions are fulfilled, we can rely on the a priori guarantees of control theory.
Conversely, in the scenarios where they are not fulfilled, the behaviour of the CPS is unpredictable and empirical evaluation through testing is needed.

We use the falsification of control design assumptions to define the problem of stress testing control-based CPS software.
In this testing problem, the objective is to generate test cases that falsify, to different degrees, the design assumptions (i.e., stress test cases) and identify when such falsification prevents the control algorithm from providing guarantees.
We use knowledge from the control engineering domain to identify the different types of design assumptions that engineers make during the control design process.
By excluding the assumptions that can be addressed with existing testing approaches, we focus this work on the assumptions related to the use of \emph{linearised physics models}~\cite{Astrom:2008}.
For generating test cases that falsify the linearised models, we use frequency-based control-theoretical models to qualitatively describe the input space of the control layer.
In order to leverage this qualitative input space characterisation, we propose a novel approach for the definition of test cases (test case parametrisation) and a novel metric for identifying the stress test cases for control-based CPS.
More specifically, we propose to define test cases according to 
\begin{enumerate*}[label=(\roman*)]
    \item a shape function,
    \item an amplitude scaling coefficient and
    \item a time scaling coefficient.
\end{enumerate*}
Using such parametrisation and metric, we develop a complete stress testing approach for the linearised model assumption.

We assessed our testing methodology by applying it to three different case studies: the altitude control of a drone, the position control of a DC servo (a continuous current motor), and aircraft altitude control.
Our results suggest that our approach and metrics are capable of generating and identifying test cases that falsify to different degrees the control design assumptions.
This in turn allows engineers to observe the behaviour of the SUT at the bounds of its design scope where a priori control-theoretical guarantees become uncertain.
Most notably, we generate tests that expose, across our case studies, actuator saturation time ranging from $0\%$ to $95\%$, thus allowing the evaluation of the CPS under a large variety of saturation conditions.
Actuator saturation is arguably the most common source of non-linear behaviour that can falsify the linear models design assumption in CPSs~\cite{Hu:2001}.
This is due to actuators being one of the most expensive components in CPS and therefore being sized to the minimum to reduce costs.
For this reason, they often saturate during CPS operation.
Hence, assessing the ability of the CPS to operate in the presence of actuator saturation is of prime importance.
In this article, based on case studies, we showcase how our tests can be used to gain insights into design choices that limit the ability of the SUTs to perform safely in different scenarios.
For example, how actuator saturation can limit the trajectories that a drone can follow during flight.

To summarise, this article makes the following contributions:
\begin{enumerate}[label=(\roman*)]
\item We use control-theory domain knowledge to develop an input space qualitative characterisation (i.e., qualitatively describe the expected behaviour of the SUT according to different input features) for individual reference values, i.e., the real-valued inputs of the control layer (Section~\ref{sec:input-characterisation-qualitative}).
    \item We define different metrics to quantify the relevant aspects of the proposed input space characterisation (Section~\ref{sec:input-space-quantification}).
    \item We propose a novel test case parametrisation for CPS control layers that enables the use of the characterisation (Section~\ref{sec:parametrisation}).
    We use such parametrisation to develop metamorphic relations for control-based CPSs and a control-based CPS stress testing approach (Section~\ref{sec:approach-steps}).
\end{enumerate}

The rest of the article is structured as follows. First, given the multidisciplinary nature of this problem, we present in Section~\ref{sec:context} the relevant background on the control-based CPS development and the control algorithms design process.
Section~\ref{sec:context-ctrl-perspective} provides a control engineering perspective of CPS stress testing. 
In Section~\ref{sec:input-characterisation} we describe our characterisation of the reference values input space. We illustrate our testing approach in Section~\ref{sec:approach}.
Section~\ref{sec:experiments} reports on our empirical evaluation.
Section~\ref{sec:experimental-discussion} discusses the applicability and limitations of the approach.
Section~\ref{sec:related} presents the related work on the testing process for cyber-physical systems and control systems in particular.
Section~\ref{sec:conclusion} concludes the article and outlines directions for future work.
 \begin{figure}[tb]
    \centering
    \includegraphics{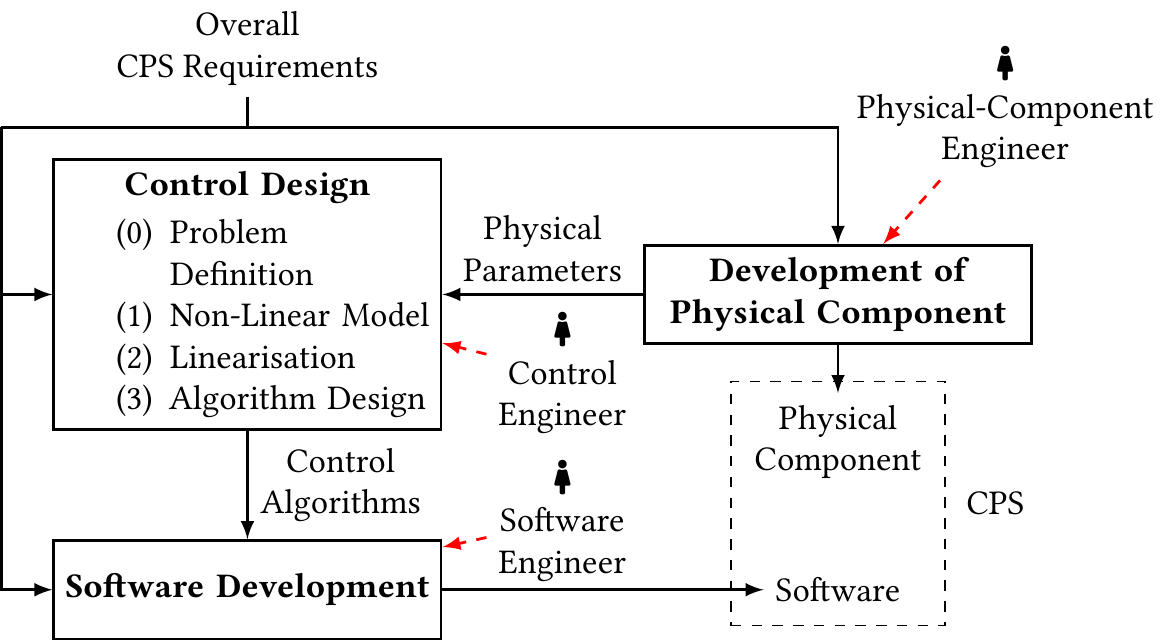}
    \caption{Simplified high-level description of the control system development; the red dashed arrows represent development steps, while the black solid arrows represent data flow.
    The development flow is simplified (e.g., neglecting iterations) to focus on the role of control engineering.
    }
    \label{fig:development-flow}
\end{figure}

\section{Context and Background}
\label{sec:context}
This section provides the relevant background on the development of control-based CPSs;
we use the position control of a drone as running example to exemplify the different concepts.
In such CPSs, the control layer receives desired values for the position of the drone, for example, the desired $x$, $y$, and $z$ coordinates.
It uses sensors to estimate the drone's current position and actuators to bring the drone to the desired one.
The section is divided in two parts.
First, in Section~\ref{sec:development-flow} we present the engineering process for developing a control-based CPS, and define the key roles involved in this process.
Then, in Section~\ref{sec:control-engineering-primer} we provide the relevant control engineering background needed to understand the remainder of this article.

\subsection{Development of Control-Based CPS}
\label{sec:development-flow}
Figure~\ref{fig:development-flow} provides a graphical overview of the typical development workflow of control-based CPSs.
This overview is simplified and focuses on the role of control engineers in the development of the software that handles the interaction between the physical and the cyber components.

The development of any engineering system starts with the definition of the system's requirements, denoted by \emph{Overall CPS Requirements} in the figure.
In the case of control-based CPSs, the requirements usually describe the desired behaviour of the \emph{physical} part of the system.
In our running example, the overall CPS requirements describe, at a high level, how the drone is supposed to move in space.
A concrete example of such a requirement could be that the drone must plan and follow a prescribed trajectory (at every point in time) with no discrepancy higher than a given threshold (e.g., one meter), and reach the desired position within a given time duration (e.g., three minutes).

These requirements are then made available to the engineers involved in the development of the CPS.
We identify three types of engineers involved, each having different duties.
In Figure~\ref{fig:development-flow}, we use red arrows to highlight their role in the development.
\begin{enumerate}[label=(\roman*)]
    \item \emph{Physical-Component Engineers} design the \emph{physical part} of the system. These could be for example mechanical or aerospace engineers depending on whether the considered CPS is a car or a drone.
    In our running example, an aerospace engineer may size up the propellers and draw the mechanical structure of the drone's body.
    \item \emph{Control Engineers} select and design the control algorithms, managing interactions between the physical and cyber parts.
    In the case of our running example, this means designing the algorithms in charge of estimating the drone's current position based on the sensors' readings, and computing the commands to be sent to the drone motors so that it can fly.
    \item \emph{Software Engineers} are in charge of the cyber part of the CPS.
    They implement the control algorithms, and design the functions that are needed to integrate the different software components (e.g., the high-level decision making in Figure~\ref{fig:cps-structure}).
    In our running example, there might be an aggressive and faster control algorithm when the drone is flying outdoors, compared to a safer one when the drone is flying indoors in a constrained space.
\end{enumerate}

In Figure~\ref{fig:development-flow}, we use black arrows to highlight the data and information flow during the development workflow. 
For instance, in our running example, the physical-component engineer communicates the data related to the propeller thrust to the control engineer, who then uses it in the \emph{control design step}, further described in Section~\ref{sec:control-engineering-primer}.
The result of the control design process is a set of algorithms that are provided to the software engineers, so that they can implement them and integrate them with the rest of the software necessary to fly the drone (e.g., the functions that perform initial checks, sensor data acquisition, communication with the motors).
When software engineers implement and integrate control algorithms, they make different choices that can alter the behaviour of control algorithms.
For example, a control algorithm can behave differently depending on how frequently and how much its inputs values (the references) are changed.
Vice versa, the design of the control algorithm (e.g., how it reacts to reference changes) will affect which implementations of the software fulfil the requirements and which do not.

\partparagraph{Consequences of Multidisciplinarity on the Testing Process}
The development workflow highlights the multidisciplinary nature of control-based CPSs.
More specifically, it highlights that control engineers play an essential role in developing control-based CPSs to bridge software implementation (i.e., software engineers' concern) and physical components development (i.e., physical-component engineers' concern).
This multidisciplinarity impacts the testing of such systems~\cite{Mandrioli:2023}.
In particular, when testing the CPS control layer, we must account for the control algorithms implemented in the control layer.
Indeed, the control layer includes control algorithms designed by the control engineers which are however implemented and integrated with other functionalities (e.g., mode switches) by the software engineers.
Furthermore, since requirements of CPS usually concern the behaviour of the physical component, testing must necessarily include the latter in the setup.
In other words, testing the software in isolation has limitations since it does not account for the interactions between the software and physical components, which  limit the design scope.
Therefore, engineers must account not only for the software implementation but also such interactions, which requires multidisciplinary considerations, including both control and software engineering.

\subsection{Control Engineering Primer}
\label{sec:control-engineering-primer}
In this section, we introduce the definition of a control design problem and the control design process.\footnote{The content of this section is mostly based on the book ``Feedback Systems: an Introduction for Scientists and Engineers''~\cite{Astrom:2008}.}
We illustrate the frequency-domain description of signals and systems (alternative to the time-domain description), and introduce the basic concepts used in the remainder of this article.

\partparagraph{Definition of Control Design Problems}
As mentioned in Section~\ref{sec:introduction}, the objective of the control layer in a CPS is to steer physical quantities to track desired values.
More rigorously, the input of the control layer is the vector $r$ of desired values (also called \emph{references}).
The control objective is to ensure that the actual values in the physical system are as close as possible to the corresponding reference values.
Using the control terminology, we say these physical quantities constitute an \emph{output} vector $y$, and the control objective is $y\approx r$ (i.e., $y$  tracks the reference values in $r$).

To achieve its objective, the control layer uses sensors to iteratively measure signals from the physical part of the system, and actuators to steer it.
Based on measurements and reference values, the control algorithm computes the commands to be sent to the actuators.
The control algorithm is executed repeatedly, at constant time intervals, resulting in a continuous interaction between cyber and physical components.
This interaction is called \emph{control loop}.
The control loop creates a cause and effect cycle for which the output of the control algorithm (the actuation) affects the physical components and therefore its own input (the sensor readings).
Because of this cause and effect cycle, the cyber and physical components are said to be in \emph{closed loop}.

In the drone example, the control objective is to use the propellers to move the drone following a reference trajectory.
The software iteratively (i) uses sensors to measure quantities like its own acceleration every millisecond, (ii) executes the control algorithm, and (iii) actuates the motors spinning the propellers by sending voltage commands.
On the physical side, the propellers generate forces that cause the drone movement, and in turn affect the future acceleration readings, leading to the closed-loop interaction.

As $r$ and $y$ are vectors, the engineers usually define multiple control loops, often one for each element of the vectors.
In a drone, we can expect to find one loop for each of the three dimensions in the space that the drone can move in: forward or backward, left or right, and up or down.
Furthermore, the CPS requirements might call for different control modes, such as fast (but risky) flight mode, and a safe (but slower) one.
In the most common control design approaches (like proportional-integral-derivative control, also known as PID control), the design of the different control loops and control modes is addressed separately: engineers develop a dedicated control algorithm for each mode.
The implementation can switch between the algorithms of the different modes  during execution.

The identification of the control loops and modes is the \emph{control problem definition}, and constitutes a preliminary step in the control design process.
For each of the identified loops and modes, the control design problem is reduced to the design of the control algorithms needed to control the physics.
Each control algorithm, when executed in closed loop with the physical component, is expected to guarantee (to the best degree possible) that the output $y$ tracks the reference $r$ when operating in the mode it is designed for.
We now discuss the control design process for one individual loop and mode.

\partparagraph{Control Design Process}
As highlighted inside the control design block in Figure~\ref{fig:cps-structure}, the design of a control algorithm comprises of three main steps.
The first step is to define an equation-based model of the physical component.
The role of this model is to provide a representation of how the actuators affect the measurements and output.
For example, in the drone, this is a model that represents how the thrust generated by the propellers affects the position and orientation (called attitude) of the drone itself.
These models can be obtained using either first-principle approaches, through the laws of physics, or with data-driven approaches through system identification~\cite{bittanti:2019}.
Either way, the models are typically in the form of non-linear differential equations, i.e., non-linear equations that contain both signals and their derivatives (rate of change).
Non-linear models are difficult to analyse, as small changes in the input can cause significantly different behaviours and hence they do not allow for approaches that apply with generality to a variety of systems~\cite{khalil:2002}.

To overcome the complexity of non-linear models, the second step of the control design is to approximate them using linearised models.
Retrieving such approximated models is called \emph{linearisation}, and restricts the model scope to the surroundings of an expected operating point.\footnote{
    Linearising a non-linear equation means approximating its non-linear relations (e.g., if a variable is squared, like in the case of aerodynamic drag) with a linear function.
    The linear function is based on the first derivative of the non-linear relation, and more specifically, on the value of the first derivative in the chosen operating point.
}
Accordingly, the operating point is chosen as the physical state around which we expect the system to operate most of the time.
In the drone, the operating point would be the horizontal state in which the drone is parallel to the ground and not tilted in any direction.
Through linearisation, we then obtain a model that is still valid for small variations in the attitude angles around the operating point.
The model is now a set of ordinary linear differential equations and therefore it can be handled in a simpler way thanks to a large variety of analytical tools~\cite{Astrom:2008}.

The third step is finally the control algorithm design.
To design such algorithm, control engineers use control theory.
To this end, they make different assumptions about the system (e.g., that the physical component is sufficiently close to the assumed operating point).
For models based on linear differential equations, control theory provides numerous tools to perform exact analyses and design control algorithms with formal performance guarantees.
Such tools are based on a frequency-domain description of the physical system.
Frequency-domain descriptions are well-suited for treating ordinary differential equations because they provide a compact description for the derivative of a signal with respect to the signal itself.
This makes it easier to analyse the physics and draw conclusions on the system's properties.
The control algorithms obtained using control theory are also in the form of linear differential equations.

\partparagraph{Frequency-Domain Descriptions}
We now provide a high-level description of the frequency-domain for both signals and systems, together with the intuition of why the frequency-domain is well suited for analysing and manipulating differential equations.

The frequency-domain description of signals is based on the fact that signals can be decomposed and treated as the sum of sinusoidal functions with different frequencies.
The description in the frequency-domain specifies which sinusoidal components are present in the signal and what their amplitude is.
This is in contrast to the time-domain, where signals are represented as a sequence of values over time.
The frequency-domain sinusoidal components are commonly called \emph{frequency components}: for example, a fast-changing signal is mostly composed of fast (i.e., high-frequency) sinusoids.
On the contrary, a signal that does not change much is mostly composed of slow (i.e., low-frequency) sinusoids.

\begin{figure}[tb]
    \includegraphics{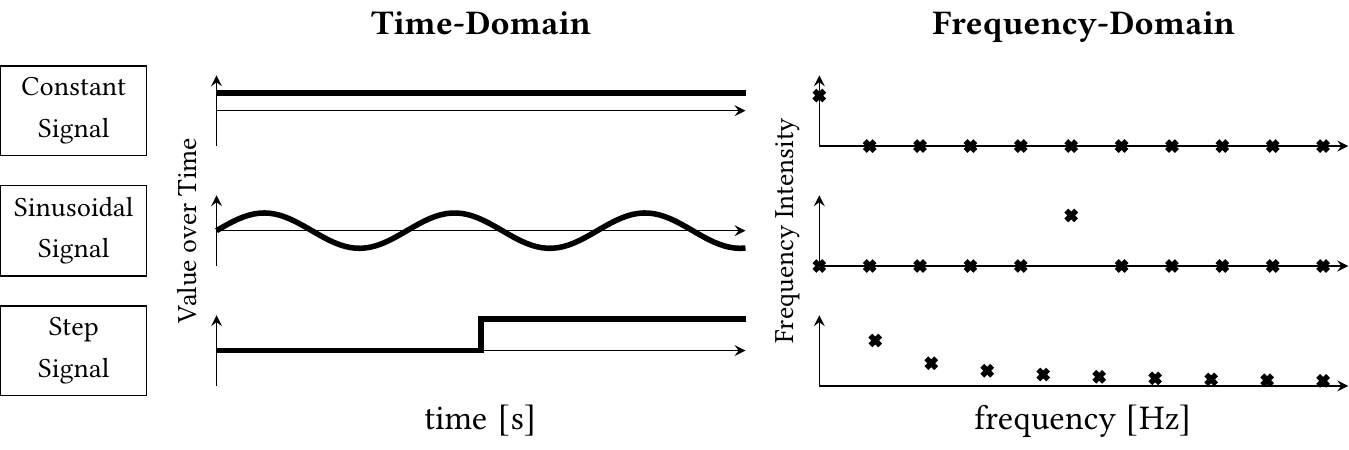}
    \caption{Examples of DFT spectra (on the right) of different signals defined in the time-domain (on the left).
    A constant signal is described by only a zero frequency component, a pure sinusoidal maps to one single frequency component, and a non-periodic step maps to multiple frequencies.
    }
    \label{fig:dft-examples}
\end{figure}

The translation of a signal from the time-domain representation to its frequency-domain one uses the \emph{Fourier Transform}---or its time-sampled equivalent \emph{Discrete Fourier Transform} (DFT~\cite{fourier-algo}), which we will use in the remainder of this article.
Figure~\ref{fig:dft-examples} shows three examples of time-domain signals and their frequency-domain representations obtained with the DFT.
The first row shows a constant signal, whose frequency representation consists of a single sinusoidal wave at $\SI{0}{\hertz}$.
The second row shows a pure sinusoidal signal that is mapped by the DFT into a single frequency component.
More complex signals, like the step function in the third row, include a larger number of frequency components.

The frequency-domain representation provides a description of signals according to their rate of change, or frequency content.
The derivative of a signal is another signal that also describes its rate of change.
This similarity can be seen as the intuitive reason why the frequency-domain description is convenient for analysing differential equations.

\begin{figure}[t]
    \includegraphics{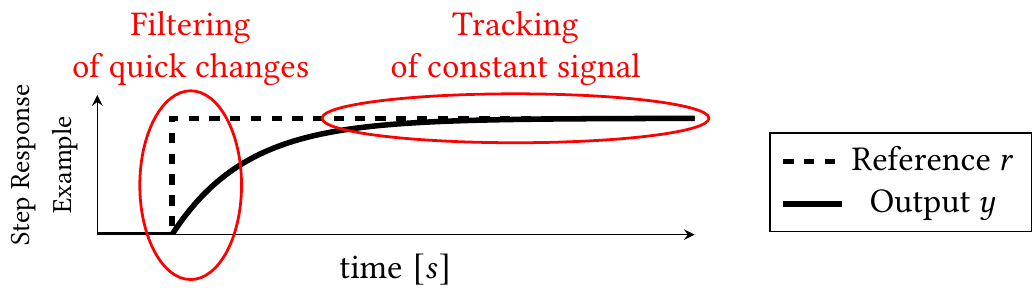}
    \caption{Example of how we can expect the CPS output $y$ (solid line) to track the desired value $r$ (dashed line).
    In the figure we intuitively highlight how a control system usually behaves like a low-pass filter, by filtering the fast-changing components of the input and tracking the slow-changing ones.
    }
    \label{fig:step-response-example}
\end{figure}

In the frequency-domain, systems (i.e., entities that take an input signal and generate an output signal) are described by how much they react to an input according to its frequency content, and more specifically by how much they amplify or reduce every frequency component of the input signal.
This is measured by the ratio between the intensity of a given frequency component in the output and in the input.
If the output over input ratio is smaller than $1$ we observe a reduction; conversely, if it is greater we observe amplification.
When such ratio is equal to $1$, we have unit amplification, which corresponds to a frequency component not being altered by the system.
For example, many physical systems behave like a low-pass filter, transmitting or amplifying slow signals (i.e., low-frequency components), while reducing quickly changing components (i.e., high-frequency components).
This behaviour of reducing a frequency component is called \emph{filtering}.

In the case of control systems, we ideally want the physical quantity $y$ to track the reference $r$ at every time instant.
In the frequency-domain this corresponds to unit amplification between the reference and the output at every frequency.
This is apparently infeasible, since a change in $r$ (a quantity in the cyber component that can be arbitrarily set) requires some time for $y$ (a quantity of the physical world that has to obey the physics laws) to follow.
In other words, reference changes that are too rapid cannot be tracked by the output of a system.
In Figure~\ref{fig:step-response-example} we show an example of how we could expect the drone to follow a step-like change in the desired position along one direction. The output $y$, denoted by the solid line, does not follow instantly the reference $r$, denoted by the dashed line.
On the contrary, the quick change of reference value is smoothed in the output signal, which gradually reaches the new reference value.
The frequency-domain concept of filtering can be used to describe this phenomenon.
In fact, we could rephrase this as ``the output only tracks the slowly changing (low-frequency) components of the reference, while it filters the fast-changing (high-frequency) ones''.
We illustrate this intuitive interpretation in Figure~\ref{fig:step-response-example} with red ellipses.
The ellipse on the left highlights the filtering behaviour that occurs when the reference signal has a rapid change, while the one on the right highlights the tracking behaviour when the reference signal does not change.

Given this behaviour consisting of tracking the low frequencies and filtering the high frequencies, for a given control loop, we can identify the so-called \emph{closed-loop bandwidth}, denoted by $f_b$.
The closed-loop bandwidth is the threshold frequency below which we expect to have a tracking behaviour (i.e., $y\approx r$) and above which we have a filtering behaviour.
This quantity therefore corresponds to the fastest frequency components of the reference that the control system is able to track.
Accordingly, it is also considered a quantification of the speed of the control system: the higher the closed-loop bandwidth, the higher the rate of change of reference signals that it can track.
We remark that such threshold, the closed-loop bandwidth, is a property of the system and should not depend on the specific input.

While it may seem intuitive that the control engineer wants to design a control algorithm that maximises $f_b$, to obtain tracking of fast-changing references, there are other factors to account for.
As an example, a high value of $f_b$ usually comes at the cost of a control actuation with a high value possibly leading to hardware damages, e.g., a fast-moving drone will generate high forces that can ruin the actuators.
Furthermore, noise can be found in the measurements at high frequencies,e.g., an accelerometer that measures a drone's acceleration is usually affected by high-frequency electrical noise.
Therefore, if the control system reacts to input signals in the high-frequency range (providing high speed), it will also react to noise.
In turn, this will reduce the system performance making its behaviour unpredictable.
Such considerations lead to a trade-off in the control algorithm design between speed of the system and noise rejection.
 \section{Control Engineering Perspective on CPS Stress Testing}
\label{sec:context-ctrl-perspective}
In this section, we first motivate and define the problem of stress testing the control layer in a CPS.
Second, based on the development workflow of control-based CPSs illustrated in Section~\ref{sec:context}, we identify the types of assumptions that the engineers make in the different control software development stages.
For each of these types, we discuss which techniques are already available for testing the corresponding assumptions, and which assumptions require an application-specific solution; we exclude the latter from the scope of this work.
For the remaining types we discuss which assumption types should be tested first.
We conclude this section by defining the problem addressed in this article.

\subsection{Problem Motivation}
\label{sec:problem-motivation}
As discussed in Section~\ref{sec:control-engineering-primer}, when applying control theory, control engineers make \emph{design assumptions} about both the physical part of the CPS and the control algorithm to be developed.
The role of these assumptions is to abstract away the details of the system that are not necessary for developing control algorithms, and to define the fundamental building blocks (e.g., linear models) used by the theory.
For example, when designing the control algorithm of a drone, engineers assume that the generated thrust is proportional to the voltage command and that it does not saturate when the maximum power of the motors is reached.
This allows engineers to use a linear model of how the voltage, when applied to the motors, affects the drone movement and position.
This linearity assumption is necessary to apply traditional control theory~\cite{Astrom:2008}.

As discussed in Section~\ref{sec:development-flow}, software engineers are provided with the control algorithms from control engineers.
These algorithms are only one component of the control layer.
In fact, when implementing the latter, software engineers address the implementation of other functionalities such as the flight mode changes, the interaction with sensors and actuators (e.g., filtering and sanity checks), the parallel execution of the different control loops and the discretisation of the equations.
The implementation and integration of other software functionalities can render the design assumptions made by control engineers invalid.
For example, the linear model used for the control design is an assumption that is falsified when the drone motors saturate. This happens because the motors are requested to produce more thrust than their capacity, as in the cases when the reference value changes too much or too fast.
In such scenarios, the drone control algorithm will be operating in conditions different from the ones assumed during design.

When testing the CPS implementation, software engineers have to be aware of possibly unpredictable software behaviour due to falsified design assumptions.
For example, when testing the ability of a drone to perform a delivery, software engineers have to consider the possibility that the motors may saturate, thus leading to unpredictable behaviour.
This implies that the flight performance of the drone might be impaired, potentially affecting the overall fulfilment of the requirements concerning the safe execution of the delivery.
Conversely, software implementation choices can determine for which scenarios the design assumptions hold or not.
For example, the update mechanism of the drone position reference values (e.g., based on external inputs or regular periods) can result in faster or slower reference changes, which can cause the saturation of the motors (and the consequent loss of guarantees that hold under design assumptions).
Therefore, when generating test cases for the CPS control layer (as well as for the CPS overall) and evaluating the tests outcomes, software engineers have to consider the potential impact of falsified design assumptions.

Control algorithms are usually robust (at least to some degree) to the falsification of the different design assumptions; this property is one of the reasons for the successful adoption of control theory~\cite{Astrom:2008}.
Control theory provides metrics to quantify the algorithm robustness to the deviation from assumptions, e.g., ``stability margins''.
However, those metrics are also based on the control design models and are still subject to the validity of the assumptions.
Hence, \emph{the quantification of the extent to which a CPS can be pushed outside of its design scope cannot be provided a priori by control theory and intrinsically requires empirical approaches or, in other words, testing.}
This evaluation can be obtained through stress testing of the software that implements the control layer, by targeting the control design assumptions.
Therefore, such a type of testing is ideally performed on the final implementation of the CPS.
However, this is generally expensive in both time and resources.
As a common alternative, implementation details (of both software and physical component) can be added to the simulation models and those can be used in place of the final implementation for testing their impact.
The more exhaustive and adherent the additions are to the real implementation, the more the testing on the simulation model will be relevant to the CPS implementation.

\subsection{Design Assumptions in Control Algorithms}
\label{sec:design-assumptions}
Before defining our stress testing problem, we need to identify the types of design assumptions that control engineers make at design time. These assumptions are made at the different development stages of a control-based CPS.
In Section~\ref{sec:development-flow} we identified three main development stages:
\begin{enumerate}[label=(\roman*)]
    \item control problem definition,
    \item control algorithm design, and
    \item control algorithm implementation.
\end{enumerate}
We now discuss the design assumptions made at each stage.

\partparagraph{Assumptions at Control Problem Definition Time}
At this stage the engineers identify the different control loops and modes for which they will develop a control algorithm.
As a consequence, when designing the individual control loops, they assume that (i)~the different loops do not interfere with each other and (ii)~the mode changes do not impact the control design that follows~\cite{Astrom:2008}.
For example, for a drone, the design of the altitude controller may not account for the horizontal controllers (and vice versa).
Similarly, the design of the ``aggressive flight'' controllers is done independently from the ``safe flight'' controllers.
Such assumptions significantly simplify the design of the control algorithms, allowing, among others, to independently design the response to a change in each element of the vector $r$.
However they do not always hold in practice.
For example, when the drone tilts to move horizontally, it also loses vertical thrust, affecting the altitude controller.
Another case of assumptions not holding is when a mode change command is issued during the flight.
This can cause a sudden change in the motors' commands, thus affecting the CPS performance.

\partparagraph{Assumptions at Control Design Time}
During the control design, the engineers develop a non-linear model of the physical part of the CPS, based in part on the information received from the physical-component engineers that designed it.
Such a model (like any model) is only an approximation of reality and will not consider or only approximate certain aspects of the problem.
For example, a drone model assumes a given mathematical relation between the rotational speed of the propellers and the generated vertical thrust.
However, this type of aerodynamic phenomena is difficult to quantify. Moreover, there could be some inconsistency between the mathematical model and the real physical system.
By using such models, the engineers implicitly assume that they are a sufficiently accurate representation of the physical reality.

As mentioned above, the models of the physics also need to be linearised in order to use the tools from control theory.
The linearised version of the model is only valid in the surroundings of the operating point chosen for the linearisation.
Practically, by using the linearised model, the engineers implicitly assume that, during operations, the CPS stays sufficiently close to the operating point so that the linearised model is an accurate enough representation of the physical part.
For example, the propellers cannot generate more thrust than the motors can provide: the motors saturate (max-out) once they reach their maximum capacity.
To linearise this relation, the engineers assume that the motors are not in the saturated state, and that they always provide a thrust proportional to the voltage command.
When, during the actual flight, the motors saturate, this proportional relation loses validity, as well as the model assumed during the control algorithm design.

\partparagraph{Assumptions at Control Algorithm Implementation Time}
Control algorithms are generally specified as linear differential equations.
Such equations are defined with the use of continuous mathematics.
However, they are implemented on computers which are discrete machines.
Hence they have finite precision in the representation of the parameters and execute the algorithms in discrete steps over time.
Accordingly, the engineers, when designing the control algorithm with continuous mathematics, are implicitly assuming that the discretisation happening during the implementation does not significantly alter the algorithm.
More specifically, they assume both that the finite precision does not significantly alter the computed values, and also that the discrete execution does not alter the frequency properties (meaning the properties of the algorithm execution over time).

\partparagraph{Design Assumptions Summary}
To summarise, we identify the following types of design assumptions that are made by engineers during the development of control algorithms:
\begin{enumerate}[label=\textbf{A\arabic*}]
    \item \label{asm-loop} they ignore the interaction between different control loops;
    \item \label{asm-mode} they ignore the impact of mode changes on the control algorithms' performance;
    \item \label{asm-modl} they assume that the initial non-linear model of the physics is a sufficiently accurate representation of the real system;
    \item \label{asm-linr} they assume that the system stays sufficiently close to the operating point chosen for the linearisation so that the linearised model is valid;
    \item \label{asm-finite} they assume that the finite precision of the representation of the equation variables and parameters is adequate; and
    \item \label{asm-time} they assume that the execution in discrete time steps does not significantly affect the expected execution time properties of the algorithm.
\end{enumerate}
When performing stress testing for a control-based CPS, engineers should aim at falsifying each of these assumptions.

We note that there are branches of control engineering that aim at mitigating each of those simplifying assumptions, such as multivariable control (targeting~\ref{asm-loop}) and robust control (targeting~\ref{asm-modl}).
However, like the stability margins mentioned above, such approaches are still subject to design assumptions and do not exclude the need for empirical verification.
Furthermore, those are rather advanced theories and, as of now, find limited application in practice~\cite{Desborough:2002}.
We now discuss which types of assumptions can be already stress tested with existing software testing techniques and which ones require an application-specific solution.

Testing numerical properties of numerical algorithms (Assumptions~\ref{asm-finite} and~\ref{asm-time}) is not a  novel problem; there is a significant literature corpus~\cite{Yi:2017, He:2020}, also targeting control algorithms~\cite{Sanchez:2018, Magnani:2021}.
Similar considerations can be made about testing execution timing properties.
A number of works can be found in the literature for testing real-time software~\cite{Bozhko:2021,Lu:2012}.
Furthermore, we note recent works dedicated to the verification and testing of the robustness of control algorithms to execution timing faults~\cite{Ghosh:2022, Vreman:2021}.
Given the above previous works, we leave the testing of numerical and timing properties out of the scope of this article.

Testing the validity of the physical model (Assumption~\ref{asm-modl}) is a highly application-specific problem.
To test the aspects of the physics model that are unknown, one must know the aspects that were uncertain when developing it.
For example, for a drone, two assumptions of the model can be on the aerodynamic properties of the propellers (needed to evaluate the vertical thrust that can be generated) and on the rigidity of the drone body (to simplify the equations describing the motion of the drone in space).
Among those, the former is likely to be associated to a higher degree of uncertainty because the aerodynamic phenomena are generally hard to characterise.
In contrast, the assumption on the rigidity is more likely to be valid: intuitively, we do not expect the drone motor supports to bend.
Such considerations are clearly application-specific and require an understanding of the specific model that is being used.
Accordingly, the generation of test cases that falsifies this type of assumptions cannot be treated in a general fashion.
Given its application-specific nature, we leave the testing of this type of assumptions out of the scope of this work.

We are therefore left with the assumptions regarding non-interactions between control loops~(\ref{asm-loop}) and control modes~(\ref{asm-mode}), and about the sufficiently large range of validity of the linear models~(\ref{asm-linr}).
Among those we note that the first two depend on the latter.
In fact, if an individual control loop does not have a sufficiently large range of validity when operating independently (i.e., without mode switches and in the absence of reference changes for the other loops), then, the switching across different modes and the interaction between loops are unlikely to improve its range.
For example, if we have an altitude control loop for a drone that is not very robust when operating alone, then it is unlikely to perform better when the control loops of the horizontal directions are also active and can disturb it.
Given this dependency, we argue that \emph{testing the validity of linear models should occur before testing the interactions between control modes and control loops}.
In light of this discussion, this work focuses on the testing of the linearised model control design assumptions, for which we give our problem statement below.

\subsection{Problem Statement}
\label{sec:problem-statement}
In this article, we address the problem of stress testing the implementation of a CPS control layer.
In the control layer of a CPS there are several control loops; we focus on the problem of stress testing the linearised model design assumptions for the implementation of the individual loops.
Our SUT is therefore an individual loop characterised by its reference value input and physical quantity output.
We assume available basic information about the control-loop design and implementation, namely an estimate of the closed-loop bandwidth from the control design and a range of valid values for the reference.
The \emph{objective} is to generate test cases that falsify the linearity design assumption, and identify when the falsification pushes the control algorithm out of its design scope (and hence causes the loss of the control-theoretical guarantees).
To falsify the linearity assumption we have to generate test cases where the physical component behaves differently from the (linear) model used during the control algorithm design.
This difference in the behaviour should occur at various degrees and should increasingly make the control algorithm unable to provide the control-theoretical guarantees.
The tests thus expose the robustness of the control algorithm to the different degrees of  falsification of design assumptions.
Accordingly, our \emph{test inputs} exercise the control layer and consist of sequences of reference values over time.
Our \emph{test outputs} are the traces of the physical quantity that has to track the reference value.

To address the generation of stress test cases we use control-domain knowledge to qualitatively characterise the input space of a control loop.
In order to make the qualitative characterisation usable in practice, we propose a novel approach to the parametrisation of test cases for control-based CPSs based on the frequency-domain, in contrast to traditional time-domain approaches.
Using such parametrisation, we can then introduce a novel metric to identify the stress test cases as well as different metamorphic relations that describe the expected behaviour patterns across different test cases.
Leveraging the proposed parametrisation, metric, and metamorphic relations, we then develop a stress testing approach for the different control loops of a CPS control layer.
 \section{Control Loop Input Space Characterisation}
\label{sec:input-characterisation}
In this section, we use domain knowledge from control theory to provide a \emph{qualitative characterisation} of a single control loop input space.
The proposed characterisation maps frequency and amplitude features of the input to the expected behaviour, i.e., the expected relation between the (scalar) output $y$ of a control loop and the (scalar) input reference $r$.
In the first part, we present the qualitative characterisation based on the validity boundaries of the linearised model and insights from control theory.\footnote{
    We note that a similar qualitative characterisation of the input space of a control loop is found only in one book on control engineering from $1959$~\cite{Gille:1959}.
    However, the treatment of the topic is brief and high-level and has not been investigated further in later literature.
}
We use a minimal example (a simplified model of the altitude control of a drone) to exemplify the different system behaviours highlighted by the characterisation.
In the second part of this section, we list the qualitative aspects of the characterisation and propose approaches to quantify them.
Such a quantification enables the practical use of the characterisation, and constitutes the basis for the test case generation approach proposed in the following section.
We conclude the section by discussing our problem statement in the context of the proposed characterisation.

\begin{figure}[t]
    \centering
    \includegraphics{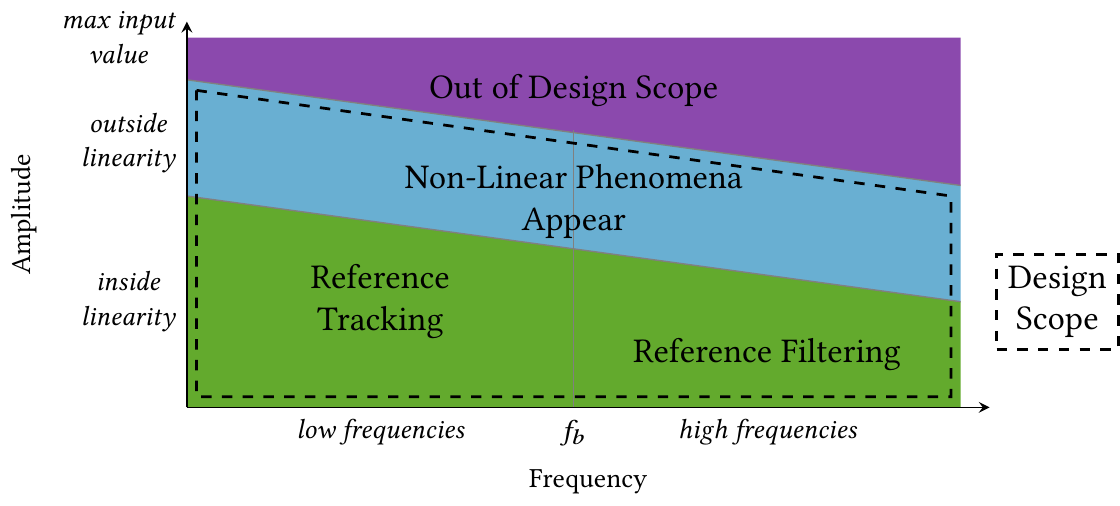}
    \caption{Qualitative frequency-amplitude characterisation of the input space of a control loop.
    The colours highlight the validity of the linearised model with respect to the input frequency content and amplitude.
    The green area corresponds to the input signals for which the system remains within the assumptions of control theory.
    The azure area corresponds to the input signals that trigger non-linear phenomena but not enough to cause significant performance degradation in the system.
    The purple area corresponds to input signals for which control theory assumptions are not fulfilled and the behaviour of the system becomes unpredictable.
    We highlight the design scope (i.e., the union of the green and azure areas) with a dashed box.
    }
    \label{fig:input-space}
\end{figure}

\subsection{Qualitative Input Space Characterisation}
\label{sec:input-characterisation-qualitative}
In order to leverage domain knowledge from control theory (Section~\ref{sec:control-engineering-primer}), we base our characterisation on a frequency-domain description of the input sequences.
This means describing the input space with two dimensions: one capturing the input frequency content and the second one capturing its amplitude.
Being a two-dimensional space, the input space can be represented as a \emph{frequency-amplitude plane}.
We provide a graphical representation of this input space plane in Figure~\ref{fig:input-space}: one input sequence corresponds to one or more points on the plane according to its frequency content and its amplitude.
We can use knowledge from control theory to identify different areas in the input plane according to the expected behaviour of the control loop, depicted by colours and boundaries in Figure~\ref{fig:input-space}.
We identify these areas by checking where control-theoretical guarantees apply (i.e., checking the validity of the linear model), and where they exhibit a tracking and filtering behaviour within the applicability boundaries of control theory.
In the figure, we draw the areas with {\em simplified} boundaries represented as straight lines (the dashed lines in the figure).
Such lines are apparently a simplification and, in practice, the boundaries might not be represented as straight lines nor provide a clear division between the different areas.
In the last part of this section (Section~\ref{sec:input-characterization-benefits}) we discuss in more detail the benefits and limitations of the proposed characterisation.

\begin{figure}[tb]
    \centering
    \includegraphics{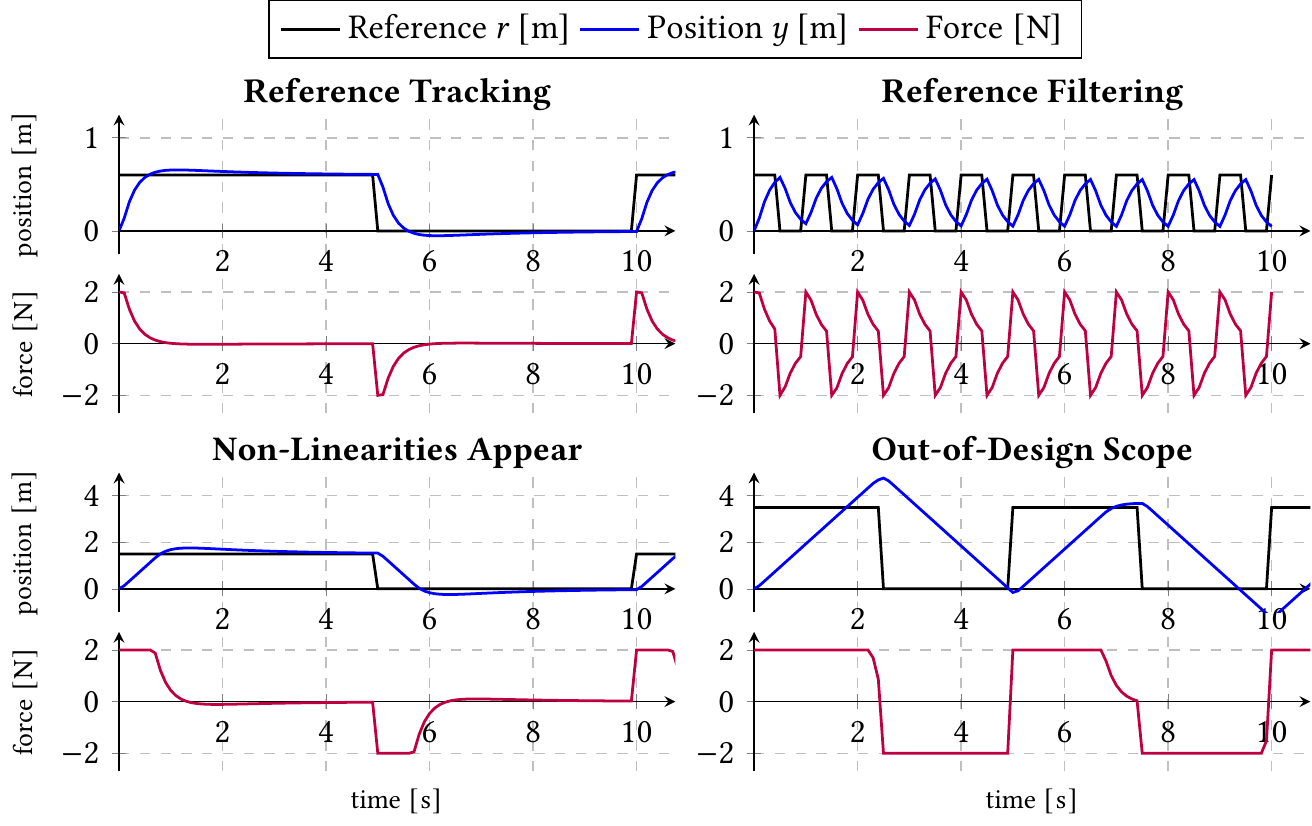}
    \caption{
    This figure exemplifies the four different behaviours highlighted in the qualitative characterisation of the input space of a control loop.
    The plots are based on the simulation of a minimal model of the altitude control of a drone.
    The only source of non-linearity in this example is the saturation of the motor that limits (between $\pm\qty{2}{\newton}$) the thrust that can be generated by the motors.
    The blue lines show the altitude of the drone (i.e., the output $y$), the black line shows the desired altitude (i.e., the input $r$), and the red line in the bottom plots shows the thrust generated by the motors.
    Appearance of non-linear phenomena can be detected when the red line saturates at $\pm\qty{2}{\newton}$.
    }
    \label{fig:bh-examples-time}
\end{figure}

In order to exemplify the different behaviours highlighted in the characterisation, we use a minimal example of the altitude control of a drone.
To enable the easy detection of the limitations of the linear model, we use a simulator based on a linear model and introduce one single source of non-linearity: the saturation of the thrust.
This saturation limits the force that can be applied by the motors to move up and down the drone.
In Figure~\ref{fig:bh-examples-time}, we report four executions showing the response to square waves with different amplitude and period values.
For each execution, the top plot shows the desired altitude (the black line) and the actual altitude (the blue line).
As discussed, the desired altitude is the input reference, and the actual altitude is the output physical quantity.
The bottom plot shows instead the command sent to the motor that is used to accelerate or decelerate the drone (the red line).
The saturation of the motor (and hence the validity of the linear model) can be detected in these plots when the force becomes fixed at $\pm\qty{2}{\newton}$ (see the bottom-left and bottom-right plots).

We first discuss the upper plots that concern flight simulations where motor saturation does not occur and that therefore are within the CPS design scope.
We then discuss the bottom plots, where saturation occurs.
Such tests are at the boundaries or outside of the CPS design scope.
We conclude the description of Figure~\ref{fig:input-space} discussing its connection with our definitions of design scope and stress tests.

\partparagraph{Behaviours Within the Design Scope}
Inputs consisting of small amplitude values will not push the CPS far away from its operational point.
Accordingly, for lower amplitude values we are within the validity bounds of the linear model: this is represented by the green area in Figure~\ref{fig:input-space}.
Within this area, we expect the system to be able to track the slow-changing inputs: those sequences correspond to low-frequency inputs inside the ``\emph{\textbf{Reference Tracking}}'' area.
Fast-changing signals map instead to high-frequencies and are not expected to be tracked: those belong to the ``\emph{\textbf{Reference Filtering}}'' area.
In the figure we highlight the closed-loop bandwidth $f_b$ that separates the tracking and filtering areas.
To exemplify the reference tracking and filtering behaviours, in the upper plots of Figure~\ref{fig:bh-examples-time}, we feed the controller of the drone with a slower (in the left-hand side plot) and a faster square wave (in the right-hand side plot), with an amplitude value equal to $\qty{0.6}{\meter}$.
In the former we can see that the reference is successfully tracked within seconds after a step change.
In the latter the reference changes are too fast and the drone cannot follow it successfully: we say that it is filtered.

\partparagraph{Behaviours at the Boundary and Outside of the Design Scope}
When we consider input signals that are larger in amplitude, the CPS moves further away from the operational point used for the linearisation and non-linear phenomena start to appear.
In the altitude controller example this corresponds to hitting the motor saturation.
Accordingly, in the bottom-left plot of Figure~\ref{fig:bh-examples-time}, we feed the drone with a larger square wave with an amplitude equal to $\qty{1.5}{\meter}$.
As we can see from the plot of the control action, the motor saturates for some time after the occurrence of the  step in the reference.
This, however, does not significantly affect the way that the actual altitude of the drone follows the desired reference, i.e., the reference is still successfully tracked.
Since the reference is successfully tracked, we can consider this test to be still within the design scope of the CPS, despite the occurrence of motor saturation.
In other words, the control algorithm is showing some robustness to the motors being saturated over a limited amount of time.
Accordingly, in Figure~\ref{fig:input-space}, such signals correspond to the azure area ``\emph{\textbf{Non-Linear Phenomena Appear}}'' which we consider part of the design scope as the CPS behaviour is not impaired.
When the system moves even further away from the design scope, the linearised models are falsified even more and there is no way to predict the system behaviour.
This is the ``\emph{\textbf{Out-of-Design Scope}}'' purple area.
For example, in the bottom-right plot of Figure~\ref{fig:bh-examples-time}, we can see that the drone is not only unable to track the square wave with an amplitude value of  $\qty{3.5}{\meter}$, but it also exhibits a new behaviour, a triangular wave.

Finally, we note that either large (high amplitude) or fast-changing (high frequency) inputs can lead the system out of its design scope.
For example, in the drone altitude control, either a fast-changing input or a large input can require high thrust and hence can cause motors saturation.
Furthermore, fast change combined with a large inputs lead to a compounded effect on the validity of the linear model. 
Accordingly, in Figure~\ref{fig:input-space}, we depict the thresholds for which non-linear behaviours appear and the bound of the design scope (i.e., the green area) as decreasing when frequencies increase.

To summarise, given the qualitative plot of Figure~\ref{fig:input-space}, the design scope is identified as the union of the green and azure areas; we highlight this area  using a dashed box.
The system is instead considered out of the design scope when the occurrence of non-linear phenomena affects the CPS behaviour, which corresponds to the purple area.
Accordingly, the stress tests are the tests that cover the azure area and especially its boundary with the purple area.
Those are the tests where the non-linear phenomena appear and become large enough to affect the CPS behaviour, thus requiring empirical verification.

\subsection{Qualitative Aspects of the Characterisation and their Quantification}
\label{sec:input-space-quantification}
We aim to use the qualitative characterisation proposed in the previous section to generate test cases that falsify the linearised model used for the algorithm design, and hence push the control layer to its performance limits.
In other words, we want to sample (test) points in the frequency-amplitude plane and identify the behaviour that the test results expose in various areas of the plane.
In order to stress test the CPS, we want to sample around the border of the  ``out-of-scope'' area to understand when the control algorithm is no longer able to provide the performance guarantees.
Furthermore, we also want to identify the border between ``tracking'' and ``filtering'' behaviours in order to characterise the fastest signals that the control algorithm can track.

We note that the latter distinction between the ``tracking'' and ``filtering'' areas (i.e., the closed-loop bandwidth) is not related to the falsification of the design assumptions (both areas are in fact coloured green).
However, it represents how fast the control loop can track a reference and hence represents a performance limit of the system.
If we want to push the system to its performance limits, then we have to ensure that the test cases cover both behaviours.

Accordingly, in order to make our qualitative characterisation usable for the generation of stress test cases, we have to make quantitative the following qualitative aspects:
\begin{itemize}\item One input sequence generally contains more than one frequency, and hence can map to more than one point in the frequency-amplitude plot.
    Accordingly, we need to define a \emph{mapping between a test input sequence to a corresponding set of frequency-amplitude coordinates} in the input plane.
    This enables the identification of which points in the frequency-amplitude plane are sampled by a test (Section~\ref{sec:mapping}).
    \item Detecting when a test trace shows a behaviour altered by the falsification of the linearised model---when a test belongs to the purple area---does not have a formal definition in the existing literature.
    Accordingly, we need to define a ``\emph{degree of non-linearity}'' observed in a given test result.
    Such a degree of non-linearity should capture the impact of non-linear phenomena on the CPS behaviour.
    In other words, we want to detect test cases that belong to the purple area and that are outside of the scope of the design assumptions (Section~\ref{sec:dnl-def}).
    \item Since a test can map to multiple frequency-amplitude points, it can expose multiple behaviours simultaneously.
    Accordingly, we need to define a \emph{mapping between the different behaviours observed in a given test and its frequency-amplitude points} (i.e., the different coordinates mentioned above).
    This enables the distinction of the different behaviours (tracking, filtering, out of scope) that might appear in the same test (Section~\ref{sec:bh-map}).
    \item It is practically impossible to know a priori (1)~the actual shape of the threshold for which non-linear phenomena start to appear, and (2)~the threshold for which they falsify the linear models to a sufficient degree to impair the system performance.
    Hence, the bounds between the areas of different colours can have an arbitrary shape.
    However, we can \emph{use the relative positioning of the behaviour areas to define a set of properties that are expected to hold when comparing the behaviours exposed by test cases with different frequency and amplitude content} (e.g., filtering will appear at higher frequencies than tracking).
    This enables the definition of test case generation strategies in the frequency-amplitude plane that explore the different behaviours of the control loop (Section~\ref{sec:properties}).
\end{itemize}

\begin{figure}[t]
    \includegraphics{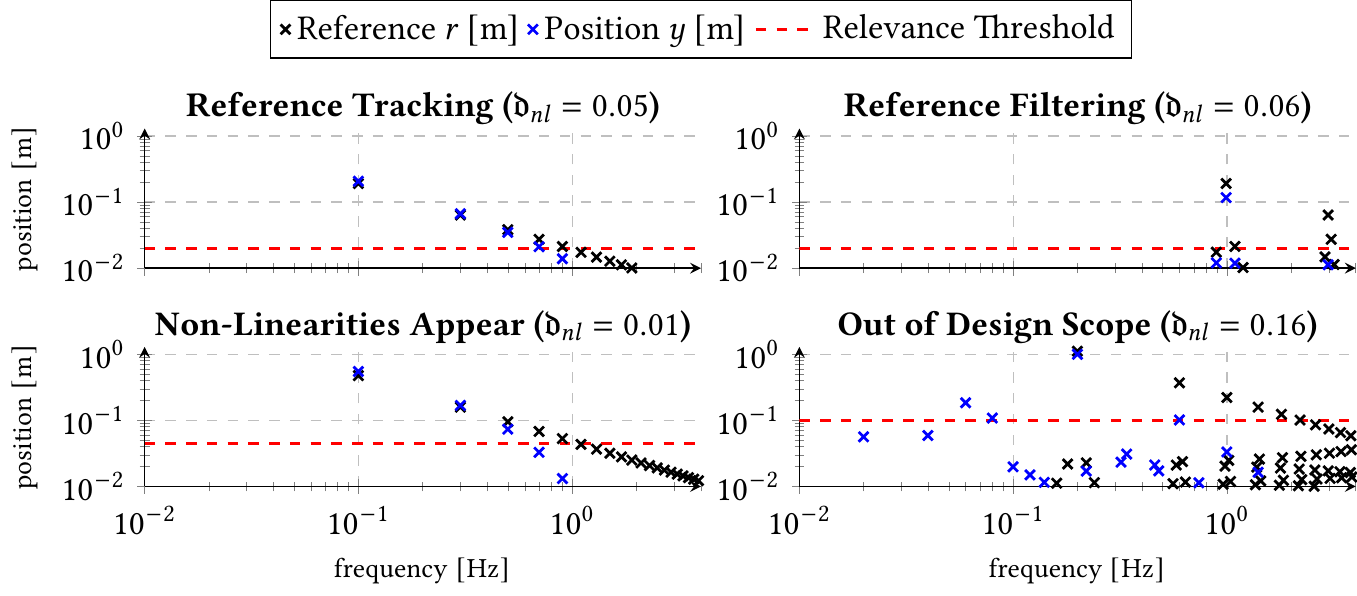}
    \caption{
    DFT of the input (black crosses) and output (blue crosses) traces from the tests on the minimal example of the drone altitude control shown in Figure~\ref{fig:bh-examples-time}.
    These plots exemplify how we use the frequency content of the input and output to detect the different behaviours of control systems.
    The filtering behaviour is detected when input components are not found in the output (top-right plot).
    Non-linear behaviour is detected when new frequency components are found in the output (bottom-right plot).
    Furthermore, the red dashed line highlights the threshold that we adopt to identify the relevant input components used for the mapping of the test case to the frequency-amplitude plane.
    }
    \label{fig:bh-examples-freq}
\end{figure}

We now address the definition and quantification of each of these qualitative aspects.
Since we leverage the DFT of the input and output of the tests,  in Figure~\ref{fig:bh-examples-freq} we show the frequency-domain representation for the tests of Figure~\ref{fig:bh-examples-time}.
The figure uses the same colour convention as its time-domain equivalent: blue crosses represent the frequency components of the trace of the actual position of the drone, and black crosses represent the frequency components of the input sequence.
We use the plots of the DFT to exemplify the different definitions and explain the underlying intuitions.
We remark that, based on common practice in the frequency-domain, we use a logarithmic scale on both the axes of all the plots in Figure~\ref{fig:bh-examples-freq}.
This enhances the readability of the plots.

\subsubsection{Mapping of Tests to Frequency-Amplitude Points}
\label{sec:mapping}
Given an input reference sequence $r\funof{t}$, we want to define the frequency-amplitude coordinates sampled with the associated test.
We define this mapping according to the frequency spectrum (the DFT) of the input reference.
In practice, inputs are signals sampled over time: therefore, the spectrum computed with the DFT is also discrete~\cite{fourier-algo}.
More specifically, the time-domain samples are mapped to an equal number of evenly spaced frequency components, like in the DFT examples in Figure~\ref{fig:dft-examples}.
The number of the frequency components is therefore large: for example, a $5$ seconds trace sampled every millisecond is mapped to $5000$ frequency components.
However, most of those components are usually equal to zero or close to it, meaning that only few of the frequency samples actually carry information about the signal.\footnote{
    The reason for such excess of samples is that signals are usually oversampled (sampled more frequently than strictly necessary) for redundancy and robustness.
    This oversampling introduces extra frequency components in the higher part of the spectrum that do not carry much information about the signal and are therefore zero or close to it.
}
Accordingly, among all of the frequency components computed with the DFT, we consider as relevant only the ones with larger amplitudes.
Formally, given an arbitrary input $r\funof{t}$, we map it to a set of frequency-amplitude coordinates $\langle f,A\rangle$:
\begin{equation}
    \mathit{fAmap}[r\funof{t}] = \{ \langle f,A\rangle:\, A=|\DFT[r\funof{t}]\funof{f}| \land A>\rho \cdot \max_f\{|\DFT[r\funof{t}]\funof{f}|\} \} ,
    \label{eq:famap}
\end{equation}
where $\DFT[\cdot]$ denotes the DFT, $|\cdot|$ denotes the modulus,
and $\rho$ is a parameter in the range $[0,1]$ that we use to select the relevant components (i.e., the larger ones) in a relative way to the largest one: $\max_f\{|\DFT[r\funof{t}]\funof{f}|\}$.
We exemplify the definition of this relative threshold (used for selecting the relevant input components) with the red dashed line (for $\rho=0.1$) in the plots of Figure~\ref{fig:bh-examples-freq}.
We can then observe in Figure~\ref{fig:bh-examples-freq} that: the faster square waves of the two right-hand side plots map to points further to the right on the frequency axis than the other tests associated with slower square waves.
Furthermore, the larger amplitude values of the square waves from the bottom plots map to points higher in the amplitude axis than the other tests associated with smaller square waves.
This exemplifies how the size and the speed of the inputs are captured in the frequency-domain.

\begin{figure}[tb]
    \includegraphics{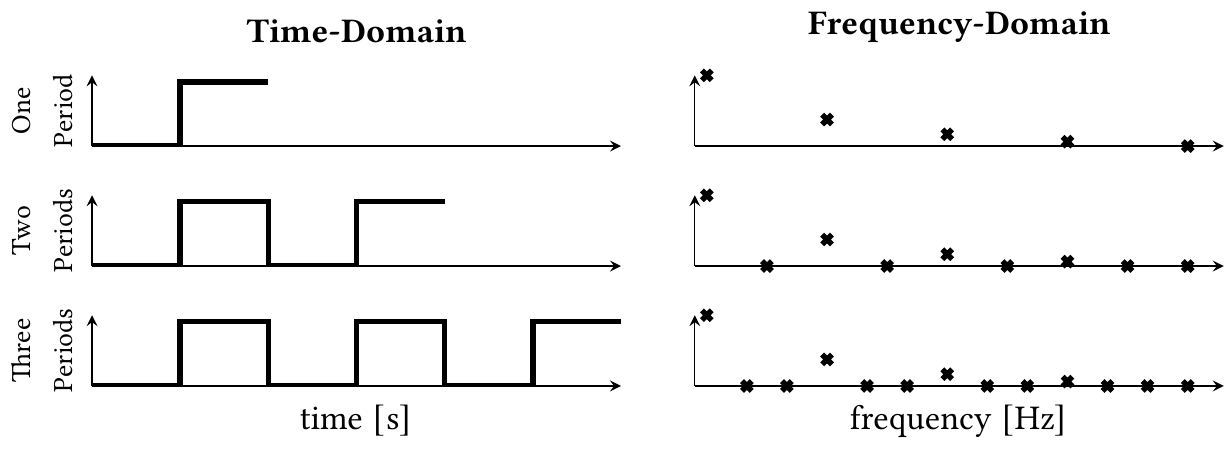}
    \caption{Graphical representation of how the repetition of the same input sequence increases the resolution in the frequency-domain.
    The repetition of the same sequence does not introduce new information to the input signal, whose non-zero frequency components remain unchanged.
    However, when the input is repeated, the output signal is also sampled in frequencies outside of the main input components (the ones that have a value equal to zero in this plot) hence enabling the detection of frequency spectrum broadening because of non-linear behaviour.
    }
    \label{fig:fft-example}
\end{figure}

\subsubsection{Degree of Non-Linearity Definition}
\label{sec:dnl-def}
With the degree of non-linearity, we want to detect when non-linear phenomena impact the CPS behaviour.
As exemplified in the bottom-right plot of Figure~\ref{fig:bh-examples-time}, when non-linear phenomena push the CPS out of its design scope, they introduce unexpected behaviours in the output.
This kind of behaviour can be harmful as it implies that the control algorithm is introducing some new behaviour in the system that was not part of the reference.
For example, in the bottom-right plot of Figure~\ref{fig:bh-examples-time} the altitude of the drone reaches $\qty{4.5}{\meter}$ when the reference is at $\qty{3.5}{\meter}$.
The bottom-right plot of Figure~\ref{fig:bh-examples-freq} shows the DFT of the input and output of the test exhibiting non-linear behaviour in our drone altitude control example.
In this example, we remark the presence in the output (the blue crosses) of components in the frequency range $[0.02, 0.4]\unit{\hertz}$ that were not present in the input (the black crosses).

According to this intuition, we define the degree of non-linearity on the base of new output frequency components that appear outside of the input spectrum.
This requires \emph{sampling the output in between the frequencies of the input components}.
To increase the number of samples in the frequency spectrum without altering the frequency content of the input we can repeat the input sequence (i.e., make it periodic).
This  does not alter the information contained in the input (since it is just repeated), and the non-zero frequency components do not change.
However, for each repetition we double the samples in the time-domain, resulting in  doubling the samples in the frequency-domain.
The new samples obtained in this way are found on the frequency axis \emph{between} the samples previously available (rather than only in the higher part of the spectrum, as for the samples introduced by the oversampling).
We exemplify graphically the sampling of new frequencies in Figure~\ref{fig:fft-example}.
The figure shows how repeating the steps in the time-domain (plots on the left-hand side) increases the number of samples in the frequency-domain (plots on the right-hand side) by adding new zero-valued frequency samples in between existing samples, hence increasing the sampling resolution.

When identifying the impact of non-linear phenomena on the CPS, we are interested in detecting \emph{frequency components that were small in the input and become large in the output}.
Hence, we define the degree of non-linearity according to the \emph{maximum amplitude} of the output spectrum outside of the relevant components of the input spectrum.
We look therefore at all the frequencies sampled in the DFT of the output $y$, excluding the relevant ones found in the input (i.e., $f\mathit{Amap}[r\funof{t})]$).
Formally, given a reference sequence $r\funof{t}$, we can define the set of frequencies to check as $f_{\mathit{new}}=\{f:\exists \DFT[r\funof{t}]\funof{f} \land \neg\exists(\langle f,\ast\rangle \in \mathit{fAmap}[r\funof{t}])\}$, where $\ast$ can be any amplitude value.
For example, in the tests of Figure~\ref{fig:bh-examples-time} it means that we are looking at the frequencies that are \emph{not} associated with any input frequency component (black cross) above the red dashed line. Such an amplification of frequency components only appears in the bottom-right plot, where the saturation time is sufficiently long to affect the CPS behaviour.
Differently, in the bottom-left plot, saturation occurs but not for long enough to alter the CPS behaviour.
Furthermore,  to obtain comparable results across tests with different amplitudes, we normalise our metric with respect to the amplitude of the input $\max_f\{|\DFT[r\funof{t}]\funof{f}|\}$ (remarking that the sampled frequencies are necessarily the same for input and output across the tests).
The intuition is that a deviation from a large reference change matters less than the same deviation from a small reference change.
We then obtain the following definition for the \textbf{degree of non-linearity} of a given test $i$
\begin{equation}
    \dnl\funof{i}=\frac{\max_{f \in f_{\mathit{new}}} \{|\DFT[y_i\funof{t}]\funof{f}|\}}
                      {\max_f\{|\DFT[r_i\funof{t}]\funof{f}|\}},
\label{eq:dnl}
\end{equation}
where $r_i\funof{t}$ and $y_i\funof{t}$ are respectively the input and output associated with the test and the other elements follow the same conventions as in previous equations.
When we apply the $\dnl$ formula to the examples in Figure~\ref{fig:bh-examples-freq}, we obtain the values reported in the titles of the different plots.\footnote{
    In this example, we used ten repetitions of the step sequence to compute the $\dnl$.
}
By comparing the numbers, we see that the new frequencies that appear in the bottom-right plot cause the $\dnl$ to be one order of magnitude higher than the $\dnl$ of the other tests.
This example also shows that having a new frequency component with $16\%$ of the maximum input amplitude (caused by non-linear phenomena) can significantly impact the system's performance.

We exemplify the use of this metric with our examples in Figure~\ref{fig:bh-examples-freq}.
As indicated in the plots titles, for the tests where the saturation is not triggered, we measured a $\dnl$ of $0.05$ and $0.06$.
In the bottom-right test that shows misbehaviour, we desirably obtain a higher value of $0.16$ for our metric.
Finally, in the bottom-left plot we measure a $\dnl$ value of $0.01$, hence lower than the one in the upper plots.
Such a low value might seem counter-intuitive given that in this test the saturation is triggered but, thanks to the algorithm robustness, in this case the saturation does not alter the behaviour of the CPS by introducing new potentially dangerous components in the output.
Therefore, such a low $\dnl$ value allows us to distinguish when the occurrence of non-linear phenomena, actually pushes the CPS out of its design scope (bottom-right test) from when it does not (bottom-left test).
Instead, the fact that the $\dnl$ value is even lower than the value of the upper plots can be attributed to numerical noise.
Indeed, in each of the three tests, no large amplitude is found in the $f_{\mathit{new}}$ set of frequencies: amplitude values are small or close to zero and small variations caused by numerical noise can alter the values.
However, when non-linear behaviour appears and larger amplitudes are found in $f_{\mathit{new}}$, the numerical noise becomes less relevant.

To conclude, we note that the idea of repeating the input sequence comes with a trade-off between the test duration and the frequency resolution.
Higher frequency resolution increases the chances of detecting new frequency components, hence non-linear behaviours; however, more repetitions require longer tests.
The number of frequency samples needed to detect new frequencies in the output  depends on the specific application.
Accordingly, such a number can be evaluated empirically using a manual test, with several input repetitions, that shows non-linear behaviour (this test can be obtained using high amplitude and frequency content, and an arbitrary shape).
Then, by computing the $\dnl$ using different trace lengths (corresponding to different numbers of repetitions), we can evaluate how many periods are needed so that the frequency samples are sufficient to detect new undesired output components.\footnote{
    We exemplify the procedure of selecting the number of input repetitions in the experimental part of this work, specifically in Section~\ref{sec:experimental-settings}.
}

\subsubsection{Mapping of Behaviour to Frequency-Amplitude Points}
\label{sec:bh-map}
When we analyse the degree of non-linearity, we obtain a metric that characterises all the input frequency-amplitude components.
In fact, by looking at the $\dnl$ metric, there is no way to identify which input component causes the non-linear behaviour.

Differently, when the SUT behaves linearly, some frequency components of the same input are tracked and pass through to the output, while other components are filtered.
For example, in the top-right plot of Figure~\ref{fig:bh-examples-freq} we can observe that the input component at frequency $\qty{1}{\hertz}$ is found, albeit reduced in amplitude, also in the output, while the component at frequency $\qty{3}{\hertz}$ has a much lower amplitude value in the output.
Therefore, when we quantify the filtering behaviour of tests exposing linear behaviour, it is necessary to analyse the different frequency components individually.

For each of the frequency-amplitude component of the input, we define a \emph{degree of filtering} based on the ratio of the output and input amplitudes.
If a frequency component is perfectly tracked, its amplitude in the input and output are equal, hence yielding a ratio equal to $1$.
Instead, if the ratio of the input over the output is below the unit, it corresponds to filtering since part of the signal is lost.
Analogously, values above $1$ indicate amplification: while a small amplification can be expected in real-world systems, large amplification can be dangerous (for the very same reason as the risks of broadening the frequency spectrum).

Hence, given a test $i$, we define the degree of filtering \emph{for a given input frequency component} $f$ as the difference between $1$ and the mentioned output-input ratio:
\begin{equation}
    \dof\funof{i,f}=1-\frac{|\DFT[y_i\funof{t}]\funof{f}|}{|\DFT[r_i\funof{t}]\funof{f}|},
\end{equation}
with the same conventions as for the equations above, and the caveat that this definition is valid only for tests that expose linear behaviour.
Given the absolute values at the numerator and denominator, this metric takes values in the range $[-\infty, 1]$.
A value of $1$ describes complete filtering, while $0$ describes perfect tracking.
Negative values correspond to input amplification.

\subsubsection{Expected Properties of the Characterisation}
\label{sec:properties}
Due to the qualitative nature of the characterisation presented in Figure~\ref{fig:input-space}, we cannot use it to predict the exact shape of the different behaviour areas.
However, we can use the relative positioning of the areas to define properties over the tests.
More specifically, the areas' relative positioning shows that
\begin{enumerate*}[label=(\arabic*)]
    \item tests with higher frequency content and larger amplitudes are expected to push the CPS further away from the operation point used for the linearisation and cause non-linear behaviour, and
    \item tests with higher frequency content are expected to expose more filtering behaviour than ones at lower frequencies.
\end{enumerate*}
We therefore formulate the following expected properties:\footnote{
    In Section~\ref{sec:metamorphic-relations}, we formalise the properties into metamorphic relations.
}
\begin{enumerate}[label=\textbf{PR\arabic*}]
    \item \label{prop:dnl} The degree of non-linearity should increase when the amplitude and frequency values increase.
    In fact, the further we move away from the origin of the frequency-amplitude plane the closer we should be to the input area outside of the design scope.
    \item \label{prop:dof} For tests within the design scope (i.e., tests that show linear behaviour), the filtering degree should increase as the frequency increases.
    In other words, faster signals should always be harder to track than slower ones.
    \item \label{prop:bnd} The closed-loop frequency bandwidth should be independent of the specific test.
    In fact, when the system behaves linearly, the threshold between the tracking and filtering areas should not depend on the specific input and be instead a property of the system~\cite{Astrom:2008}.
\end{enumerate}

We now exemplify an evaluation of these properties on the tests of Figures~\ref{fig:bh-examples-time} and~\ref{fig:bh-examples-freq}.
Property~\ref{prop:dnl} is fulfilled by all tests except for the bottom-left one.
In fact, the test with the highest amplitude (the bottom-right one) is the one with the largest $\dnl$ and the one with the second highest $\dnl$ is the one with the fastest input (the upper-right one).
However, when we compare the two plots on the left-hand side, we would expect the bottom one to have a higher $\dnl$ than the upper one, since the latter receives an input with the same frequencies but smaller amplitude.
The exception of the bottom-left plot showcases that, when the linear model loses validity (in this case because of saturation), the behaviour of the system becomes unpredictable, though not necessarily worse as illustrated by the low $\dnl$ value.
This unpredictability further underlines the importance of testing in the areas of the input space that are close to the design assumptions validity bounds.
Property~\ref{prop:dof} is fulfilled: looking at the tests exhibiting linear behaviour, we can observe that the blue crosses move further down in the position axis from the black ones as we move to higher frequencies.
This means that the reference signal is found increasingly less often in the output.
Concerning Property~\ref{prop:bnd}, we can observe, in the tests shown in the top part of the figure, that the frequency above which the input is filtered in the output (i.e., the closed-loop bandwidth) is similar for both tests (around $\qty{0.9}{\hertz}$).
In other words, the $f_b$ measured from the two tests in the upper plots is similar and thus the two tests comply with the third property.
For the bottom-left test, the frequency at which the input is filtered is lower  (around $\qty{0.6}{\hertz}$).
This is possibly due to the appearance of the saturation that limits how large and fast references the drone can track (hence practically decreasing the closed-loop bandwidth for larger amplitudes).
Given that in the real world we would not necessarily know that the bottom-left test is triggering some non-linear phenomenon, this discrepancy from the two upper tests can be used to highlight that this test might require a more detailed analysis (even though the control performance is possibly still acceptable).

\subsection{Benefits and Limitations of the Frequency-Amplitude Characterisation}
\label{sec:input-characterization-benefits}
Our testing objective is to generate and identify stress test cases that push the system around the limits of validity of the linearised model.
In test cases where non-linear phenomena appear, the control-theoretical guarantees are gradually lost (depending on how far the behaviour of the CPS deviates from the one described by the model) and new unpredictable frequency components appear in the output.
Accordingly, we propose the use of the $\dnl$ metric, as a measure of the new frequency components that appear in the CPS output, to identify the stress test cases and therefore whether a test is inside or outside of the design scope.
In other words, $\dnl$ captures and quantifies non-linear behaviour.
In order to specifically explore the \emph{validity limits} of the design assumptions, we are interested in test cases where the $\dnl$ is non-zero (hence not being fully within the design scope) but also not too large (hence not being far outside of the design scope).

However, the identification of a boundary in the input space where the $\dnl$ transitions from zero to non-zero values is not possible for arbitrary input sequences.
In fact, as discussed above, the $\dnl$ metric equally characterises all of the frequency-amplitude components of a given test.
Consequently, a given point in the frequency-amplitude plane can lead to different $\dnl$ values depending on the other components of the input.
For example, in both the two right-hand side tests presented in Figure~\ref{fig:bh-examples-freq}, we can identify a main component around $\qty{1}{\hertz}$ and amplitude $\qty{0.2}{\meter}$.
The two components are located very close in the plane but are associated with two tests that yield different $\dnl$ values (and accordingly different behaviours).
The test depicted in the upper plot shows linear behaviour and is within the design scope, while the test in the bottom plot is outside of the scope.
This discrepancy appears because the two components belong to different tests and are therefore coupled with different other components.
This further highlights that Figure~\ref{fig:input-space} is a simplified representation of the input space with clear-cut boundaries.
Depending on the other frequency-amplitude components that one point in the frequency-amplitude plane is paired with in a given test, it can map to different behaviours (i.e., $\dnl$ values).
The exposed behaviour of a point in the plane can in fact change according to which other frequency components are present in the same input.

To overcome the above limitation, in our testing approach we propose a novel test-case parametrisation for the control layer of CPSs.
With this parametrisation we separate, in a test case, the definition of
\begin{itemize}\item the frequency content,
    \item the amplitude content and
    \item the combination (i.e., the relative positioning) of the different frequency-amplitude components.
\end{itemize}
As shown in the next section, thanks to this separation, we can leverage the proposed characterisation to control the expected behaviour of the tests.
 \section{Testing Approach}
\label{sec:approach}
In the first part of this section, we propose a novel test case parametrisation for the control layer of CPSs.
The objective of the proposed parametrisation is to enable the definition of test cases according to the concepts presented in the previous section.
More specifically, we design our parametrisation with the following objectives:
\begin{itemize}\item it allows the use of the degree of non-linearity $\dnl$ (Equation~\ref{eq:dnl}),
    \item it distinguishes possible choices for the different input features (frequency content, amplitude content and the combination of different frequency-amplitude components), and
    \item it enables the definition of the expected properties from Section~\ref{sec:properties} as metamorphic relations~\cite{Ayeredi:2021,Chen:2019}.
\end{itemize}
In the second part of this section we define the mentioned metamorphic relations. In the third part, we use our test case parametrisation and the metamorphic relations to develop our testing approach.
We use the $\dnl$ and the expected property~\ref{prop:dnl} to obtain a preliminary optimistic bound on the input amplitudes that the control layer can track.
Using the preliminary bound, we use our heuristic sampling method to generate test cases that explore the area of the frequency-amplitude plane where stress test cases are likely to appear.
We then execute the test cases, and use the $\dnl$ and the $\dof$ metrics, together with the metamorphic relations, to identify test cases outside of the design scope.

\subsection{Test Case Parametrisation}
\label{sec:parametrisation}
One test case $i$ corresponds to a sequence $r_i\funof{t}$ of reference values over time $t$.
In order to achieve the parametrisation objectives listed above, we define $r_i\funof{t}$ as a function defined by three elements: an amplitude gain, a time scaling coefficient, and a periodic shape function.
More specifically, we use the expression
\begin{equation}
    r_i\funof{t}=\AGain_i\,\shape_i\funof{\tGain_i t},
\label{eq:tc-parametrisation}
\end{equation}
where:
\begin{itemize}[itemsep=1mm]
    \item[$\shape\funof{\cdot}$] is a periodic function defining the input shape.
    For example, it can define a square wave or a triangular wave (arbitrary sequences can be chosen as long as they are periodic).
    Without loss of generality, we assume that shape functions are normalized to have a unit period (i.e., $\forall t : \shapefcn{t}=\shapefcn{t+1}$) and also have a unit amplitude range (i.e., $\max_t\shapefcn{t}-\min_t\shapefcn{t}=1$).
    \item[$\AGain$] is a gain used to scale the input amplitude.
    \item[$\tGain$] is a time scaling coefficient that changes how quickly we go through the input shape.
\end{itemize}
One test case $i$ is therefore fully defined by a triplet $\langle\shape_i,\AGain_i,\tGain_i\rangle$.
We note that, as long as $\shape$ is a periodic function, the sequence $r_i\funof{t}$ is also periodic.
Hence it is a repeated sequence and enables the use of the proposed $\dnl$ (Equation~\ref{eq:dnl}).

\partparagraph{Sampling of Frequency-Amplitude Coordinates}
We now use the reference square waves from the examples in Figures~\ref{fig:bh-examples-time} and~\ref{fig:bh-examples-freq} to exemplify how the proposed parametrisation works and how it enables intuitive sampling of the frequency-amplitude plane.
More specifically, we describe how the parametrisation separates the choice of the relative positioning of the different main frequency-amplitude components and of their frequency and amplitude values.
The four tests from the figures can be defined using our parametrisation with the $\shape$ function being a square wave with unit period, and the reference position switching between $0$ and $1$.
The $\AGain$ value corresponds to the amplitude of the square wave and, hence, the tests in the upper plots both have an amplitude gain of $0.6$, while the bottom plots have an amplitude gain of $1.5$ and $3.5$, respectively.
Concerning the values of $\tGain$, the left-hand side plots' period is $10$ and hence have a time scaling of $0.1$ to make the signal slower.
The right-hand side plots have instead periods equal to $1$ and $5$, respectively, and hence corresponding time scaling values of $1$ and $0.2$.

The shape function $\shape$ defines the relative positioning of the different $\langle f,A\rangle$ test coordinates.
In other words, it defines their pattern (e.g., combination of larger and smaller components) in the frequency-amplitude plane.
However, thanks to the linearity of the DFT, this pattern is independent of the $\AGain$ and $\tGain$ coefficients.
For example, if we look at the main frequency-amplitude components of the inputs in Figure~\ref{fig:bh-examples-freq} (the black crosses above the red dashed line), we can see that the different square waves all map to components along a decreasing straight line.\footnote{
    Square waves map to a decreasing straight line in logarithmic scale.
    In linear scale they map to a hyperbole, as exemplified in Figure~\ref{fig:fft-example}.
}
This showcases that the relative positioning of the main components is defined only by the shape function and is independent of the scaling coefficients $\AGain$ and $\tGain$.

The $\AGain$ coefficient enables the movement of the $\langle f,A\rangle$ test coordinates along the amplitude axis.
Increasing (decreasing) its value makes the frequency-amplitude components map to components further up (down) in the plane.
Thanks to the linearity of the DFT, multiplying the value of $\AGain$ proportionally increases the main components' amplitude coordinate $A$~\cite{fourier-algo}.
For example, comparing the two left-hand side plots of Figures~\ref{fig:bh-examples-time} and~\ref{fig:bh-examples-freq}, we can see that the two square waves with the same period map to components with the same frequency but scaled amplitude.

The $\tGain$ coefficient enables the movement of the $\langle f,A\rangle$ test coordinates along the frequency axis: i.e., the choice of the horizontal coordinate $f$.
Analogously to the amplitude, increasing (decreasing)  its value increases (decreases) the speed with which we go through the shape function and thus its frequency content.
Due again the linearity of the DFT, a scaling of this coefficient corresponds to an equivalent scaling of the main components' $f$ coordinates.
For example, comparing the two upper plots of Figures~\ref{fig:bh-examples-time} and~\ref{fig:bh-examples-freq} we can see that the two square waves that have the same amplitude but different periods map to points with the same amplitude but respectively lower and higher frequency.

\subsection{Definition of Metamorphic Relations}
\label{sec:metamorphic-relations}
The properties introduced at the end of Section~\ref{sec:input-space-quantification} describe relations between different tests.
For example, we used such properties to describe the relation between the four tests in Figures~\ref{fig:bh-examples-time} and~\ref{fig:bh-examples-freq}.
Properties that concern the inputs and outputs of multiple test cases are known as metamorphic relations (MR)~\cite{Chen:2019}.
We define one MR for each property.
We define the first two MRs as implications over tests with the same shape.
Accordingly, each one contains a \emph{condition} and an expected \emph{implication}.
The conditions concern relations between the amplitude and frequency content of different test inputs as well as whether the output shows non-linear behaviour or not.
The implications concern relations between the $\dnl$ and $\dof$ observed in the tests outputs.
In contrast, we define the third MR as a property across tests with different shapes.

Using our test case parametrisation, for tests based on the same shape, we can directly identify relations in the input amplitude and frequency content using parameters $\AGain$ and $\tGain$.
More specifically, given two tests $i$ and $j$ with the same shape (i.e., $\shape_i=\shape_j$), the relation between $\AGain_i$ and $\AGain_j$ ($\tGain_i$ and $\tGain_j$) identifies their relation in terms of amplitude (frequency) content.
For example, $\AGain_i>\AGain_j$ implies that test $i$ maps to points higher in the frequency-amplitude plane with respect to points for test $j$.
Analogously, $\tGain_i>\tGain_j$ implies that test $i$ maps to points further to the right in the frequency-amplitude plane with respect to points for test $j$.

Property~\ref{prop:dnl} states that an increase in amplitude or frequency content of the input should cause an increase in $\dnl$.
Thanks to our characterisation, for two tests $i$ and $j$ based on the same shape ($\shape_i=\shape_j$), an increase in amplitude and frequency are identified with $\AGain_i>\AGain_j$ and $\tGain_i>\tGain_j$.
Their conjunction can be used to define the MR condition.
The MR implication instead concerns an increase in the degree of non-linearity.
This can be expressed as $\dnl\funof{i} > \dnl\funof{j}$.
Hence, we can write the following MR:
\begin{equation}
    \text{\textbf{MR1:}}\,\,
        \funof{\shape_i=\shape_j} \land \funof{
            \funof{\AGain_i>\AGain_j} \land
            \funof{\tGain_i\geq\tGain_j} \lor
            \funof{\AGain_i\geq\AGain_j} \land
            \funof{\tGain_i>\tGain_j}
        }
    \implies
        \dnl\funof{i} > \dnl\funof{j}.
\end{equation}
As described in Section~\ref{sec:properties}, this MR is, for example, satisfied by the upper-left and bottom-right tests of Figure~\ref{fig:bh-examples-freq}, where the increase of $\tGain$ from $0.1$ to $0.2$ and the increase of $\AGain$ from $0.6$ to $3.5$ lead to an increase in $\dnl$ from $0.05$ to $0.16$.
Differently, it is not satisfied by the tests of the left-hand side plots, where an increase of $\AGain$ from $0.6$ to $1.5$ and same $\tGain$ correspond to a decrease in $\dnl$ from $0.05$ to $0.01$.

Property~\ref{prop:dof} states that, as long as the tests show linear behaviour, an increase in the frequency content will correspond to an increase in the filtering behaviour.
To identify linear tests, we use a threshold $\dnlth$ on our degree of non-linearity.
Tests below such threshold ($\dnl\funof{i}<\dnlth$) are therefore considered to show linear behaviour and we limit the definition of this MR to those.
Like for the previous MR, we identify an increase in the frequency content for two tests with the same shape as a greater time scaling parameter $\tGain_i>\tGain_j$.
The conjunction of the $\dnl$ threshold and the increase in frequency content constitute the MR condition.
For the MR implication, an increase in the filtering behaviour can be identified with an increase of the $\dof$.
However, different from the $\dnl$, the $\dof$ applies to each main frequency-amplitude component of the test.
Hence we need to evaluate it for each component of the inputs, i.e., $\forall f | \langle f,A\rangle \in \mathit{fAmap}[r_i\funof{t}]$, and compare it with the corresponding component of the other test, i.e., the one with frequency $f\cdot\frac{\tGain_j}{\tGain_i}$ (leveraging the linearity of the DFT, the proportion in the $\tGain$ coefficients matches the proportion in the frequencies).
Using this quantification of the implied  increase in $\dof$, we obtain
\begin{equation}
    \text{\textbf{MR2:}}\,\,\,\,
\begin{array}{cc}
        \funof{\shape_i=\shape_j}    \land
        \funof{\tGain_i>\tGain_j}    \land
        \funof{\dnl\funof{i}<\dnlth} \land
        \funof{\dnl\funof{j}<\dnlth}
    \implies \\
        \forall f | \langle f,A\rangle \in \mathit{fAmap}[r_i\funof{t}],
            \dof\funof{i,f}>\dof\funof{j,f\cdot\tGain_j/\tGain_i} .
\end{array}
\end{equation}
For example, this MR is satisfied by the upper tests of Figure~\ref{fig:bh-examples-freq}, where the increase of $\tGain$ from $0.1$ to $1$ corresponds to an increase in $\dof$ for the main frequency component (respectively the ones at $\qty{0.1}{\hertz}$ and $\qty{1}{\hertz}$).
In fact, the blue cross (representing the output frequency component) moves further down along the amplitude axis with respect to the input component, hence highlighting filtering action.
More precisely, the $\dof$ value of the main component of the upper-left plot is $0$ at $\qty{0.1}{\hertz}$ and the $\dof$ value of the upper-right plot is $0.45$ $\qty{1}{\hertz}$.

Property~\ref{prop:bnd} states that, as long as the tests show linear behaviour, the closed-loop bandwidth should not depend on any specific test.
We can use the $\dof$ metric to identify the closed-loop bandwidth.
According to its control-theoretical definition, the closed-loop bandwidth $f_b$ corresponds to a ratio of $0.5$ between output and input~\cite{Astrom:2008}.
Therefore, the bandwidth can be identified as the threshold below (above) which the frequency components show a $\dof$ lower (higher) than $0.5$, determining the reference tracking (filtering) area.
To complement the two previous MRs that concern tests having the same shape, we use this property to define an MR across different shapes.
We denote the closed-loop bandwidth estimated with the tests of a given shape $\shape$ as the threshold $f_{b,\shape}$.
Therefore the third MR can be defined as the expectation that the $f_b$ estimated using two tests $i$ and $j$ with different shapes are similar
\begin{equation}
    \text{\textbf{MR3:}} \,\, |f_{b,\shape_i}-f_{b,\shape_j}|<\epsilon ,
\end{equation}
where $\epsilon$ is a small discrepancy that can be accepted.
As described in Section~\ref{sec:properties} when exemplifying~\ref{prop:bnd}, the two upper tests of Figure~\ref{fig:bh-examples-freq} satisfy this MR as they show filtering behaviour for frequencies above $\qty{0.9}{\hertz}$, while the bottom-left plot shows a bandwidth of $\qty{0.6}{\hertz}$, hence not satisfying this MR.
As discussed above, this is due to the appearance of the saturation non-linear phenomenon, that however has a limited impact on the CPS and does not push it outside of its design scope.

\begin{figure*}[t]
    \centering
    \includegraphics{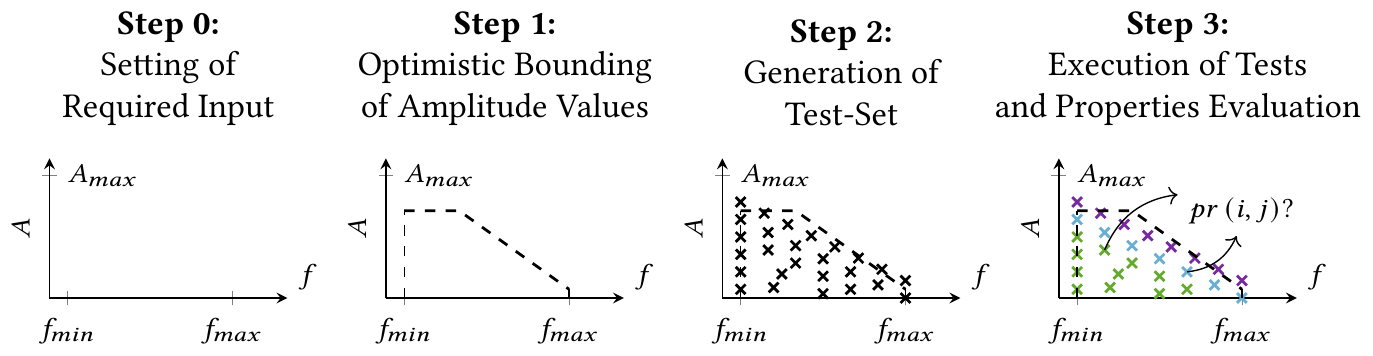}
    \caption{Graphical representation of main steps of proposed testing approach.
    For each step we describe the contribution to the testing in terms of the information that it adds in the frequency-amplitude plane.
    In each plot we show the bounds provided by the engineers as required input.
    The dashed line represents the optimistic bound on the amplitude values obtained in the Step~1.
    The crosses represent how the tests generated in Step~2 sample the frequency-amplitude plane.
    In Step~3 the colouring of the crosses and $\mathit{pr}\funof{i,j}?$ respectively exemplify the evaluation of the behaviour observed in the tests and the evaluation of the metamorphic relations.
    }
    \label{fig:overview-approach}
\end{figure*}

\subsection{Approach Steps}
\label{sec:approach-steps}
In this section we describe the main steps of our testing approach.
We use our test case parametrisation to generate test cases that cover ranges of amplitudes and frequencies.
We then rely on the proposed metrics and MRs enabled by the parametrisation to identify stress test cases.

In our approach, we require the engineers to provide ranges of frequency and amplitude values that are relevant for the SUT.
Starting from these ranges, we propose a testing approach based on three main steps.
We use Figure~\ref{fig:overview-approach} to provide an overview and depict how the required input and steps contribute to the generation and identification of stress test cases:
\begin{enumerate}[label=\emph{Step \arabic*.}, leftmargin=1.3cm]
    \setcounter{enumi}{-1}
    \item \textbf{Setting of Required Input.}
    In this preliminary phase, we ask the engineers to provide bounds on the relevant frequency and amplitude ranges as well as the desired resolution.
    We use the ranges ($f_{\min}$, $f_{\max}$, and $A_{\max}$ in Figure~\ref{fig:overview-approach}) to limit the scope of the testing to practically relevant values.
    Furthermore, a resolution $\delta A$ is also needed to identify the smallest amplitude variations in the reference value that are expected to impact the control behaviour.
    \item \textbf{Optimistic Bounding of Amplitude Values.}
    This iterative step provides an optimistic evaluation of amplitude values that cause non-linear behaviour in the SUT (the dashed line in Figure~\ref{fig:overview-approach}).
    The objective is to obtain a bound that excludes inputs that have such a large amplitude that they are ``trivially'' outside of the design scope.
    To obtain such values, we use tests with the sinusoidal shape.
    The sinusoidal shape maps to a single frequency-amplitude component and does not account for the combination of different components, hence making this step optimistic.
    \item \textbf{Generation of Test-Set.}
    This step uses different input shapes to generate the actual test set.
    We use the optimistic amplitude bounds determined by the previous step to limit the target area of the input space during test case generation.
    For each shape, we use uniform sampling of frequencies and sampling of an exponential distribution over amplitudes to cover the target area.
    Each of the black crosses in Figure~\ref{fig:overview-approach} corresponds to a pair of frequency-amplitude coordinates, and hence a generated test.
    \item \textbf{Execution of Tests and Properties Evaluation.}
    In this step we execute the tests and quantify the observed behaviours.
    The crosses' colours in Figure~\ref{fig:overview-approach} depict $\dnl$ values, green being a low value and purple a high value.
    We use the degree of non-linearity and proposed MRs ($pr\funof{t_i,t_j}$ in Figure~\ref{fig:overview-approach}) to identify test cases that push, to various degrees, the control algorithm out of its design scope.
\end{enumerate}
We now delve into the details of each step.

\subsubsection{Step 0: Setting of Required Input}
In order to practically initialise the approach we ask the engineers to define different quantities specific to the control system.
Such quantities initialise the ranges of the amplitudes and frequencies that are relevant for the SUT, and define the desired resolution across tests.
The required quantities are:
\begin{itemize}
    \item A bound on the maximum amplitude value $A_{\max}$ that can be used for the reference\footnote{
        The lower boundary is always zero since the amplitude describes the absolute value of the signal.
    }: this is an upper boundary to practically limit the exploration of high values along the amplitude axis.
    For example, for drone altitude control, it can be set to the maximum altitude that the drone is expected to fly at.
    \item A frequency range $[f_{\min}, f_{\max}]$ around the expected closed-loop bandwidth $f_b$.
    Such an expected value can be obtained from the control design process or from the speed requirements.
    As discussed above, it is important to cover the closed-loop bandwidth in order to push the CPS to its performance limits.
    As a rule of thumb, a factor of $10$ around the expected $f_b$ should be sufficient to make sure to include the actual one.
    With this rule of thumb, if the tests show that the $f_b$ is not included in the used frequency range, it implies that there is an error of more than one order of magnitude on the expected system bandwidth.
    Such large error in the expected value suggests the presence of some issue in the system development.
    \item An amplitude resolution $\delta A$ that is used to define the smallest variation in the input size that is expected to have an impact on the system behaviour.
    Ideally, this should be a range within which differences in performance are expected to either not matter or be indistinguishable because of practical limitations (e.g., sensor resolution).
    However, this is likely to result in a fine-grained resolution that is neither practically needed nor achievable, and larger values can be used.
\end{itemize}
As we discuss later, we propose a heuristic approach to retrieve a desired frequency resolution from the results of Step~1.
In fact, it can be difficult to intuitively define a desired frequency resolution.
In case of specific needs, the engineers can set the desired frequency resolution manually (e.g., to control the number of tests given a testing budget).
We remark that both the ranges and the resolution can be defined in a conservative way (i.e., using larger ranges and smaller resolution steps) at the cost of increasing the number of tests.

Concerning the approach itself, we require the definition of an upper bound $\dnlth$ on the degree of non-linearity.
This quantity defines a $\dnl$ value above which tests are considered to be outside of the design scope of the SUT.
Being a threshold on the $\dnl$, it can be interpreted as the maximum relative amplitude of new frequency components that we can accept in the CPS output.
Accordingly, it can be chosen based on the maximum relative accepted deviation from the reference value.
For example, for drone altitude control, we might be interested in tests that show up to $\qty{0.15}{\meter}$ of deviation from the reference when hovering around $\qty{1}{\meter}$, hence setting $\dnlth=0.15$.\footnote{
    Note that, $\dnl$ being based on the DFT, it has a linear proportion with time-domain values.
    Hence relative amplitudes in the two domains are equivalent.
}
Furthermore, its value does not need to be strict.
In fact, as far as test case generation is concerned, it is used only for the preliminary optimistic bounding of the amplitude values (Step~1).
Hence, a larger value can be selected without compromising the approach effectiveness.
However, in the same way as for the ranges and resolutions, a conservative choice for this parameter results in a larger test set.

\subsubsection{Step 1: Optimistic Bounding of Amplitude Values}
The purpose of this step is to obtain, for the different frequency values, an optimistic evaluation of the input amplitude values $\AGain$ that push the control algorithm out of its design scope.
With this optimistic evaluation we can restrict the sampling of the frequency-amplitude plane and avoid trivially large amplitudes that expose non-linear behaviour (i.e., $\dnl>\dnlth$).
By excluding such amplitude values, we can focus the test case generation on the area of the input space where the non-linear phenomena start to affect the CPS ability to track reference signals.
Intuitively, this restriction of the amplitude values helps avoiding sampling the purple area of Figure~\ref{fig:input-space} and target the azure area.

To obtain this optimistic bound, we test the SUT using sinusoidal inputs ($\shape=\sin$).
As described in Section~\ref{sec:control-engineering-primer} (Figure~\ref{fig:dft-examples}), sinusoidal inputs sample a single point in the frequency-amplitude plane and therefore avoid interactions between different components.
Since sinusoidal tests do not account for such interactions, they provide only an \emph{optimistic} evaluation of the SUT behaviour for that frequency-amplitude combination.
In other words, even if a sinusoidal test with a given value of $\AGain$ and $\tGain$ exposes linear behaviour ($\dnl\leq\dnlth$), it could happen that other shapes, paired with the same values, expose non-linear behaviour ($\dnl>\dnlth$).
On the other end, if a sinusoidal test shows non-linear behaviour, the use of a different shape (with the same values for $\AGain$ and $\tGain$), is unlikely to show linear behaviour.
In fact, by changing to a more complex shape, we are adding new frequency components on top of a component that was already sufficient to push the control algorithm out of its design scope.

\begin{figure}
    \centering
    \includegraphics{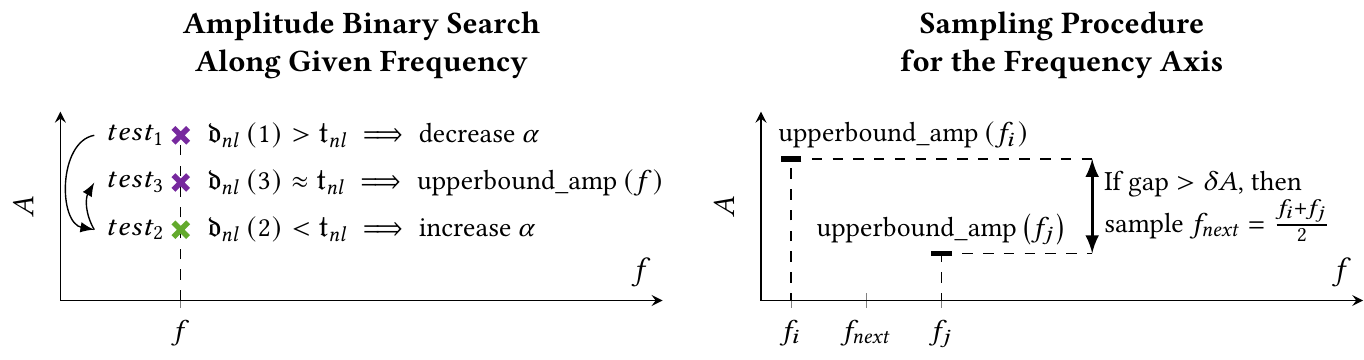}
    \caption{Graphical examples of the sampling strategies used for the amplitude and frequency axes in Step~1 of the approach (respectively the $\text{binary\_search\_upperbound}$ and $\text{find\_gap}$ functions in Algorithm~\ref{alg:nl-optimistic-bound}).
    For exploring the amplitude axis (left-hand side plot) we use a binary search starting from $A_{\max}$ until we find the amplitude for which sinusoidal tests show a $\dnl$ close to the chosen threshold $\dnlth$.
    In this example the search takes $3$ steps, and performs one test that shows too high $\dnl$ ($\mathit{test}_1$),one that shows too low $\dnl$ ($\mathit{test}_2$), and a final one close to the threshold ($\mathit{test}_3$).
    For exploring the frequency axis (right-hand side plot) we exemplify the choice of sampling a new frequency ($f_{\mathit{next}}$) between two that have already been explored ($f_i$ and $f_j$).
    If the upper bound of the two frequencies shows a difference higher than the desired amplitude resolution (i.e., $\text{upperbound\_amp}\funof{f_i}-\text{upperbound\_amp}\funof{f_j}>\delta A$ then we chose to sample a new frequency between them (and perform the binary search along the amplitude axis).
    }
    \label{fig:sampling-examples}
\end{figure}

We show the pseudocode implementing this step in Algorithm~\ref{alg:nl-optimistic-bound}.
The algorithm takes as inputs the frequency range, the desired amplitude resolution, the upper bound on the maximum amplitude value, and the non-linearity upper bound $\dnlth$.
It then uses tests with sinusoidal inputs to sample the frequency-amplitude plane and obtain a frequency-dependent bound of amplitude values over the frequencies.
We represent this bound on the amplitude values with a map object $\text{upperbound\_amp}$ that maps each frequency $f$ to a maximum amplitude value $A_\mathit{bound}$.

\begin{algorithm}
    \caption{Step 1: Optimistic Bounding of Amplitude Values}
    \label{alg:nl-optimistic-bound}
    \begin{algorithmic}[1]
        \Function{OptimisticAmplitudeBound}
                 {$f_{\min}$, $f_{\max}$, $\delta A$, $A_{\max}$,$\dnlth$}
            \State $A_{\mathit{bound}} \gets \text{binary\_search\_upperbound}\funof{f_{\min}, \delta A, A_{\max}, \dnlth}$
            \State add\_sample $\funof{ \text{upperbound\_amp}, f_{\min}, A_{\mathit{bound}}}$
            \State $A_{\mathit{bound}} \gets \text{binary\_search\_upperbound}\funof{f_{\max}, \delta A, A_{\max}, \dnlth}$
            \State add\_sample $\funof{ \text{upperbound\_amp}, f_{\max}, A_{\mathit{bound}}}$
            \While{max\_gap $\funof{\text{upperbound\_amp}}>\delta A$}
                \State $f_{\mathit{next}} \gets$ find\_gap $\funof{\text{upperbound\_amp}, \delta A}$
                \State $A_{\mathit{bound}} \gets \text{binary\_search\_upperbound}\funof{f_{\mathit{next}}, A_{\max}, \delta A, \dnlth}$
                \State add\_sample $\funof{ \text{upperbound\_amp}, f_{\mathit{next}}, A_{\mathit{bound}}}$
            \EndWhile
            \State\Return upperbound\_amp
        \EndFunction
    \end{algorithmic}
\end{algorithm}

In our algorithm, we sample the amplitude and frequency dimensions ($\tGain$ and $\AGain$ values) with different strategies.
We graphically exemplify such strategies in the two plots of Figure~\ref{fig:sampling-examples}.
For the frequencies, we start by sampling the minimum $f_{\min}$ and maximum frequency values $f_{\max}$ provided in the required input.
For each frequency, the algorithm performs a binary search of $\AGain$ values between $0$ and $A_{\max}$ (Lines~$2$ and~$4$) until it finds the minimum value for which the SUT shows $\dnl>\dnlth$.
This binary search is exemplified in the left-hand side plot in Figure~\ref{fig:sampling-examples}, where we fix the frequency $f$ and vary the amplitude gain $\AGain$ until we find a test that shows a degree of non-linearity close to $\dnlth$.\footnote{
    In order to perform a binary search, we assume that MR1 holds, i.e., that $\dnl$ increases for increasing values of $\AGain$.
    Thanks to the simple nature of the sinusoidal input, MR1 is more likely to hold for inputs based on this shape.
    However, as demonstrated in the experimental evaluation (Section~6.4), MR1 does not always hold in practice.
    The invalidation of MR1 in the preliminary search can lead to overestimate the upper bound; however, such overestimation is still a valid upper bound.
    In fact, the search stops only around tests with a high $\dnl$ value (hence, tests that are in the purple area of Figure~5) or around the maximum allowed value $A_{\max}$.
    An overestimation implies that the bound is not as tight, thus rendering the approach less efficient.
    Specifically, it increases the number of tests needed to cover the area below the bound.
}
In practice, we stop the binary search when the difference between the upper and lower amplitude values identified by the binary search is smaller than the desired amplitude resolution $\delta A$ (i.e., the next amplitude update is smaller than half the resolution).
We then take as $A_{\mathit{bound}}$ the lower amplitude values obtained with the binary search and add it, together with the sampled frequency value, to the map object $\text{upperbound\_amp}$ (Lines~$3$ and~$5$).

Then the iterative part of the algorithm starts with the loop at Line~6.
In this iteration we sample the frequency axis until the amplitude value difference between the $A_{\mathit{bound}}$ values of subsequently sampled frequencies are smaller than $\delta A$.
Such gap is graphically exemplified in the right-hand side plot of Figure~\ref{fig:sampling-examples}.
In the plot we have already explored frequencies $f_i$ and $f_j$ and we evaluate if we need to sample a new frequency between them according to the difference $\text{upperbound\_amp}\funof{f_i}-\text{upperbound\_amp}\funof{f_j}$.
If there is a gap larger than $\delta A$ between two subsequent frequencies, then we sample a new frequency $f_{\mathit{next}}$ between them (Line~$7$).
For the new frequency we perform again a binary search over the amplitude values and add the obtained $A_{\mathit{bound}}$ to the $\text{upperbound\_amp}$ map object (Lines~8 and~9).
Once all the amplitude gaps are smaller than $\delta A$ we stop the sampling of new frequencies. 
Finally, the algorithm returns the obtained map object (Line~$11$). 
The map object represents the optimistic bound of the amplitude values leading the SUT out of its design scope and we use it to avoid testing trivially large inputs.

For some systems, it may happen that none of the amplitude values in the range between $0$ and $A_{\max}$ causes the SUT to show non-linear behaviour.
In such cases, the binary search converges to a value close to $A_{\max}$ and the use of the $\text{upperbound\_amp}$ will not be able to reduce the number of test cases in the next steps of the approach.
Such occurrence does not hinder the effectiveness of the approach: it only suggests (but does not imply) that the $A_{\max}$ value might be set too low.

\subsubsection{Step 2: Generation of Test-Set}
In this step we generate the actual test set.
Given our test case parametrisation, we need to define a set of shape functions and different amplitude and time scaling parameter values.
We now discuss each of them.

Concerning the shape functions, we propose to use a set of shapes inspired by common practice in control engineering.
In control engineering, the most common inputs used to evaluate a control algorithm are the step (instantaneous change of the reference) and ramp (linear change in the reference)~\cite{Astrom:2008}.
Accordingly, we propose the use of shape functions that resemble such inputs: square, sawtooth, triangular, and trapezoidal waves.
If other patterns are available from use cases in a specific CPS application, engineers can expand this set.

For each of the chosen shapes, we generate a set of $\langle\tGain,\AGain\rangle$ pairs.
We aim at exploring the area of the input space delimited by the upper-bound threshold identified in Step 1.
Since one shape corresponds to more than one frequency-amplitude $\langle f,A\rangle$ point, we take as reference the point associated with the largest amplitude, which we call the \emph{main component} of a shape.
Accordingly, we sample values for $\AGain$ and $\tGain$ and thus move the main component of each shape to obtain test cases that cover the area delimited by the upper-bound threshold.

Regarding the frequency axis, in the general case, we assume that there is no reason to test a specific range.
Therefore, we aim at uniformly covering (with the main component of the shape) the frequency range $[f_{\min},f_{\max}]$.
Practically, we sample frequencies at equal intervals in the given range.
Since it can be difficult to intuitively set the frequency resolution, we propose to compute such resolution $\delta f$ from the results of Step~1.
Specifically, we suggest to use the average frequency gap (i.e., $\mathit{avg}_i\{f_{i+1}-f_{i}\}$) obtained in the amplitude upper-bound.
Intuitively, these gaps are obtained by imposing a maximum difference $\delta A$ in the amplitude threshold for which the SUT shows non-linear behaviour.
Therefore, they should correspond to frequency variations for which the $\dnl$ does not change significantly.
In specific cases, the engineers can adapt the sampling of the frequency axis according to application-specific needs.
For example, if a frequency range is known to be particularly relevant for the specific SUT, a biased random sampling can be applied.

Differently, along the amplitude dimension, we are generally more interested in exploring the area with large amplitudes.
For the amplitude values we can leverage the upper bound obtained at Step~1 and avoid sampling large amplitudes that, already with a sinusoidal input, would provide a very high $\dnl$.
Practically, we limit the sampling range with the optimistic upper bound obtained in each sampled frequency.
Within this range, we propose to sample different amplitudes $\AGain$ according to a beta distribution skewed toward the higher values~\cite{Sinharay:2010}.\footnote{
    A beta distribution is a version of the more classical exponential distribution with bounded support.
}
The number of samples is determined according to the desired amplitude resolution $\delta A$.
Practically, we compute the number of tests for each frequency $f$ with $\text{upperbound\_amp}\funof{f}/\delta A$, to obtain a number of samples compatible with the desired resolution.
Similarly to the frequency dimension, application-specific sampling strategies might be adopted if areas of particular interest are given.

\subsubsection{Step 3: Execution of Tests and Properties Evaluation}
\begin{figure}
    \centering
    \includegraphics{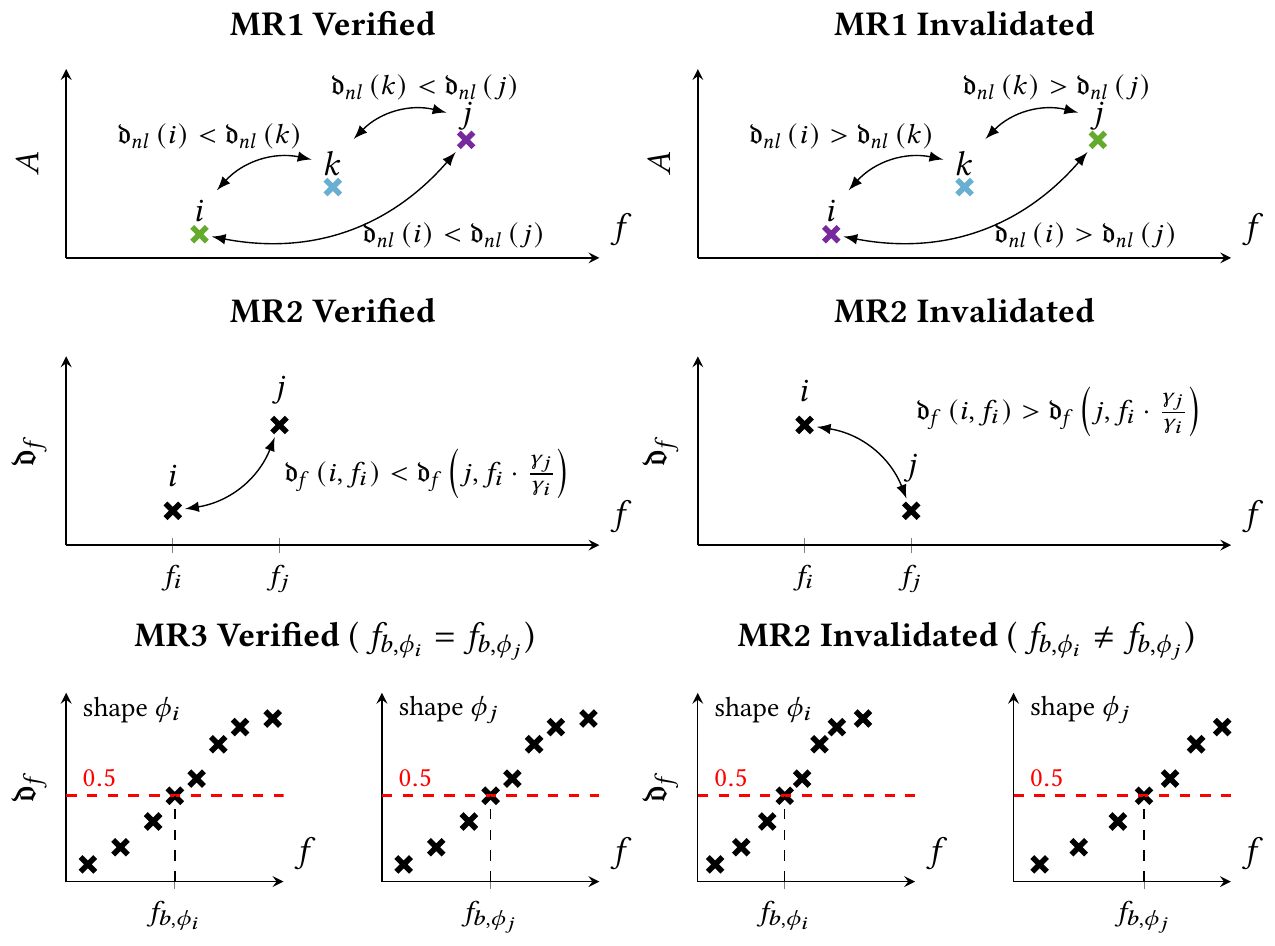}
    \caption{This figure shows how to manually evaluate the proposed MRs.
    For each MR (the different rows) we report an example in which it is verified (left column) and one in which it is invalidated (right column).
    To evaluate \textbf{MR1} we compare the $\dnl$ measured in different tests (tests $i$, $j$, and $k$) and their relative position in the frequency-amplitude plane.
    We compare the $\dnl$ for each pair of tests (highlighted by the arrows in the plots).
    If the $\dnl$ is higher for tests further away from the origin (left-hand side plot), then MR1 is verified.
    If the $\dnl$ decreases for tests further away from the origin (right-hand side plot), then MR1 is invalidated.
    To evaluate \textbf{MR2} we compare the $\dof$ in frequency components of tests (tests $i$ and $j$) with different time scaling coefficients ($\tGain_i>\tGain_j$).
    If a larger time scaling coefficient corresponds to a higher $\dof$ (left-hand side plot), then MR2 is fulfilled.
    If a larger time scaling coefficient corresponds to a smaller $\dof$ (right-hand side plot), then MR2 is invalidated.
    To evaluate \textbf{MR3} we compare the closed-loop bandwidth (the frequency for which the $\dof$ becomes greater than $0.5$) using tests with different shapes (shapes $\shape_i$, and $\shape_j$).
    If the closed-loop bandwidth evaluated with tests that use different shapes are equal (left-hand side plot $f_{b,\shape_i}=f_{b,\shape_j}$) then MR3 is verified.
    If the two closed-loop bandwidths evaluated do not correspond (right-hand side plot $f_{b,\shape_i}\neq f_{b,\shape_j}$) then MR3 is invalidated.
    }
    \label{fig:mrs-evaluation-examples}
\end{figure}

In this final step we proceed to executing the tests and evaluating their outputs.
The evaluation is based on the $\dnl$ and $\dof$ metrics defined in Section~\ref{sec:input-space-quantification} and includes the verification of the MRs and the identification of the stress test cases.
We first discuss the identification of stress test cases and evaluation of MR1 using $\dnl$.
We then discuss the evaluation of MR2 and MR3 using  $\dof$.
We use Figure~\ref{fig:mrs-evaluation-examples} to exemplify the manual evaluation of the MRs.
In the experimental section, we showcase in detail the visualisation and analysis step for our case studies.

We execute each test and store its output trace.
We then compute, according to the definitions given in Section~\ref{sec:input-space-quantification}:
\begin{enumerate*}[label=(\roman*)]
    \item the set of relevant $\langle f,A\rangle$ components,
    \item the degree of non-linearity, and
    \item the degree of filtering of each $\langle f,A\rangle$ component.
\end{enumerate*}
Such information can then be analysed by leveraging plots on the frequency and amplitude dimensions.

First, we discuss how to use $\dnl$ to identify stress test cases and evaluate MR1.
To do so, the main component of each test (i.e., the component with the largest amplitude) can be displayed in a scatterplot on the frequency-amplitude plane using its $\langle f,A\rangle$ coordinates.
The markers can then be coloured with a gradient that represents the measured $\dnl$ (e.g., purple for $\dnl\geq\dnlth$, and green for $\dnl=0$, with the colour gradients for representing the values in the range $[0, \dnlth)$).
We exemplify such scatterplot with the mentioned colour choice in the upper plots of Figure~\ref{fig:mrs-evaluation-examples}.
The engineers can then use the colours to distinguish the input space regions within the design scope of the CPS (green regions characterised by a low $\dnl$) and regions outside of the CPS design scope (purple regions characterised by a high $\dnl$).
The test cases at the borders between the regions where we observe a transition between colours are the stress test cases.
The inspection of the stress test cases provides information about the phenomena in the CPS that can falsify the design assumptions and push the system out of its design scope.
\begin{itemize}
    \item \textbf{MR1} can be evaluated analysing the relative positioning of regions characterised by low and high $\dnl$.
    In fact, MR1 states that $\dnl$ should increase with  increasing frequency and amplitude values.
    Practically, regions with low $\dnl$ (green regions) should be closer to the origin of the axes than regions with high $\dnl$ (purple regions).
    Accordingly, a pattern similar to the one shown in Figure~\ref{fig:input-space} should appear.
    For example, in the upper left-hand side plot of Figure~\ref{fig:mrs-evaluation-examples} where the tests $i$, $k$ and $j$ show increasing $\dnl$, consistent with their distance from the origin (respectively green, azure, and purple colouring).
\end{itemize}
If the test results show such patterns, then they fulfil MR1.

We now discuss the evaluation of MR2 and MR3.
As discussed in Section~\ref{sec:metamorphic-relations}, MR2 and MR3 describe expected $\dof$ properties for tests within the design scope (i.e., tests for which $\dnl<\dnlth$).
Accordingly, we use these MRs to validate the identification of tests that are within the design scope: if such identification is correct, then the tests should fulfil the two MRs, otherwise there is likely some issue with the testing process.

In order to discuss their evaluation, we note that MR2 and MR3
\begin{enumerate*}[label=(\roman*)]
    \item concern only tests that show linear behaviour,
    \item do not depend on the input amplitude, and
    \item address $\dof$.
\end{enumerate*}
Accordingly, to evaluate them, we first select only the tests that fulfil $\dnl<\dnlth$, i.e., the tests that are considered within the design scope.
From such tests, we extract the individual $\langle f,A\rangle$ components and associated $\dof$ value.
We then plot, in separate plots for each shape, $\dof$ of each frequency-amplitude point as a function of its frequency coordinate $f$.
We exemplify such plotting in the middle and bottom plots of Figure~\ref{fig:mrs-evaluation-examples}, where the crosses represent a frequency-amplitude component with the horizontal coordinate being the corresponding frequency value ($f_i$ and $f_j$) and the vertical coordinate representing $\dof$ measured for that component.
Such plots can then be used to evaluate MR2 and MR3:
\begin{itemize}
    \item \textbf{MR2} states that an increase in the frequency values corresponds to an increase in the degree of filtering. Accordingly, the proposed frequency-$\dof$ plot should show an increasing trend irrespective of the used shape.
    For example, in the middle plots of Figure~\ref{fig:mrs-evaluation-examples} the left-hand side one shows an increasing pattern that fulfils MR2, while the right-hand side one shows a decreasing pattern that invalidates MR2.
    \item \textbf{MR3} states that $f_b$ should not depend on the specific shape of the input.
    We recall that $f_b$ is defined as the frequency below which $\dof<0.5$ and above which $\dof>0.5$.
    For example, in the bottom plots of Figure~\ref{fig:mrs-evaluation-examples}, we use shapes $\shape_i$ and $\shape_j$ and observe the closed-loop bandwidths $f_{b,\shape_i}$ and $f_{b,\shape_j}$.
    Accordingly, the trend of the degree of filtering values for the different shapes should cross the $0.5$ value around the same frequency.
    In Figure~\ref{fig:mrs-evaluation-examples}, the left-hand side plots show $f_{b,\shape_i}=f_{b,\shape_j}$ and fulfils MR3, while the right-hand side plots show $f_{b,\shape_i}\neq f_{b,\shape_j}$ and thus invalidates MR3.
\end{itemize}
If the above properties are not verified by some frequency-amplitude components, the tests associated to such components should be inspected for identifying possible faults in the testing process.
 \section{Empirical Evaluation}
\label{sec:experiments}
In this section, we empirically evaluate the proposed testing approach.
We aim to assess its effectiveness in using the frequency and amplitude dimensions for generating test cases that falsify the linearity assumptions made during the control algorithm design.
Furthermore, we want to evaluate the ability of the MRs to highlight test cases that fall outside of the design scope and provide insights about the specific non-linear phenomena that can lead to the SUT behaving unexpectedly.
Accordingly, we define two research questions (RQ).

\partparagraph{RQ1 - Generation of Stress Tests}
We aim to evaluate how effective is the use of the frequency and amplitude characterisation of the test inputs to trigger non-linear phenomena that limit the design scope.
With RQ1, we ask: \emph{How effective is our test case generation approach in pushing a CPS control-loop away from the linearisation point and triggering non-linear phenomena around the boundary where they appear?}

\partparagraph{RQ2 - Relevance of Metamorphic Relations}
With the second research question, we aim to evaluate the possibility of using the proposed MRs to identify inconsistencies with the qualitative input space characterisation and thus identify which tests stress the system such as to induce non-linear behaviour.
Hence, we ask: \emph{How the proposed MRs identify stress test cases at the bounds of the SUT design scope?}

\vspace{1mm}

This section is structured as follows.
We first present and motivate the case studies (Section~\ref{sec:experimental-suts}) as well as the settings choices required for our testing approach (Section~\ref{sec:experimental-settings}).
Then, for each question, we present our empirical methodology, followed by the evaluation results, that we use to answer the RQs (Sections~\ref{sec:experimental-rq1} and~\ref{sec:experimental-rq2}).
We conclude this empirical evaluation with a discussion of the threats to validity (Section~\ref{sec:experimental-limitations}), and information about code and data availability (Section~\ref{sec:experimental-data}).

\subsection{Systems Under Test}
\label{sec:experimental-suts}
To answer our research questions, we apply our testing approach to the simulation models of three CPSs: a Python model of the Crazyflie drone, a Simulink model of a DC servo (a continuous current motor), and a Simulink model of a lightweight aircraft.
We choose these CPSs based on three main criteria.
First, we aim at covering the two most common types of algorithms found in traditional control theory: PID control and state feedback~\cite{Astrom:2008,Desborough:2002}.
The drone and the lightweight aircraft include PID control algorithms, while the DC servo uses state feedback.
Intuitively, the difference between the two is that PIDs compute the actuation based on the distance between the physical output and the desired value, while state feedback bases the computation on the whole system state estimation.

Second, we need SUTs that include non-linear phenomena as well as detailed knowledge about these phenomena (e.g., saturation bounds), and access to their development documentation.
This knowledge is necessary to evaluate the effectiveness of our approach in triggering such phenomena.
Accordingly, for the Crazyflie drone and the DC servo, we use simulation models that we built based on their development documentation and actual CPS implementations, which are available to us.
In this way, we can also build models behaving consistently with the real systems.
However, in order to gain further insights and higher confidence in the general validity of our approach, we also include a CPS model that was not built by us.
We chose the publicly available model of the altitude control of a lightweight aircraft.\footnote{
    \url{https://nl.mathworks.com/help/aeroblks/lightweight-airplane-design.html}
}
This model fits our needs as it includes both a detailed model of the aircraft dynamics (hence including the non-linear phenomena) as well as the control algorithm implementation.\footnote{
    As discussed at the end of Section~\ref{sec:problem-motivation}, in the scope of this article, we need more detailed models than the ones used by control engineers.
    This limits our choice of publicly available models as many of the publicly available ones do not have such level of detail (e.g., \url{https://nl.mathworks.com/help/simulink/slref/aircraft-longitudinal-flight-control.html}).
}

Third, we want our case studies to capture the variability that can appear among SUTs when applying our testing approach.
In fact, when applying our testing approach, engineers cannot know a priori which non-linear phenomena affect the SUT design scope.
For this reason we want to include different types of non-linear phenomena in our case studies.
In this regard, the DC servo, being an electric motor, has a wide variety of applications, and depending on the specific application it can expose different non-linear phenomena.
This feature of the DC servo allows us to inject different non-linearities that are plausible in a specific application context.
In this way we can obtain the desired SUT variability.\footnote{
    Conversely, we decided not to manually inject non-linearities in the model of the Crazyflie drone in order to keep the SUT identical to the one used during the actual development of the drone.
}
We now briefly describe the three CPSs as well as their simulation models, including information on how we built the Crazyflie drone model and the different versions of the DC servo.

The Crazyflie drone is developed by Bitcraze\footnote{
    \url{https://www.bitcraze.io/}
} and is completely open-source.
This gives us full access to the source code, as well as software and design documentation~\cite{Greiff:2017}.
We use the documentation to build a detailed Python model of the physics.
In addition, we use the source code together with the design documentation to build the model of the control software\footnote{
    \url{https://github.com/bitcraze/crazyflie-firmware}
}.
In previous research, we compared the flight of the model with the flight of the real drone and observed that they show similar behaviour~\cite{Mandrioli:2023}.
This gives us confidence on the real-world validity of the experiments performed with this model.
The control layer of the drone includes three separate control loops for the three directions.
Among those, we chose to test the altitude control, which controls the drone's vertical position as, different from the other control loops, it allows for the quantification of the non-linear phenomena occurrence that we leverage in answering RQ1.
Despite being a small drone (size of $6\times\qty{6}{\centi\meter}$), we note that the Crazyflie drone implements standard drone control algorithms that are also used in larger drone applications~\cite{Mueller:2016}.

The Simulink model of the DC servo is based on laboratory hardware used for educational purposes at Lund University.
Hence, like for the Crazyflie drone, we have access to the CPS implementation and its documentation.
The control layer of the DC servo receives from the users a desired angular position of the motor axis (the input reference $r$).
Given such input references and the motor axis position measurements (the output $y$), the control layer generates a voltage signal to be fed to the motor in order to make it move (the actuation signal).
We built the Simulink model to execute the very same C code that is used in the microcontroller of the DC servo.
This gives us confidence that the simulation model behaves similarly to the actual implementation.
While the considered DC servo is used for educational purposes, it is analogous to electrical motors found in many industrial applications such as the joint of a robot or the steering axis of an autonomous car.
As mentioned above, we chose this SUT because electrical motors can expose, depending on the specific application, different types of non-linearities that can affect the behaviour of the control algorithm.
This allows us to develop different but still practically relevant versions of the DC servo in which we manually inject non-linearities to study their impact.
To ensure these non-linearities are commonly found in real applications, we use the following non-linearities defined in the ``Discontinuities'' folder of the Simulink library:\footnote{
    \url{https://www.mathworks.com/help/simulink/discontinuities.html}
}
\begin{itemize}
    \item \emph{Saturation} represents the limitation of variables in the system.
    It generally affects actuators (which have limited capacity) and sensors (which typically have a limited range of values they can read).
    \item \emph{Quantisation} is the effect of the finite precision in the analog-to-digital conversion of the sensors signals.
    \item \emph{Pulse-Width-Modulation} is a common technique used for digital-to-analog conversion, which can distort the actuation signals.
    \item \emph{Coulomb Friction} is always found in the physical world but is however hard to model and non-linear and, for these reasons, often neglected or approximated.
    \item \emph{Dead-Zone} and \emph{Backlash} are always found in mechanical gearing as they are caused by the play between the different cogs.\footnote{
        Among the blocks listed in the Discontinuities folder, we excluded the ``rate limiter'' as it is just a special case of saturation, and the ``relay'' as it is a very simple and old control approach, used only for non-critical systems.
    }
\end{itemize}
We add to this list a quadratic friction model as, in practice, it is a very common source of non-linear behaviour, e.g., aerodynamic drag in a car or drone.
Among these non-linearity sources, we consider saturation, quantisation, and pulse-width-modulation to be always present as they are part of the interface between the cyber and physical parts of the system.
In contrast to these sources, we inject one by one the quadratic friction, the coulomb friction, the dead-zone and backlash.
We therefore obtain the following five versions of the DC servo according to the included non-linearities:
\begin{enumerate}[label=\textbf{DC\arabic*}]
    \item \label{dc} saturation (on actuation and sensor), quantisation, pulse-width-modulation.
    \item \label{dc-cf} saturation (on actuation and sensor), quantisation, pulse-width-modulation, and Coulomb friction.
    \item \label{dc-qf} saturation (on actuation and sensor), quantisation, pulse-width-modulation, and quadratic friction.
    \item \label{dc-dz} saturation (on actuation and sensor), quantisation, pulse-width-modulation, and dead-zone.
    \item \label{dc-bl} saturation (on actuation and sensor), quantisation, pulse-width-modulation, and backlash.
\end{enumerate}

Finally, in the altitude control of the lightweight aircraft, the objective is to reach and maintain a desired altitude.
To achieve this, the software is responsible for setting the elevator angle (one of the moving surfaces of the aircraft tail) based on various sensor data such as air pressure and gyroscope readings.
This CPS, as noted above, was not developed by us and we do not have detailed knowledge on the non-linear phenomena that it entails.
This limits our ability to quantify the occurrence of non-linear phenomena (RQ1) and of identifying root causes for the misbehaviour observed in the tests.
However, it does not impact our findings in terms of identification of the test cases where non-linear phenomena impact the CPS ability to track references (RQ2).
To summarise, we apply our approach to seven different SUTs: the two altitude control models for the drone and the aircraft, and the five different versions of the DC servo.

\subsection{Testing Settings}
\label{sec:experimental-settings}
In this subsection, we present the settings needed for the application of our testing approach to the selected SUTs.
We first report on the required input (defined in Section~\ref{sec:approach-steps}) and the number of input repetitions needed to evaluate the $\dnl$ metric (defined in Section~\ref{sec:input-space-quantification}).
We conclude by reporting the resulting number of executed tests.

\partparagraph{Selection of Required Inputs}
We summarise in Table~\ref{tab:required-input} the values chosen, based on our domain knowledge, for the required inputs of our testing approach.

Concerning the frequency range, the drone and the DC servo should be able to track the reference within a time span in the order of seconds.
Accordingly, we place the expected closed-loop bandwidth for these SUTs around $\qty{1}{\hertz}$.
Intuitively, this corresponds to taking around $\qty{2}{\second}$ to track a unit change in the reference value (e.g., $\qty{1}{\meter}$ for the drone and $\qty{1}{\radian}$ for the DC servo).
For the drone, we define the frequency range to be $[\qty{0.1}{\hertz},\qty{2}{\hertz}]$.
For the DC servo, we do not have a clear estimate from its development documentation.
Hence, we choose a broader range $[\qty{0.005}{\hertz},\qty{3}{\hertz}]$.
The lightweight aircraft is a slower system, and we expect the bandwidth to be around $\qty{0.05}{\hertz}$, which gives a response to a unit change in the reference of the order of tens of seconds.
Accordingly, using the rule of thumb of including one order of magnitude above and below the expected closed-loop bandwidth, we choose the frequency range $[\qty{0.005}{\hertz},\qty{0.5}{\hertz}]$.

Concerning the maximum amplitude, we reason on the largest inputs that we can expect in practice for our SUTs.
Since the Crazyflie drone is expected to mostly fly in indoor environments, we chose a maximum amplitude of $\qty{6}{\meter}$, corresponding to a large room.\footnote{
    For example, Crazyflie has been used for automated inventorying of supermarkets~\cite{Greiff:2021}.
}
The DC servo is supposed to track desired angular positions.
Hence, we limited the reference amplitude changes to one full rotation of the motor axis, $\qty{2\pi}{\radian}$.
The lightweight aircraft is expected to fly around $\qty{2000}{\meter}$ in altitude and perform manoeuvres around that height.
Hence we set a maximum amplitude range for the manoeuvres of $\qty{750}{\meter}$, resulting in references between $\qty{1250}{\meter}$ and $\qty{2750}{\meter}$.

Concerning the amplitude resolution, we reason about the variations of the input amplitude that can cause a change in the behaviour of the SUT.
For the drone, this is in the order of centimetres.
For example, asking the drone to reach $\qty{1.01}{\meter}$ or $\qty{1}{\meter}$ should not show significantly different behaviour.
Accordingly, we chose $\delta A=\qty{5}{\centi\meter}$ for the drone.
The DC servo is expected to react to reference changes in the order of rotation degrees.
Since such devices are expected to be rather precise, we chose a range smaller than the desired precision.
Specifically, we chose $\frac{1}{400}$ of a rotation for the DC servo (less than one degree): $\delta A=\qty{0.015}{\radian}$.
For the lightweight aircraft, we can expect that a difference in the reference in the order of meters does not significantly alter the system behaviour; hence, we chose $\delta A=\qty{5}{\meter}$.

\begin{table}
    \caption{Selected input values for our testing approach to test the drone, the DC servo, and the lightweight aircraft.
    }
    \label{tab:required-input}
    \begin{tabular}{lcccccccccc}
        \toprule
        \textbf{SUT}                     & $f_{\min}$  & $f_{\max}$  & $A_{\max}$  & $\delta A$  \\
        \midrule
        \textbf{Crazyflie Drone}         & $\qty{0.1}{\hertz}$   & $\qty{2}{\hertz}$   & $\qty{6}{\meter}$     & $\qty{0.05}{\meter}$   \\
        \textbf{DC servo (all versions)} & $\qty{0.005}{\hertz}$ & $\qty{3}{\hertz}$   & $\qty{2\pi}{\radian}$ & $\qty{0.015}{\radian}$ \\
        \textbf{Lightweight Aircraft}    & $\qty{0.005}{\hertz}$ & $\qty{0.5}{\hertz}$ & $\qty{750}{\meter}$   & $\qty{5}{\meter}$      \\
        \bottomrule
    \end{tabular}
\end{table}

Finally, in order to apply the testing approach to the SUTs, we have to set the non-linear threshold. 
The $\dnlth$ threshold is connected to the maximum relative deviation from the reference that the SUTs can accept.
In our experiments, across the SUTs, we observed unexpected behaviours when output signals deviate $15\%$ from the reference $r$.
Accordingly, we set $\dnlth=0.15$.
This means that we identify tests with more than $15\%$ deviation from the reference $r$ to be trivially out of the design scope and therefore not interesting in terms of stress testing.
For example, in the drone case study, this means that we are interested in tests that deviate of up to $\qty{15}{\centi\meter}$ when the drone is expected to hover at $\qty{1}{\meter}$ of height.

\begin{table}
    \caption{$\dnl$ of the preliminary tests computed using different numbers of periods for each SUTs.
    For the drone, we can see that the value stabilises when using five periods; hence, that is the number of periods that we use when applying our approach to this SUT.
    For each of the DC servo versions, the value detects the non-linear behaviour after six repetitions; hence, we use seven to allow for some margin.
    For the lightweight aircraft, we see that the $\dnl$ value steadily increases up to ten input repetitions and is greater than the chosen threshold for nine or more repetitions; hence, we chose ten input repetitions.
    }
    \label{tab:cf-preliminary-num-periods}
    \begin{tabular}{lcccccccccc}
\toprule
        {\bf Number of Periods} & {\bf1} & {\bf2} & {\bf3} & {\bf4} & {\bf5} & {\bf6} & {\bf7} & {\bf8} & {\bf9} & {\bf10} \\
        \midrule
        \textbf{Crazyflie Drone}   & $0.12$ & $0.26$ & $0.36$ & $0.40$ & $0.39$ & $0.41$ & $0.47$ & $0.54$ & $0.56$ & $0.54$\\
        \ref{dc}    & $0.05$ & $0.10$ & $0.09$ & $0.09$ & $0.08$ & $0.50$ & $0.45$ & $0.42$ & $0.37$ & $0.30$\\
        \ref{dc-cf} (Coulomb Friction)   & $0.05$ & $0.10$ & $0.09$ & $0.08$ & $0.07$ & $0.54$ & $0.49$ & $0.47$ & $0.43$ & $0.36$\\
        \ref{dc-qf} (Quadratic Friction) & $0.01$ & $0.02$ & $0.03$ & $0.01$ & $0.02$ & $0.25$ & $0.24$ & $0.24$ & $0.25$ & $0.24$\\
        \ref{dc-dz} (Dead-Zone)          & $0.05$ & $0.10$ & $0.09$ & $0.09$ & $0.07$ & $0.49$ & $0.45$ & $0.44$ & $0.41$ & $0.35$\\
        \ref{dc-bl} (Backlash)           & $0.05$ & $0.11$ & $0.10$ & $0.09$ & $0.07$ & $0.49$ & $0.45$ & $0.42$ & $0.38$ & $0.32$\\
        \textbf{Lightweight Aircraft}    & $0.00$ & $0.01$ & $0.02$ & $0.04$ & $0.05$ & $0.08$ & $0.10$ & $0.14$ & $0.17$ & $0.22$\\
        \bottomrule
    \end{tabular}
\end{table}

\partparagraph{Setting the Number of Input Iterations}
In order to apply our testing approach to the SUTs, we need to define the number of input repetitions needed to compute the $\dnl$ metric (Section~\ref{sec:input-space-quantification}).
For efficiency, we want to select the smallest number of repetitions (i.e., shorter tests) yielding sufficient accuracy when computing $\dnl$.
To evaluate this trade-off, we performed a preliminary experiment, in which we ran a single long test with the maximum amplitude and time scaling coefficients (corresponding to $A_{\max}$ and $f_{\max}$ in Table~\ref{tab:required-input}) such that the SUT exposes non-linear behaviour.
We detect non-linear behaviour by inspecting whether the output follows the input reference.
We use $10$ input repetitions and compute the $\dnl$ value using different trace lengths that correspond to different numbers of input periods.
We then evaluate after how many input periods the $\dnl$ value converges to a value detecting non-linear behaviour (i.e., it is higher than the chosen $\dnlth$).
As this evaluation is different for the different SUTs, we run it for each of them before applying our testing approach.

In order to manually bring the SUTs out of their design scope, we use tests with high frequency and amplitude.
As discussed in the $\dnl$ definition (Section~\ref{sec:dnl-def}), the specific amplitude, frequency, and shape used for this preliminary experiment are not relevant as long as the test pushes the SUT out of its design scope.
We run tests with the specified maximum amplitude and frequencies  (Table~\ref{tab:required-input}).
To cause non-linear behaviour, for the drone and the lightweight aircraft, we used a test with the sinusoidal shape, while for the DC servo, we used a test with the steps shape.

We report in Table~\ref{tab:cf-preliminary-num-periods} the $\dnl$ values obtained for each of the SUTs using different numbers of input periods.
For the drone, the $\dnl$ value exceeds the threshold already when using two periods.
We chose, however, to use $5$ periods as $\dnl$ increases until then.
For the DC servo, we observe a similar pattern in each version of the system.
The $\dnl$ value is below the threshold when using $5$ or less periods and then suddenly increases to a higher (and constant) value above the threshold for $6$ or more periods.
To allow some margin, we chose to use $7$ periods in our tests.
For the lightweight aircraft, the $\dnl$ value passes the threshold for $9$ periods; hence, for the same reason as for the DC servo, we use tests with $10$ periods.

We note that the measured $\dnl$ does not always increase when we increase the number of periods used for its computation (e.g., for $6$ to $10$ periods in the \ref{dc-dz}).
In general, the more frequency samples we have, the more likely we are to detect new frequency components in the output signal and hence having a higher $\dnl$.
However, when increasing the length of the traces, the new frequency samples move on the frequency axis, as shown in Figure~\ref{fig:fft-example}.
Since different frequencies are sampled, different amplitudes can be measured, and hence, there is no guarantee that the $\dnl$ value increases monotonically.

\begin{table}
    \caption{Number of tests executed for each SUT in Step~1 (Optimistic Bounding of Amplitude Values) and Step~3 (Execution of Tests and Properties Evaluation).
    }
    \label{tab:num-tests}
    \begin{tabular}{lccc}
\toprule
        {\bf Approach Step} & {\bf Step 1} & {\bf Step 3} & {\bf Total} \\
        \midrule
        \textbf{Crazyflie Drone}         & $112$ & $1012$ & $1124$ \\
        \ref{dc}                         & $ 12$ & $5100$ & $5112$ \\
        \ref{dc-cf} (Coulomb Friction)   & $ 12$ & $5010$ & $5022$ \\
        \ref{dc-qf} (Quadratic Friction) & $ 12$ & $5100$ & $5112$ \\
        \ref{dc-dz} (Dead-Zone)          & $ 12$ & $5100$ & $5112$ \\
        \ref{dc-bl} (Backlash)           & $ 12$ & $5100$ & $5112$ \\
        \textbf{Lightweight Aircraft}    & $172$ & $3421$ & $3593$ \\
        \bottomrule
    \end{tabular}
\end{table}

\partparagraph{Number of Tests and Duration of Experiments}
In Table~\ref{tab:num-tests}, we report the number of tests executed for each SUT.
The table reports the number of tests executed respectively for Step~1 and Step~3.\footnote{
    Recall from Section~\ref{sec:approach-steps} that Step~2 does not require the execution of test cases.
}
We note that the DC servo requires more tests as it has wider ranges and a higher resolution was selected.
For the drone and lightweight aircraft, the execution of Step~1 reduces the number of tests needed to cover their input space.
In total, for our experimental campaign, we executed \num{30187} tests.\footnote{
    In Section~\ref{sec:experimental-discussion}, we comment on what insights the number of executed tests provides on the testing approach.
}

The execution of the tests took few hours for the drone and the lightweight aircraft on a 2017 MacBook Pro, with $\qty{2.3}{\giga\hertz}$ Dual-Core Intel Core i5 processor.
The different versions of the DC servo took instead almost a full day for each version.
This is due both to the higher number of tests and also to the pulse-width modulation.
In fact, the simulation of the pulse-width modulation includes a high-frequency signal that requires a very short simulation time step, hence making the simulation significantly slower.

\subsection{RQ1 - Generation of Stress Tests}
\label{sec:experimental-rq1}
\partparagraph{Methodology}
To answer this question, we define metrics to evaluate the occurrence of the different non-linear phenomena in the performed tests.
We study tests results over the different values of the proposed frequency-amplitude input characterisation.
If our approach is effective, the tests results will show various occurrences of the non-linear phenomena depending on the frequency and amplitude content of the input.
Such tests can then be used to assess how and in which scenarios the non-linear phenomena affect system performance.
For the Crazyflie drone and the DC servo, we have complete knowledge of the non-linear phenomena involved.
Hence, we can conduct an exhaustive evaluation of their results with respect to this research question.
For the lightweight aircraft, however, we do not have such detailed knowledge about its non-linear phenomena.
Upon inspecting the Simulink model of the aircraft, we could identify non-linearity due to actuator saturation.
Therefore, we limit ourselves to evaluating the occurrence of this non-linear phenomenon.
We now discuss how we use the knowledge about the SUTs to define metrics that quantify the occurrence of non-linear phenomena.

In the case of the Crazyflie drone, we identify two non-linear phenomena: 
\begin{enumerate*}[label=(\arabic*)]
    \item the saturation of the actuators, which occurs as the electric motors have a constrained power range and 
    \item the fact that the motors cannot generate negative force, meaning that the propellers cannot generate a force that pulls the drone down.
\end{enumerate*}
Hence, downward movement is achieved only through gravity force.
Both phenomena can be detected when the voltage signal sent to the motors reaches its upper or lower limit.
Accordingly, to quantify the occurrence of non-linear phenomena for a given test, we calculate the percentage of test time during which the actuators are saturated (i.e., when they have reached their upper or lower limit).

As for the DC servos, non-linear phenomena include saturation (of both actuators and sensors), pulse-width modulation, quantisation, and injected non-linearities.
Analogously to the drone, we quantify saturation with the percentage of test time during which they occur.
However, pulse-width modulation and quantisation affect every sensor reading and actuation in the control loop.
Hence, it is not possible to distinguish tests in which they appear more than others.
Regarding the non-linearities injected in separate instances of the SUT, we quantify them as the deviation from how the model would have behaved in their absence.
For example, when we inject non-linear models of friction, we can compare them to the friction value that we would have obtained with a linear model.
Accordingly, at every time instant, we measure the difference between the two values.
We then average such values over the whole test and obtain a metric capturing the degree of observed non-linearity.\footnote{
    We note that this approach can give very different value ranges depending on the specific phenomenon considered.
    For example, the input variation caused by the play found between cogs, i.e., the backlash or the dead-zone, will be much smaller in value than the friction variation caused by the use of a quadratic model instead of a linear one.
    However, regarding the answer to RQ1, we are only interested in the definition of a value that is zero when the model is behaving linearly and that increases when the behaviour differs further from the linear model.
    We are not interested in a comparison of the absolute values.
    Hence, it is not necessary to normalise the data.
}

Finally, as mentioned above, we do not have the same level of detailed knowledge about the non-linear phenomena involved in the lightweight aircraft as we do for the Crazyflie drone and the DC servo.
However, we could identify non-linearity caused by actuator saturation based on our inspection of the model.
This saturation limits the angle that the elevator (the moving surface of the aircraft tail) can take.
In the same way as for the other case studies, we quantify the occurrence of actuator saturation in a test as the percentage of time that the elevator is at the maximum or minimum angle.

\begin{figure*}
    \centering
    \begin{subfigure}[t]{0.4\linewidth}
        \includegraphics{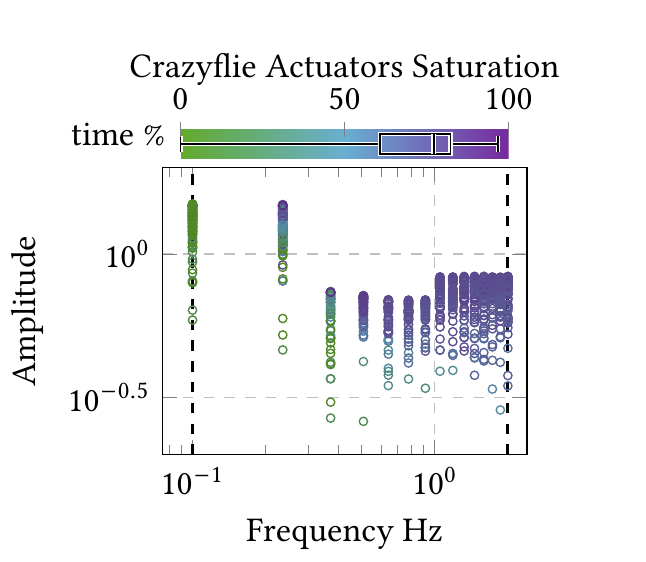}
        \vspace{-0.6cm}
        \caption{}
        \label{fig:cf-saturation-manual-quantification}
    \end{subfigure}
    \hspace{1.2cm}
    \begin{subfigure}[t]{0.4\linewidth}
        \includegraphics{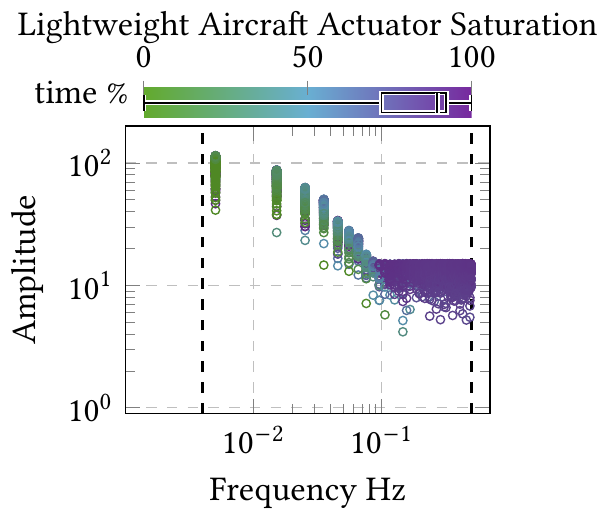}
        \vspace{-0.6cm}
        \caption{}
        \label{fig:lwac-nl-manual-quantification}
    \end{subfigure}
    \caption{These figures present the quantified occurrences of the actuator saturation respectively in the drone tests (plot on the left) and in the lightweight aircraft tests (plot on the right), measured using the defined metrics.
    On the frequency-amplitude plane, each test is captured by a single point corresponding to the main component of the input sequence.
    For each point (i.e., each circular marker), the colour gradient indicates the quantified occurrence of the actuator saturation, where green corresponds to no actuator saturation ($0\%$ of test time) and purple corresponds to complete saturation ($100\%$ of test time).
    The boxplot overlaid on the colourbar highlights the median, minimum, maximum and quartiles of the occurrences of non-linear phenomena in the tests.
    }
    \label{fig:cf-lwac-nl-manual-quantification}
\end{figure*}

\begin{figure*}
    \centering
    \includegraphics{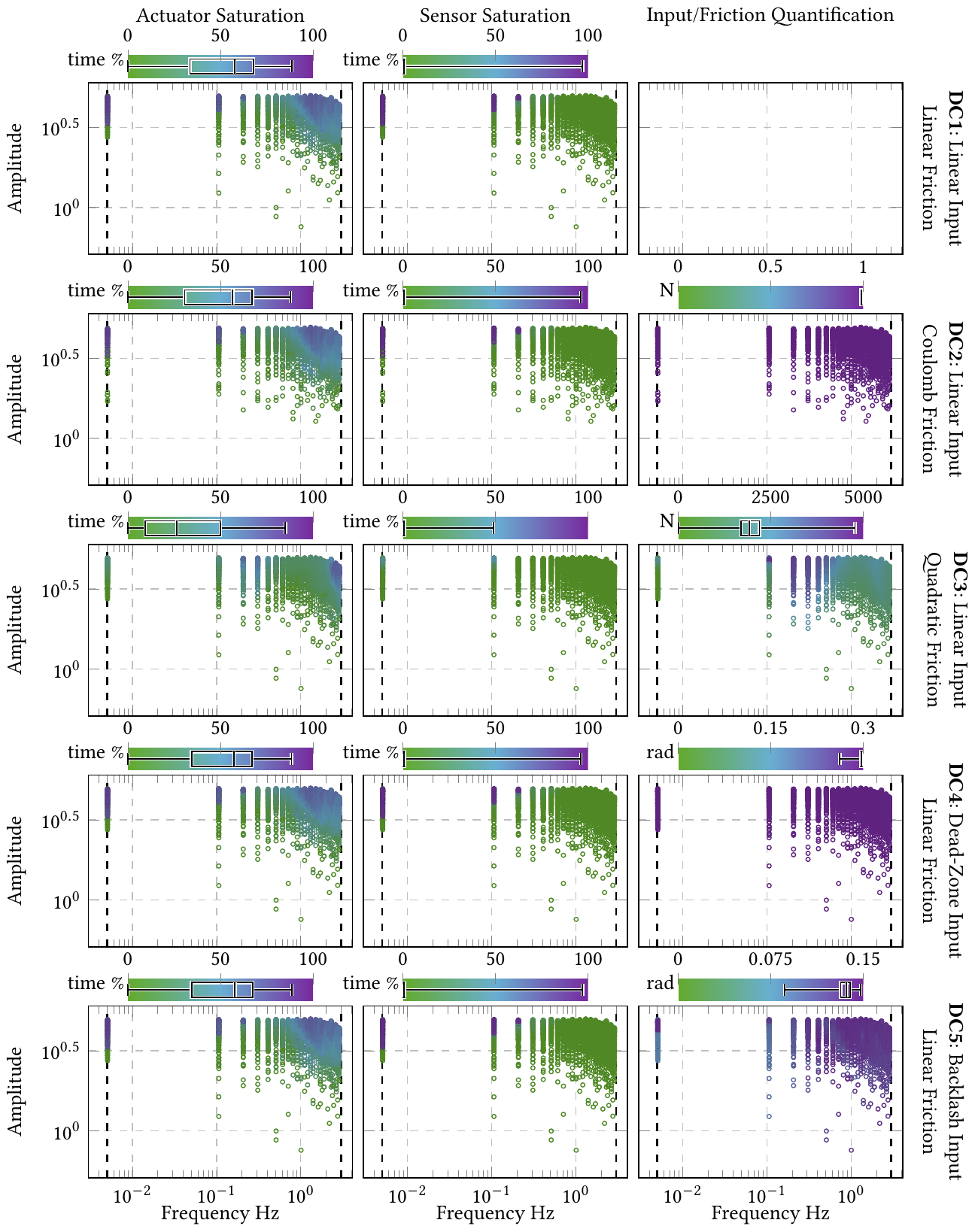}
    \vspace{-2mm}
    \caption{This figure presents the quantified non-linear phenomena occurrences in the different DC servo tests, measured using the defined metrics.
    Each row corresponds to a version of the DC servo.
    The columns correspond to the different non-linearities: the first two show the actuator and sensor saturation analogously to Figure~\ref{fig:cf-lwac-nl-manual-quantification}.
    The rightmost column shows the injected non-linearity in the friction (second and third rows) or in the actuation (fourth and fifth rows).
    The first row reports the DC servo without injected non-linearity, hence the plot is empty.
    For each marker, the colour gradient indicates the quantified non-linear phenomenon occurrence: green are zero values and purple are higher ones.
    The boxplot overlaying the colourbar highlights median, minimum, maximum and quartiles of the non-linear phenomena occurrence in the tests.
    }
    \label{fig:dc-servo-nl-manual-quantification}
\end{figure*}

\partparagraph{Reading Guide for Non-Linear Phenomena Occurrence Plots}
We report the measured occurrence of the different non-linear phenomena in Figure~\ref{fig:cf-saturation-manual-quantification} for the drone, Figure~\ref{fig:lwac-nl-manual-quantification} for the aircraft, and Figure~\ref{fig:dc-servo-nl-manual-quantification} for the different versions of the DC servo.
For the drone and the aircraft, we report plots on the occurrence of actuator saturation.
For the DC servo, in Figure~\ref{fig:dc-servo-nl-manual-quantification}, the rows correspond to the different versions of the SUT.
The columns correspond to the three non-linear phenomena: the actuator and sensor saturations, and the injected non-linearity.
Each plot shows the quantification of a non-linear phenomenon occurrence in the frequency-amplitude plane.
For each test, we use the frequency-amplitude values of the test's main component as coordinates to position the test in the plane.
We then use the colour gradient of the marker (empty circle) for representing the quantified occurrence of the non-linearity phenomenon.
Green corresponds to a zero value of the quantified occurrence, and purple corresponds to the maximum measured value.
Above each plot, we report the colour scale mapping the occurrence values with the colours.
Furthermore, we use the boxplot convention to represent the median, minimum, maximum, and quartiles of the occurrences of each non-linear phenomenon, overlaid on each of the colourbars reporting the gradient scale.
While we are mostly interested in the relationship between the occurrence of non-linear phenomena and the frequency-amplitude content of the input, the boxplots provide better visibility into the frequency of a given non-linear phenomenon in our tests.
In other words, they show how many green or purple points we have in each plot, corresponding to the  occurrence of non-linear phenomena.


\partparagraph{Results}
Using such figures, we now discuss the ability of the test approach to expose different non-linear phenomena.
The \emph{actuator saturation} non-linearity concerns all of our SUTs.
From Figures~\ref{fig:cf-saturation-manual-quantification}, \ref{fig:lwac-nl-manual-quantification}, and the left-most column of Figure~\ref{fig:dc-servo-nl-manual-quantification}, we observe that our tests cover a wide range of saturation percentages for each of our SUTs.
This can be seen from the fact that the scatter plots cover the full scale of the colour gradient and confirmed with the boxplots above them. From the boxplots for the DC servo tests, we can observe that our approach only misses tests exposing more than $90\%$ of saturation time.
This is due to the range of input values considered (one rotation of the motor) combined with the use of a powerful motor.
Such tests could be possibly covered with a higher value for $A_{\max}$: intuitively, higher reference values require higher actuation.
However, such tests would not be interesting as they would go beyond the range of possible angles.
This shows that our testing approach is able to generate test cases that trigger the saturation of actuators to diverse degrees.
Furthermore, from the frequency-amplitude scatter plots, we can highlight that, by covering different amplitude and frequency values, the approach is covering the transition areas of the frequency-amplitude plot where actuator saturation starts to appear.
As predicted, higher values of frequency and amplitude content, as well as the combination of the two, tend to cause the actuator saturation to appear.

When looking at the \emph{sensor saturation}, which concerns only the DC servos, the central column of Figure~\ref{fig:dc-servo-nl-manual-quantification} shows that our tests cover the saturation range between $0\%$ and $90\%$ for all DC servo versions.
Similarly as for the actuator saturation, higher percentages of sensor saturation could be reached for inputs with amplitude larger than $A_{\max}$, but are not interesting since they are known to be outside of the sensor range (equal to $A_{\max}$).
From the plots, we can see that the tests exposing sensor saturation (purple markers) are the ones characterised by large amplitudes and low frequencies.
Intuitively, at high frequencies, the filtering behaviour (which reduces the amount of input signal that is found in the output and hence reduces its amplitude) prevents the sensor saturation from happening.
Furthermore, it is interesting to note that the tests show a sudden transition from tests where the sensor saturation does not appear to tests where the sensor saturation appears to a large extent.
This is also confirmed by the boxplots that show a distribution biased toward the lower values of the metric used to quantify the sensor saturation occurrence.
Such a behaviour is explained by the control algorithm being unable to bring back the DC servo in a state where the sensor is not saturated once the saturation occurs, therefore keeping it in a saturated state for the rest of the test.
Therefore, we can observe that, thanks to the combination of frequency and amplitude coverage, our approach is effective at generating test cases within practically relevant ranges (i.e., smaller than $A_{\max}$) that trigger the saturation of sensors.

Lastly, we look at the other non-linear phenomena that we inject in the different versions of the DC servo.
Those are shown in the right-most column of Figure~\ref{fig:dc-servo-nl-manual-quantification}.
We observe that the coulomb friction and the input dead-zone and backlash are present in all tests to a high degree.
This can be seen both from the boxplots that are all skewed toward the right-hand side of the colorbar, but also from the scatter plots that are mostly purple.
The reason is that such non-linearities affect every movement in the DC servo and therefore are triggered in every test.
The only exception is the backlash that appears less in tests characterised by low frequencies (Figure~\ref{fig:dc-servo-nl-manual-quantification}, bottom-right plot).
In fact, the backlash affects the changes of direction of the input.
Hence, for slowly changing signals (low frequencies), there are fewer changes of direction.
Differently from the other injected non-linearities, the quadratic friction appears to various degrees across the different tests.
In Figure~\ref{fig:dc-servo-nl-manual-quantification}, we can see that this variety is achieved mostly by the coverage of the frequency axis.
In fact, the colour gradient changes mostly when moving horizontally rather than vertically.
Intuitively, friction is a phenomenon related to the rate of change of the system, which is associated to the frequency axis.
The quadratic friction is exposed the most around $0.2\unit{\hertz}$ and $0.3\unit{\hertz}$, which, as seen in Figure~\ref{fig:dc-servo-dof-per-shape}, is the closed-loop bandwidth of the system.
At this frequency the system is behaving the ``fastest'', causing friction to be the largest and hence the difference between quadratic and linear friction to be maximised.
We recall that friction is a function of the speed of motion.
Above $f_b$, the input is filtered, and the non-linearity is exposed less.
This further exemplifies the relevance of sampling the frequency axis to trigger the different non-linear phenomena.

To summarise, our experiments show that the tests generated by our testing approach effectively trigger the non-linearities present in the SUT.
This is done under the constraint of using input values within valid bounds (i.e., smaller than $A_{\max}$).
The saturation of actuators and sensors, as well as quadratic friction, are exposed to various degrees.
Notably, the tests highlight that such non-linearities appear for specific ranges of frequency and amplitude values, hence providing evidence for the relevance of such characterisation of the inputs when testing a CPS.
On the other hand, coulomb friction, dead-zone, and backlash are easily triggered in every execution of the SUT in our tests.
Such phenomena appear for every movement of the DC servo.
Accordingly, they are easily exposed and do not require a dedicated testing approach.

\subsection{RQ2 - Relevance of Metamorphic Relations }
\label{sec:experimental-rq2}
\begin{figure*}
    \centering
    \includegraphics{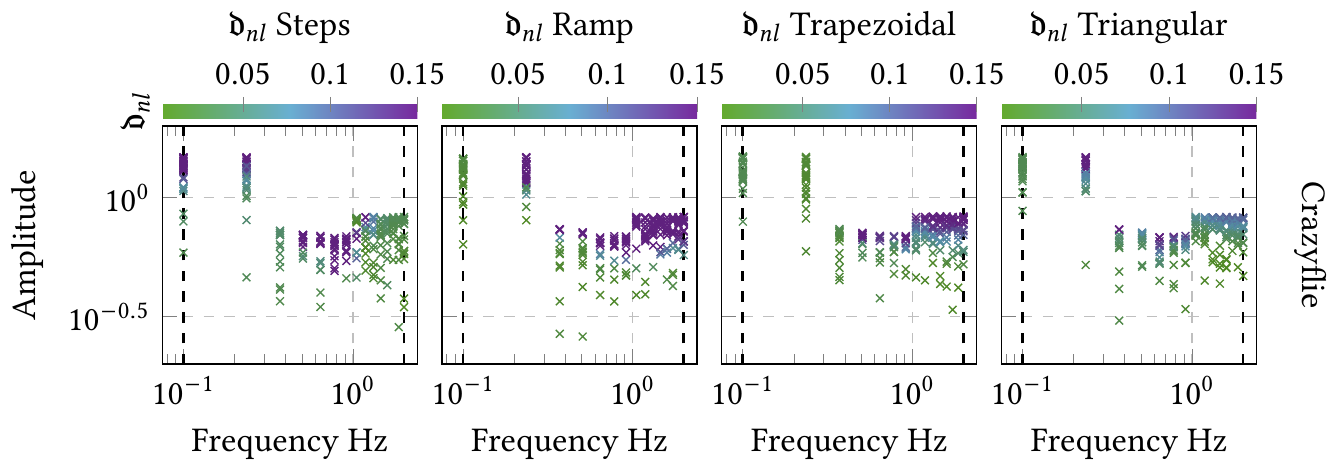}
    \caption{Degree of non-linearity of the drone test cases divided by shapes.
    For each test, the main component is plotted on the frequency-amplitude plane.
    Each point is coloured according to the measured $\dnl$: green corresponds to $\dnl=0$ and purple corresponds to $\dnl=\dnlth$ or greater.
    }
    \label{fig:cf-dnl-per-shape}
\vspace{0.5cm}
\centering
    \includegraphics{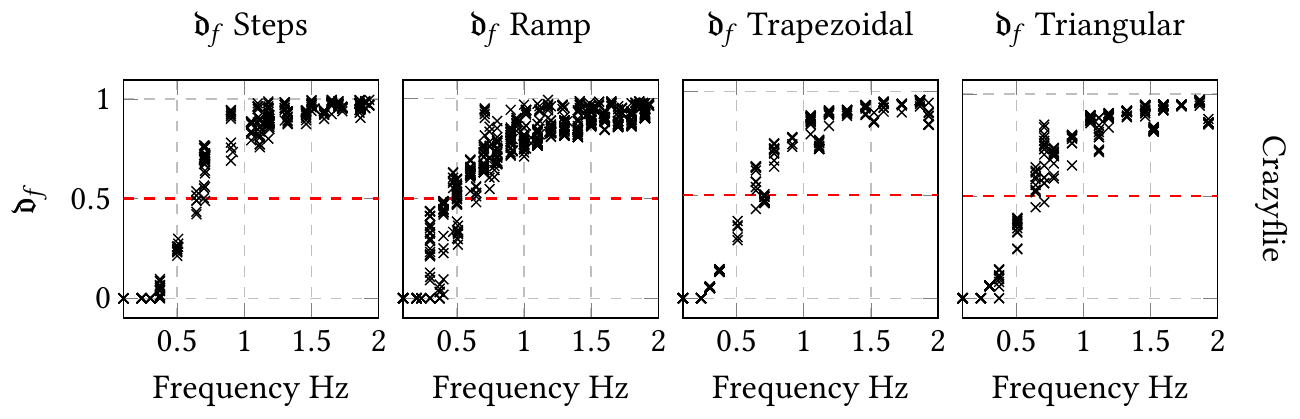}
    \caption{The figure reports the $\dof$ values for all frequency-amplitude components of the tests that show linear behaviour (identified with $\dnl<\dnlth$).
    Since MR2 and MR3 do not depend on the test amplitude for tests exposing linear behaviour, we plot the $\dof$ of each component as a function of their frequency coordinate.
    The plots are also used to identify the closed-loop bandwidth in the tests made with different shapes.
    The bandwidth is identified as the frequency where the $\dof$ becomes larger than the $0.5$ threshold, which is highlighted by the red dashed line.
    Besides some ramp tests most of the results indicate a bandwidth $f_{b}\approx\qty{0.6}{\hertz}$.
    }
    \label{fig:cf-dof-per-shape}
\end{figure*}

\begin{figure*}
    \centering
    \includegraphics{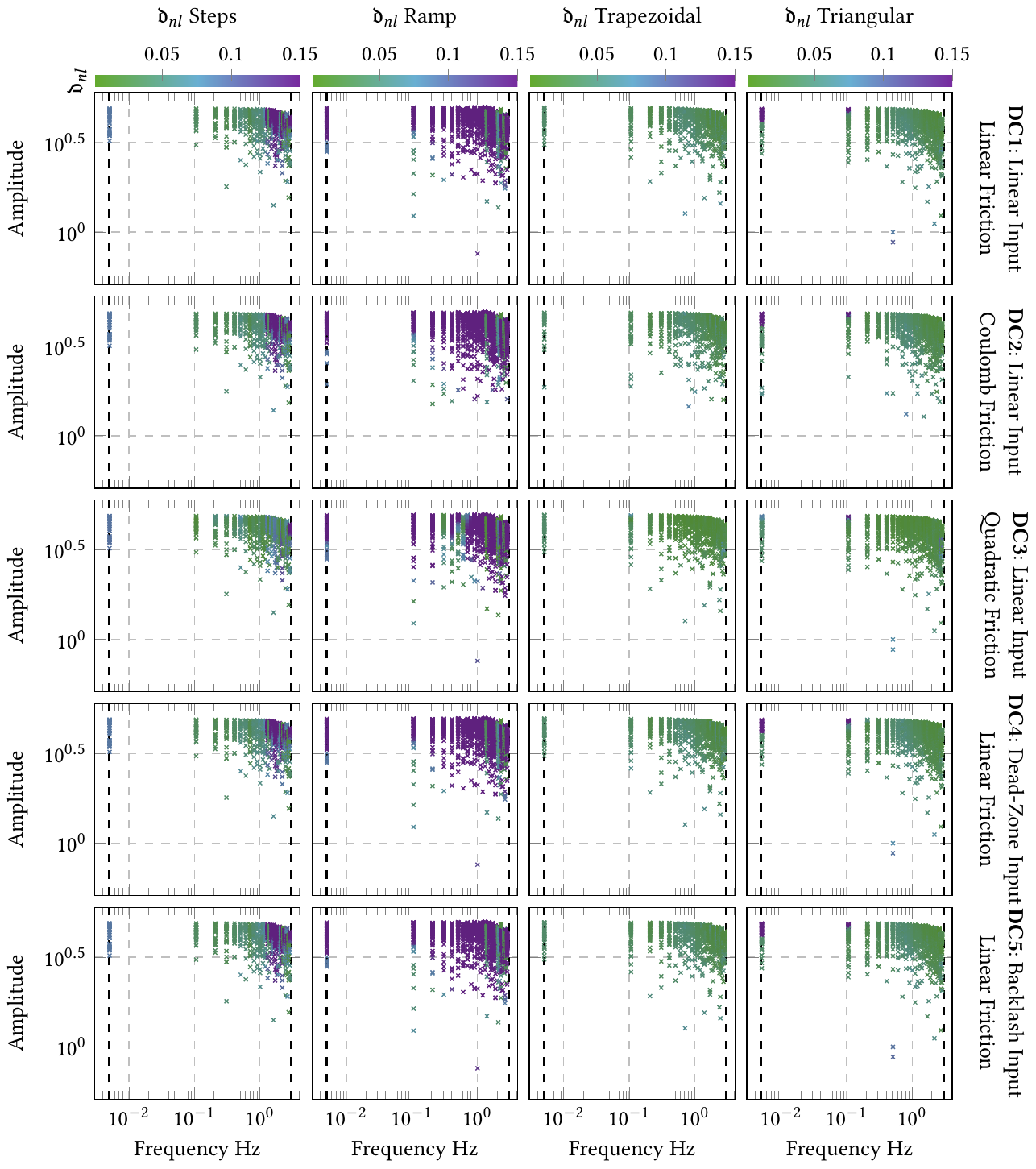}
    \caption{Evaluation of the $\dnl$ metric for the tests with different shapes (the columns) on the different versions of the DC servo (the rows).
    For each test, we plot the frequency-amplitude main component and colour it according to the measured $\dnl$.
    The colour gradient has the same interpretation in every plot and goes from green that corresponds to $\dnl=0$ to purple that corresponds to $\dnl\geq\dnlth$.
    }
    \label{fig:dc-servo-dnl-per-shape}
\end{figure*}

\begin{figure*}
    \centering
    \includegraphics{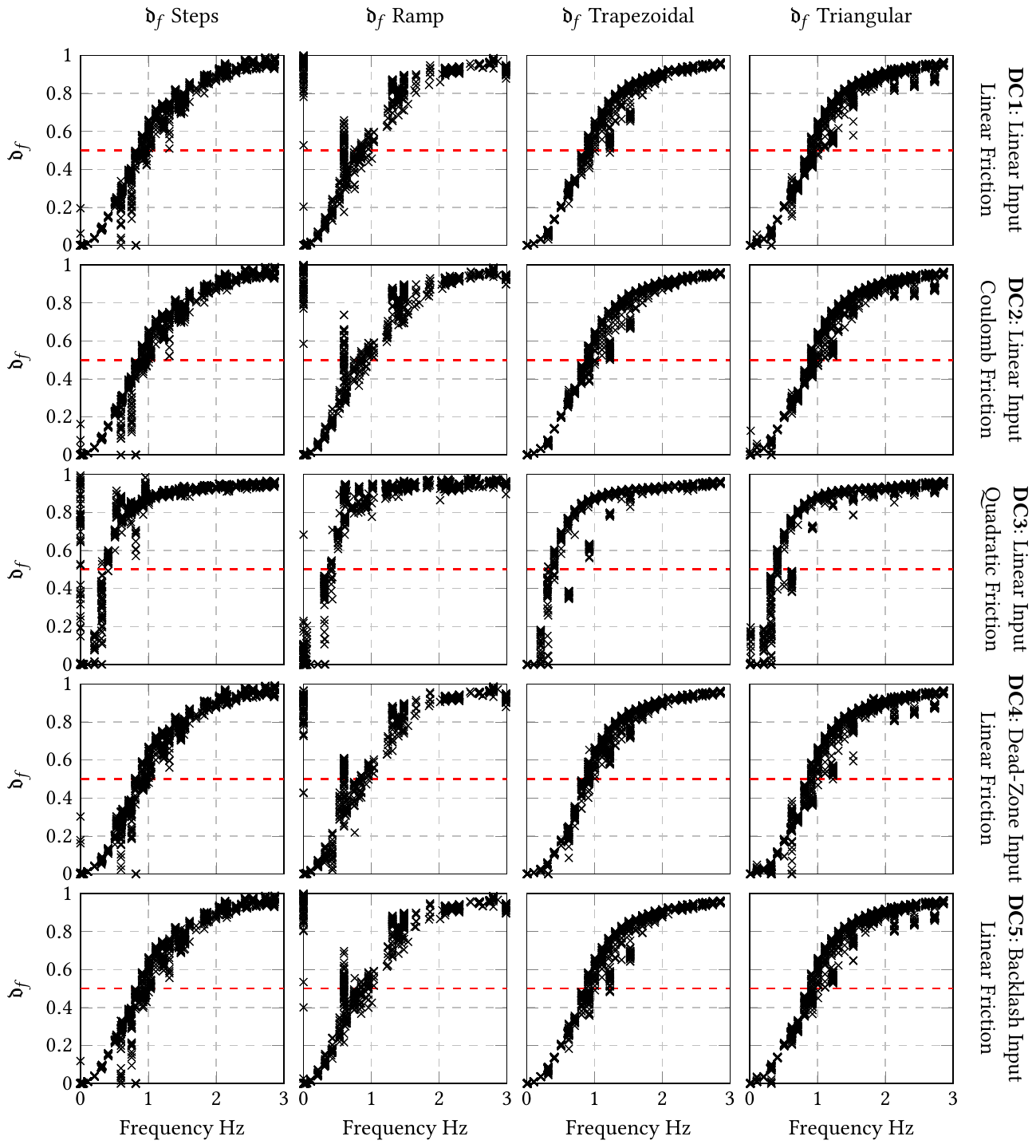}
    \caption{The figure reports the $\dof$ values for all frequency-amplitude components of the DC servo tests showing linear behaviour (identified with $\dnl<\dnlth$).
    The tests are separated into different plots according to the considered version of the DC servo (the rows) and the shape used in the tests (the columns).
    Since MR2 and MR3 do not depend on the test amplitude for tests exposing linear behaviour, we plot the $\dof$ value of each component as a function of their frequency coordinate.
    The plots are also used to identify the closed-loop bandwidth in the tests made with different shapes.
    The bandwidth is identified as the frequency where the $\dof$ becomes larger than the $0.5$ threshold, which is highlighted by the red dashed line.
    For most DC servo versions the tests indicate a bandwidth $f_{b}\approx\qty{0.9}{\hertz}$, the only exception being \ref{dc-qf} that exposes a bandwidth around $f_{b}\approx\qty{0.4}{\hertz}$.
    }
    \label{fig:dc-servo-dof-per-shape}
\end{figure*}

\begin{figure*}
    \centering
    \includegraphics{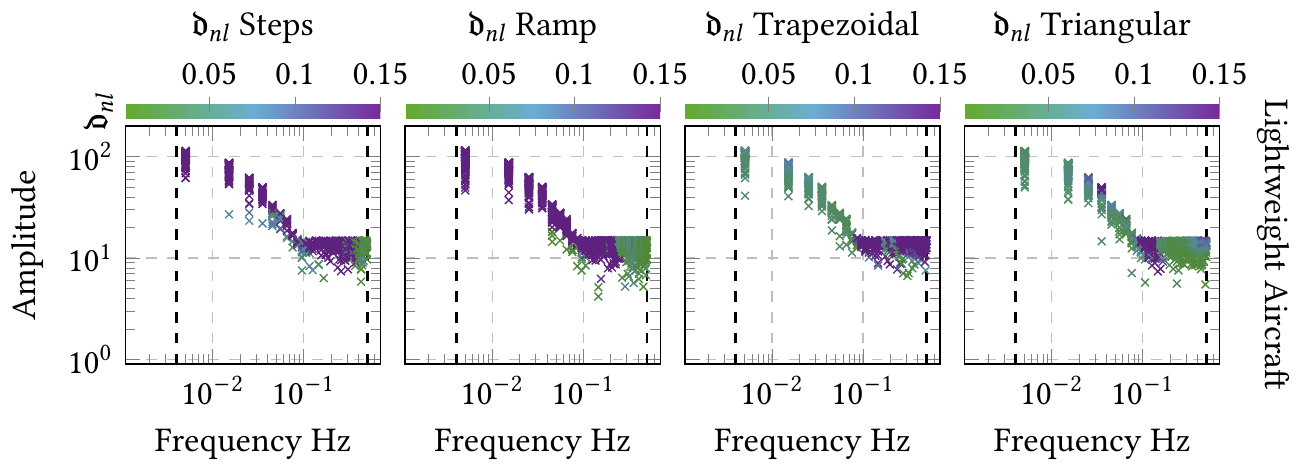}
    \caption{Evaluation of the $\dnl$ metric for the tests with different shapes on the altitude control of the lightweight aircraft.
    For each test, we plot the frequency-amplitude main component and colour it according to the measured $\dnl$.
    The colour gradient has the same interpretation in every plot and goes from green that corresponds to $\dnl=0$ to purple that corresponds to $\dnl=\dnlth$.    
    }
    \label{fig:lwac-dnl-per-shape}
\end{figure*}

\begin{figure*}
    \centering
    \includegraphics{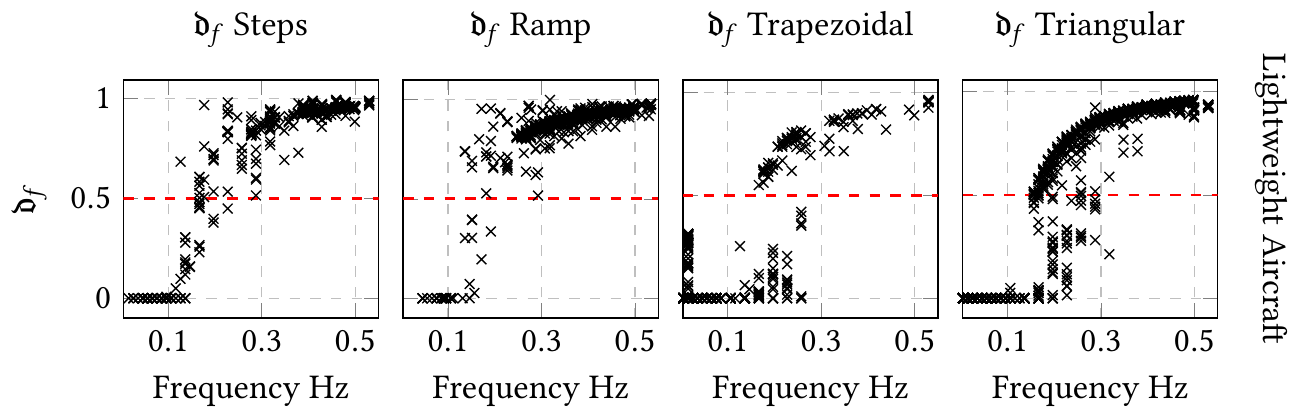}
    \caption{The figure reports the $\dof$ values for all frequency-amplitude components of the lightweight aircarft tests that show linear behaviour (identified with $\dnl<\dnlth$).
    Since MR2 and MR3 do not depend on the test amplitude for tests exposing linear behaviour, we plot the $\dof$ of each component as a function of their frequency coordinate.
    The plots are also used to identify the closed-loop bandwidth in the tests made with different shapes.
    The bandwidth is identified as the frequency where the $\dof$ becomes larger than the $0.5$ threshold, which is highlighted by the red dashed line.
    The tests indicate a bandwidth $f_{b}\approx\qty{0.15}{\hertz}$.
    }
    \label{fig:lwac-dof-per-shape}
\end{figure*}

\partparagraph{Methodology}
We discuss this RQ separately for the different MRs.
For each MR, we visually inspect the output of our testing approach (Figures ~\ref{fig:cf-dnl-per-shape} to~\ref{fig:lwac-dof-per-shape}).
We then identify the tests for which the MRs are not satisfied and discuss the causes.
For the MRs to be practically relevant, the test cases that do not fulfil them should highlight phenomena that limit the SUT design scope.



\partparagraph{Reading Guide for $\dnl$ and $\dof$ Plots}
We now present figures reporting the $\dnl$ and the $\dof$ values computed for the tests executed when applying our testing approach to the SUTs, as well as how the MRs can be evaluated in such plots.
In Figures~\ref{fig:cf-dnl-per-shape} and~\ref{fig:cf-dof-per-shape}, we present the plots for the drone.
In Figures~\ref{fig:dc-servo-dnl-per-shape} and~\ref{fig:dc-servo-dof-per-shape}, we present the plots for the DC servo.
In Figures~\ref{fig:lwac-dnl-per-shape} and~\ref{fig:lwac-dof-per-shape}, we present the plots for the lightweight aircraft.
The figures are based on the execution of Step~3 of the approach (Section~\ref{sec:approach-steps}).

Figures~\ref{fig:cf-dnl-per-shape}, \ref{fig:dc-servo-dnl-per-shape}, and \ref{fig:lwac-dnl-per-shape} report the measured $\dnl$ of every executed test.
Figure~\ref{fig:cf-dnl-per-shape} reports the $\dnl$ results for the drone, Figure~\ref{fig:dc-servo-dnl-per-shape} for the different DC servo versions (where the rows correspond to the different versions of the SUT), Figures~\ref{fig:lwac-dnl-per-shape} for the lightweight aircraft.
Since MR1 (the MR concerning the $\dnl$) discusses tests with the same shape, we separate these plots by shape in the different columns.
As suggested in the approach description, we plot the main frequency-amplitude component of each test on the frequency-amplitude plane.
This representation is analogous to the non-linear phenomena occurrence plots.
However, to distinguish them, we use here a different marker, a cross.
In fact, as discussed in Section~\ref{sec:dnl-def}, the $\dnl$ metric aims to detect the impact of non-linear phenomena on the CPS, not their sole occurrence.
Like in Figure~\ref{fig:bh-examples-freq}, we use a logarithmic scale on both axes for readability.
For each point, the colouring corresponds to the $\dnl$ value measured in the test.
Green corresponds to $\dnl=0$ (linear tests within the design scope) and purple corresponds to $\dnl\geq0.15$ (non-linear tests outside of the design scope).
Different from the non-linear phenomena occurrence plots, the colour gradient is the same in all of these plots.
MR1 states that the $\dnl$ value should increase when moving to the right or up in the frequency-amplitude plane.
Practically, this corresponds to the colour gradient transitioning from green to purple for increasing amplitudes and frequencies (i.e., when moving away from the origin).

Figures~\ref{fig:cf-dof-per-shape}, \ref{fig:dc-servo-dof-per-shape}, and \ref{fig:lwac-dof-per-shape} report the $\dof$ values measured for all the frequency-amplitude points of the linear tests, for which $\dnl<\dnlth$.
Figure~\ref{fig:cf-dof-per-shape} reports the $\dof$ values for the drone, Figure~\ref{fig:dc-servo-dnl-per-shape} for the DC servo (where again the rows correspond to the different versions of the SUT) and Figure~\ref{fig:lwac-dof-per-shape} for the lightweight aircraft.
Since MR2 discusses tests with the same shape and MR3 compares the evaluation of the bandwidth across different shapes, we present the $\dof$ plots by shape (corresponding again to the different columns).
Since MR2 and MR3 do not concern the amplitude, we plot the $\dof$ value (the vertical coordinate) of each point as a function of its frequency content (the horizontal coordinate).
Different from the previous figures, we use here a linear scale on both axes.
In order to ease the identification of the closed-loop bandwidth $f_b$, we highlight with a red dashed line the $0.5$ threshold for the $\dof$ metric.
MR2 states that high frequency content should correspond to the appearance of filtering behaviour.
Practically, this corresponds to an increasing $\dof$ for increasing frequencies.
Using the $\dof$ metric, we can also evaluate MR3.
To fulfil MR3, the frequency at which the $\dof$ value becomes larger than $0.5$ has to be similar across the different shapes.

\partparagraph{Results - MR1}
For the drone (Figure~\ref{fig:cf-dnl-per-shape}), we observe that MR1 holds for low frequencies (lower than $\qty{1}{\hertz}$) across the different shapes.
However, for frequencies higher than $\qty{1}{\hertz}$, we observe that an increasing number of tests show linear behaviour (green points) and do not comply with MR1.
This appears clearly for tests with steps and triangular shapes, and, to a lesser extent, for the remaining shapes.
Despite the quantified occurrence of non-linear phenomena depicted in Figure~\ref{fig:cf-saturation-manual-quantification} shows that the actuators are fully saturated in that frequency range (practically all markers are purple above $\qty{1}{\hertz}$), the $\dnl$ value in such tests is low.
This is due to the filtering action that mitigates the response of the control loop and prevents the non-linearity from introducing new frequency content in the output.
This statement can be further verified by observing in Figure~\ref{fig:cf-dof-per-shape} that the inputs in the frequency range above $\qty{1}{\hertz}$ all show a $\dof$ value close to $1$.

Concerning the DC servo (Figure~\ref{fig:dc-servo-dnl-per-shape}), we observe a similar pattern to the one for the drone tests with the steps shape (as well as some of the high-frequency ramp tests).
We note that across shapes, except for the trapezoidal tests, tests at the low frequencies and high amplitudes (top-left of the plots) expose a higher $\dnl$ value and do not comply with MR1.
By comparing  with the quantification of the non-linear phenomena occurrence in Figure~\ref{fig:dc-servo-nl-manual-quantification}, we can observe that the high $\dnl$ value is due to the combination of sensor and actuator saturation (first and second column).
As such tests do not comply with MR1, they provide evidence about using $\dnl$ for detecting non-linear phenomena that can limit the CPS design scope.

Among the DC servo tests, it is notable that almost all of the ramp tests show high $\dnl$ though the proposed metrics, which quantify the occurrence of non-linear phenomena (Figure~\ref{fig:dc-servo-nl-manual-quantification}), do not detect a specific non-linear phenomenon.
The likely reason for this discrepancy with the other shapes is that the ramp is the shape with the highest number of relevant frequency components (i.e., the frequency-amplitude components selected by Equation~\ref{eq:famap}).
This larger number of relevant frequency components shows that such input exercises more the SUT by feeding more frequencies at the same time to the system.
Furthermore, we recall that, in the system, there are other non-linear phenomena that could not be quantified (i.e., pulse-width modulation and quantisation).
Hence, we can suppose that these tests fail because of such phenomena, or a combination of them.
On the other hand, we note that for lower amplitudes and frequencies some ramp tests yield a lower $\dnl$.
This still partially supports the validity of MR1.

For the lightweight aircraft, we note similar patterns as for the other case studies.
First, we observe that most of the tests with steps and ramp shapes show non-linear behaviour.
As for the DC servo, this is likely due to the large number of main frequency components that characterise such signals.
We can also observe that tests in the low frequency range and low amplitudes show a low $\dnl$ value.
As for DC servo, this supports the validity of MR1.
Second, similar to the drone, we also observe that some high-frequency tests retain a lower $\dnl$ despite the occurrence of actuator saturation (Figure~\ref{fig:lwac-nl-manual-quantification}).
This can be attributed to the filtering action that reduces the stress on the SUT and helps retain linear behaviour.

To conclude, our tests results show that MR1 does not always hold.
According to our experiments, the two things that can invalidate MR1 are the unpredictable nature of non-linear phenomena (like the sensor saturation for low frequencies in the DC servo) and the filtering action at high frequencies (as observed for the drone and lightweight aircraft).
The falsification of MR1 contradicts the qualitative characterisation in Figure~\ref{fig:input-space}.
Practically, this implies that a testing strategy leveraging the frequency and amplitude dimensions to expose non-linear behaviours cannot only aim at exploring large amplitude and frequency values.
Instead, it is also important to cover the whole input range (e.g., the low frequency range tests in the DC servo that cause sensor saturation).
Most notably, we observe that the $\dnl$ metric is successful in distinguishing the scenarios in which the non-linear phenomena impact the CPS behaviour (hence bringing the CPS out of its design scope) from the scenarios where they do not (e.g., when the filtering action helps retain the linear behaviour despite the actuator saturation).

\partparagraph{Results - MR2}
In order to verify if MR2 applies to our tests results, we inspect Figures~\ref{fig:cf-dof-per-shape} for the drone, Figure~\ref{fig:dc-servo-dof-per-shape} for the DC servo, and Figure~\ref{fig:lwac-dof-per-shape} for the lightweight aircraft.
We observe that, in general, $\dof$ shows an increasing value for increasing frequencies, hence fulfilling the MR.
The exceptions are some ramp tests that show an high $\dof$ for low frequency for all the DC servo versions, as well as some step tests on the DC servo injected with quadratic friction.
Furthermore, some trapezoidal low-frequency tests for the lightweight aircraft also show high $\dof$ values for low frequency components.
Upon manual inspection, we observe for the DC servo that those tests are all characterised by large amplitudes that cause saturation (of both sensors and actuators) to occur.
However, in those tests, the output neither follows the reference (hence the high $\dof$) nor exposes new frequency components.
For this reason, those tests do not have a high $\dnl$ and were not excluded by the analysis.
For the lightweight aircraft, the trapezoidal tests associated with such points mostly expose a linear tracking behaviour, meaning that they overall follow the input reference.
However, these tests also show unexpected deviations during flight, specifically when transitioning from a constantly increasing input to a constant one.
Such deviations are not detected by the $\dnl$ metric as they are small relative to the input amplitude\footnote{
    Recall that the metric is normalised over the input amplitude.
} but they do reduce in amplitude one of the components in the low frequency range, and are hence detected using the $\dof$ metric.
Given the limited knowledge that we have about the lightweight aircraft, we are not able to identify the root cause of the failure of such tests.
However, it is apparent that such tests show misbehaviour and require further investigation by the engineers who develop the CPS.

Such tests expose a \emph{limitation of the $\dnl$ metric} regarding its capacity to highlight tests that push the system outside of the control design scope.
However, our results show that \emph{the verification of MR2 can compensate for this limitation and be used to complement the $\dnl$ metric}.
In fact, the verification of MR2 highlights test cases in which the low frequency components (i.e., frequencies below the bandwidth that should be tracked) are not tracked.

\partparagraph{Results - MR3}
To evaluate MR3, we examine the $\dof$ plots for our different case studies.
We observe that, for each of the SUTs, the crossing of the $0.5$ threshold happens around the same frequency independently of the specific input shape considered for most of the tests.
This frequency value is around $\qty{0.6}{\hertz}$ for the drone, just below $\qty{1}{\hertz}$ for all the versions of the DC servo (with the exception of~\ref{dc-qf} where it is below $\qty{0.5}{\hertz}$\footnote{
    Intuitively, the quadratic friction present in~\ref{dc-qf} makes the DC servo slower.
}) and around $\qty{0.15}{\hertz}$ for the lightweight aircraft.
For the drone, among the various input shapes, the only exceptions are some of the ramp tests that show a bandwidth lower than the other shapes.
We identify the tests that invalidate MR3 as those tests that show components with a $\dof$ greater than $0.5$ at frequencies lower than $\qty{0.6}{\hertz}$.
Upon inspection, we observe that in such tests the drone hits the ground.
However, hitting the ground limits the drone's movements but does not introduce any new components in its output.
Hence, this phenomenon cannot be detected using the $\dnl$ metric.
This is similar to what we observed in some of the ramp tests for the DC servo when evaluating MR2.

In the case of the lightweight aircraft, we observe that in some trapezoidal and triangular tests the CPS is able to track frequencies higher than the bandwidth of $\qty{0.15}{\hertz}$.
We identify such tests as having a component with $\dof$ smaller than $0.5$ for frequency higher than $\qty{0.2}{\hertz}$.
When manually inspecting such tests, we observe that they are characterised by small amplitude values (in the order of tens of meters, when the inputs can take values up to several hundreds of meters).
This suggests that the ability of the lightweight aircraft to track fast reference inputs can depend on the input amplitude, invalidating MR3 and indicating non-linear behaviour.\footnote{
    Given our limited knowledge of the SUT, we cannot identify the specific phenomenon causing this.
    Intuitively, in some systems, tracking inputs with small amplitude values is easier than tracking inputs with large amplitude values.
}
Conversely, for the steps and ramp shapes, we observe some tests that seem to show a lower closed-loop bandwidth.
We identify such tests as having a component with $\dof$ larger than $0.6$ for frequency lower than $\qty{0.15}{\hertz}$.
When manually inspecting such tests we observe that, similarly to the trapezoidal tests discussed for MR2, they are characterised by the appearance of some undesired behaviours that are however relatively small compared to the input and do not cause the $\dnl$ value to become larger than $0.15$.
Therefore, similar to MR2, we can observe that MR3 holds in most of the tests and when it is invalidated, it complements the $\dnl$ metric in detecting tests where the SUT goes out of the design scope.

To summarise, we observe from our tests that MR1 generally holds but can be invalidated by
\begin{enumerate*}
    \item the filtering action at high frequencies (that increases the robustness of the CPS to non-linear phenomena by reducing their reaction to the reference changes, like for the drone), or
    \item by specific non-linear phenomena (that cause unpredictable behaviour of the CPS, like for the sensor saturation in the DC servos).
\end{enumerate*}
This latter point supports the use of MR1 for the detection of test cases associated with the appearance of non-linear phenomena specific to the SUT.
MR2 and MR3 hold in the majority of our tests.
When they do not hold, our tests show that they can be used to complement the $\dnl$ metric to detect test cases that push the system outside of the design scope.
Hence these MRs compensate for the observed limitations of the $\dnl$ metric.

\subsection{Threats to Validity}
\label{sec:experimental-limitations}
We discuss the threats to the validity of our work in terms of external and internal validity.
Concerning the external validity, we discuss the generalisability of our observations to other CPSs.
We developed the CPS models of the drone and the DC servo used in our case studies, and hence might have biased the results.
However, we built these CPS models based on the actual documentation of the real implementations of the CPSs.
This supports the claim that our case studies are representative of real-world applications.
Furthermore, we complement these two case studies with the lightweight aircraft that was independently developed.
Notably, the observations that we can draw for this latter case study are the same as for the two previous ones.
This further supports the general validity of the findings and their applicability to other CPSs.

Concerning internal validity, we discuss the limitations of our research methodology.
It can be noted that we based significant part of the discussions of the RQs on our own understanding of what the non-linear phenomena that affect our SUTs are, especially for the drone and the DC servo.
Since we developed these simulation models based on the actual physical systems (both available to us), we are confident in our complete understanding of the behaviour of these systems.
Furthermore, having access to the physical systems allowed us to compare to the actual implementation of the CPSs when uncertain about the model design choices (e.g., the specific quantisation of the sensors and the parameters of the control algorithms).

Finally, it can be noted that we used manual inspections of frequency-amplitude plots to evaluate the satisfaction of our MRs.
However, analysing such plots is rather intuitive and helped associate the physical phenomena with the features of the input (e.g., the quadratic friction affecting more the fast-changing inputs).
This enabled a deeper understanding of how the non-linear phenomena can limit the design scope of the CPS under test.
Nevertheless, an automated inspection method of our MRs is desirable.
We leave the development of such method to future work.

\subsection{Data Availability}
\label{sec:experimental-data}
The code of the SUT models and the code implementing the testing approach are available at the following repositories:
\begin{itemize}
    \item \textbf{Crazyflie}: \url{https://doi.org/10.5281/zenodo.7994669},
    \item \textbf{DC servo and Lightweight Aircraft}: \url{https://doi.org/10.5281/zenodo.7994698}.
\end{itemize}
These repositories come with instructions to reproduce all of the experiments.
Since the tests take some time to execute, we provide the test output traces that can be used to obtain the figures of the article. 
\begin{itemize}
    \item \textbf{Crazyflie}: \url{https://doi.org/10.5281/zenodo.8043446},
    \item \textbf{DC servo}: \url{https://doi.org/10.5281/zenodo.8043260},
    \item \textbf{Lightweight Aircraft}: \url{https://doi.org/10.5281/zenodo.8043351}.
\end{itemize}

\section{Approach Applicability and Limitations}
\label{sec:experimental-discussion}
Our experiments show that the proposed testing approach is able to generate stress test cases for individual control loops of control-based CPSs.
More specifically, our testing approach generates scenarios that falsify the design assumptions associated with the linearisation of the physics models (RQ1).
Furthermore, the proposed metrics ($\dnl$ and $\dof$), together with the falsification of the associated MRs, are able to identify the tests in which the non-linear phenomena actually impact the CPS ability to track the input reference.
We now discuss the technical aspects that impact the practical applicability of the approach, as well as its applicability to different CPSs.

\subsection{Identification of Non-Linear Phenomena in Practice}
In practical scenarios, evaluating the occurrence of non-linear phenomena (e.g., Figures~\ref{fig:cf-saturation-manual-quantification} and \ref{fig:dc-servo-nl-manual-quantification}) requires detailed knowledge and understanding of the SUT, which may not always be available.
Even more importantly, even if the phenomena are known, it can be hard to know a priori which of them actually limits the CPS design scope.
For example, in the DC servo, it cannot be known a priori that the sensor and actuator saturation will appear and push the CPS out of its design scope.
Conversely, the coulomb friction, the dead-zone, and the backlash non-linearities are triggered in the tests, but they do not affect the CPS design scope.
To identify the test cases where non-linear phenomena appear and limit the design scope of the control algorithm, we showed, for our SUTs, that $\dnl$ and MR1 can be used (RQ2-MR1).
Our experiments also showed a limitation in the detection of the design scope based on the quantification of new frequency components in the output (RQ2-MR2, and RQ2-MR3).
This is because, in some specific cases, the non-linear phenomena might not introduce large enough new frequency components, causing the $\dnl$ metric to take small values.
However, such tests are detected using MR2 and MR3 since they invalidate the MRs by exposing filtering behaviour at low frequencies (where we would instead expect tracking behaviour) or an inconsistency in the observed closed-loop bandwidth.
This shows that the $\dof$ metric and the associated MRs can compensate for the limitation of $\dnl$ and support the detection of test cases that push the SUT out of its design scope.

\subsection{Campaign Duration}
Concerning test case generation, we note that the proposed rules of thumb for the required input of the approach (Section~\ref{sec:approach-steps}, Step~0) generate a high number of test cases (Table~\ref{tab:num-tests}).
To limit the number of tests, Step~$1$ of the approach aims at reducing the number of test cases needed to cover the input space by restricting the amplitudes ranges to be tested.
Across SUTs, this step is effective for the drone and the lightweight aircraft but not for the DC servo versions.
In fact, for all the versions of the DC servo, the entire amplitude input range can show linear behaviour, which is a desirable property of the system (see the tests with the trapezoidal and triangular shapes in Figure~\ref{fig:dc-servo-dnl-per-shape}).

The high number of tests can be a limiting factor when considering that we are testing only one control loop of a CPS, which typically has several control loops.
In practice, not as many tests are needed, and similar conclusions about the SUT could drawn with fewer frequency-amplitude points in Figures from~\ref{fig:cf-dnl-per-shape} to~\ref{fig:lwac-dof-per-shape}.
For example, in Figure~\ref{fig:dc-servo-dnl-per-shape} not as many green and purple points are needed to identify the areas where non-linear phenomena impact the CPS performance.
We leave to future work the challenge of minimising the number of test cases needed to stress test a control loop.
The problem of minimising the number of test cases can be seen as a testing coverage or stopping criterion problem in the frequency-amplitude plane.
In this work, instead, we focus on generating effective test cases that falsify the design assumptions of the SUT.
However, as reported above, the execution of the testing campaign for our SUTs took some hours for the drone and the lightweight aircraft, and still no more than a day for the DC servo versions.
Therefore, despite the large number of tests, for our case studies, the duration times do not hinder the practical applicability of our testing approach.

\subsection{On the use of Different Shapes}
We conclude the technical part of this discussion by noting the difference in results for the different shapes of the tests.
This is expected since these shapes can alter the way the SUT is exercised and hence expose different behaviour.
Interestingly, the ramp and step tests more easily expose non-linear behaviour.
As mentioned above, this is likely due to the large number of frequency-amplitude components that are included in a ramp and step test.
Furthermore, we can also note that there is a similarity between trapezoidal and triangular tests: in fact, they both map to one larger lower-frequency component and a number of higher-frequency components with smaller amplitude.
While the investigation of relevant patterns of frequency-amplitude components for the definition of shapes is an interesting aspect that needs further research, we can observe that, with the proposed shapes, the testing approach already shows promising results regarding the exposition of the CPS design scope.

\subsection{General Applicability of the Approach}
To evaluate the general applicability of the approach, we discuss how the answers to the RQs change across different SUTs and non-linear phenomena.
We note that our case studies include both the same CPS with different non-linear phenomena (the DC servo) as well as the same non-linear phenomenon in different CPS: the actuator saturation in fact appears in each of our CPS.
The variety of non-linear phenomena is important as we want the approach to be able to trigger each of them.
The same non-linear phenomenon in different CPS is also important as specific features of the CPS can change its impact on the design scope.
Concerning RQ1, across all three case studies, the approach has been able to generate test cases that expose non-linear phenomena.
More specifically, saturations and quadratic friction are exposed to various degrees, while the remaining non-linearities are always exposed to a high degree in all tests.
The sensor saturation was detected in only a small number of test cases (for all the versions of the DC servo).
As discussed, this is possibly a consequence of the choice of the input parameter $A_{\max}$.
Increasing this parameter value could increase the number of tests that cause sensor saturation.
However, such tests would be ``trivially'' outside of the DC servo's design scope, as they correspond to reference values outside of the range of the sensor.\footnote{
    Such test cases should be considered instead for robustness testing (with invalid inputs).
    In this work we focus instead on valid inputs.
}
The input and friction non-linearities, on the other hand, are easily triggered in every test.
However, as they are not associated with high $\dnl$ or the falsification of the MRs associated with the $\dof$ metric, we observe that they do not impair the CPS design scope.\footnote{
    It is important to note, however, that such non-linearities (coulomb friction, dead-zone and backlash) can still impair other aspects of system performance, e.g., the accuracy of the reference tracking.
}

Concerning RQ2, MR1 shows different validity and gives different types of insights across SUTs.
For the actuator saturation of the drone and the lightweight aircraft, MR1 is violated because of the robustness introduced by the filtering action.
Such robustness also appears in the DC servo studies, although to a smaller degree.
For the DC servo, the sensor saturation invalidates MR1, as $\dnl$ increases for decreasing frequencies, specifically, when going from $\qty{0.1}{\hertz}$ to $\qty{0.01}{\hertz}$ in Figure~\ref{fig:dc-servo-dnl-per-shape}.
In contrast, the verification of both MR2 and MR3 (as a complement to the $\dnl$) does not show significant differences across our SUTs.
Most notably, the ramp tests that do not satisfy MR2 were similar across all of the SUTs.

To summarise, in our case studies, the effectiveness of our stress test case generation (RQ1) and the validity of MR2 and MR3 do not depend on the specific SUT.
However, the verification of MR1 shows  differences related to the specific design bounds of the different SUTs.
Specifically, it highlights the high-frequency robustness to saturations of the drone and the issues with sensor saturation for low frequency in the DC servo.
This is desirable as it helps identify the relevant phenomena that characterise a given SUT.
As the results obtained from testing the three SUTs are consistent and promising, we believe our approach can be applied to other SUTs.

\subsection{Practical Implications of the Design Scope}
We conclude this discussion by addressing the practical implications of the proposed testing approach.
We believe that the CPS design scope information provided by our testing approach can be used both for CPS verification and validation, as well as for CPS development.
Concerning the verification and validation, the identification of the design scope can be used to distinguish relevant tests for which the requirements need to be evaluated.
More practically, if a test is outside of the CPS design scope, then it will trivially not fulfil the requirements and hence is not interesting for the verification of the requirements.
Such tests can be considered invalid inputs.
Furthermore, the design scope information can be used for runtime verification to implement checks on reference sequences that might push the CPS out of its design scope (e.g., large step changes in the case of the lightweight aircraft).
Leveraging such runtime checks, the software can either automatically re-shape the signals or raise warnings to the user.

Regarding the development of the CPS, the proposed approach can provide feedback on the design choices that limit the design scope.
For example, if the observed design scope does not contain scenarios in which the CPS is expected to perform in a predictable way, then there is a need for re-designing the system to enlarge its scope.
Using this information, the engineers can therefore assess if the system achieves the desired performance or if some of its components (e.g., actuators, senors, or sensor signal filters) need to be re-designed.
 \section{Related Work}
\label{sec:related}
The topic of testing control-based CPS software is not new in the software engineering literature.
In fact, control systems are seen as a class of CPSs in which most of the added value is placed in the software part~\cite{Broy:2007}.
Accordingly, recent works highlighted interesting research directions at the intersection of control and software engineering~\cite{Balasubramaniam:2020, Bradley:2020}.
One highlighted direction is the relation between control models and the software implementation of the CPS.
Our work explores this direction by testing the CPS implementation to identify the practical validity of the control models (i.e., the design scope).

Testing of control systems has been approached in the literature in the form of model testing, falsification, and runtime verification. Given the widespread industrial use of tools like Mathworks Simulink,
model testing has received significant attention in the last decade~\cite{Briand:2016}.
Recent work showed the complementarity of model testing and model checking for the verification of CPS requirements~\cite{Nejati:2019}.
Generation of test traces is one of the main topics in CPS model testing.
Very different types of algorithms are used for the generation of input sequences: search algorithms~\cite{Matinnejad:2014, Matinnejad:2017, Marculescu:2015,Aleti:2015,Hansel:2011}, classification trees~\cite{Lamberg:2004}, and system-identification based refinements~\cite{Menghi:2019}.
We also note a number of application-specific works in the avionics~\cite{Peleska:2002, White:2001, Samad:2003} and automotive~\cite{Bringmann:2008, Bringmann:2006} domains.
In contrast to previous work, this article is the first one to propose a test-case generation approach based on a frequency-amplitude description of the input space.
Among the previous works on model-testing, we highlight as closely related the work by Matinnejad et al.~\cite{Reza:2016}, which puts significant emphasis on the exploration of different values of the first and second derivatives (see~\cite[Fig. 2]{Reza:2016}).
The derivative values are directly connected to the frequency and amplitude dimensions of this article.
In fact, the derivative is the ratio between the change in a function input and output: changing the output amplitude or scaling the time automatically (as we do in our test case parametrisation, Equation~\ref{eq:tc-parametrisation}) directly enables the exploration of different values for the derivatives.
As improvement with respect to this previous work, our explicit use of the frequency-amplitude characterisation enables leveraging control-theoretical knowledge and specifically the definition of our proposed qualitative characterisation and MRs.

CPS falsification (as the generation of test cases that falsify a given requirement) is also an active research direction.
In a recent work~\cite{Yamagata:2021} the authors use deep reinforcement learning to perform robustness-guided falsification.
Interestingly, they note the importance of making the system internal dynamics available to the reinforcement learning algorithm, which highlights the possible benefits of leveraging control-design information during the testing.
Furthermore, a combination of model checking and path planning can be used to invalidate linear-temporal-logic formulas~\cite{Plaku:2009}.
Similar problems can be addressed with rapid exploration of random trees~\cite{Dreossi:2015}.
With respect to previous work on the falsification of CPS requirements, in this article we propose the novel perspective on the falsification of the control design assumptions.

The verification of CPSs has been addressed in the software engineering literature also from the point of view of the runtime verification.
Most notably, we remark some works that expand traditional logic (e.g., signal temporal logic) with frequency-domain descriptions of signals under the motivation of verifying data-flow languages (such as Simulink)~\cite{Chakarov:2012, Donze:2012}.
Such works are aligned with this article, highlighting the relevance of the frequency-domain description of signals for CPSs and complement it with its application to runtime verification.
Another recent work highlights the need to expand the runtime verification theory for its application to CPSs, and proposes a new Hybrid Logic of Signals to express desired CPS properties that concern both the cyber and physical components~\cite{Menghi:2021}.
This logic has been extended to include also source-code-level specifications~\cite{Dawes:2022}.
To conclude the discussion on runtime verification, we report on a first work that leverages system dynamics (expressed as differential equations, i.e., the same type of models used at control design time) to either terminate the monitoring early, or to skip samples of the signal~\cite{abbas:2022}.
While runtime verification is out of the scope of this article, the cited works highlight the relevance of the frequency description of signals, the need to extend software engineering tools for CPSs, and the potential of leveraging control models.

The control and robotics communities show growing awareness of the impact of the software implementation of control algorithms on their performance~\cite{Zimmer:2015, Silano:2018}.
Zimmer et al.~\cite{Zimmer:2015} discuss a case study on the consequences of implementation choices for the control performance.
In robotics, software-in-the-loop simulations (i.e., simulations that include the actual control software implementation) can be effective in exposing bugs in the software design of quadcopter controllers~\cite{Silano:2018, Timperley:2018}.
However, such works remain isolated efforts and the leveraging of control theory during the testing process remains limited.

We conclude the discussion of the related literature by discussing the few works that leverage control-theoretical knowledge in the testing of control-based CPSs~\cite{Aleti:2015, Menghi:2019, He:2019}.
In these works, system identification~\cite{bittanti:2019} (a field very close to, if not part of, control theory) is used to reduce the number of tests that have to be executed in order to find faults~\cite{Menghi:2019}, to reduce the parameter definition effort required by genetic algorithms~\cite{Aleti:2015}, or to define oracles~\cite{He:2019}.
In these approaches, system identification is used as part of the testing process itself, and not to obtain knowledge on the SUT.
We therefore highlight that none of these works explicitly leverages the control-theoretical design of the SUT or integrates control knowledge in the definition of the software testing problem.

In this article, we are the first to introduce a control engineering perspective in the CPS testing problem definition, with the discussion of the control design assumptions and the resulting limitation of the CPS design scope.
Furthermore, we show how to use the frequency domain to (i)~perform test case generation for CPS as well as (ii)~leverage control-theory knowledge to define metamorphic relations for CPSs.
Such contributions extend the state of the art that was mostly characterised by requirement-driven testing approaches and time-domain descriptions of the input space and behaviour of CPSs.
 \section{Conclusion}
\label{sec:conclusion}
In this article we have defined the problem of stress testing control-based, cyber-physical systems (CPS).
For this class of CPSs, developed with the use of control theory, we have highlighted the different types of assumptions that the control engineers make during the design of the CPS control layer software.
Once the CPS is implemented, such assumptions are not always valid.
Accordingly, they can limit the set of scenarios in which the SUT is able to fulfil its requirements.
To help engineers identify such scenarios, we have developed a testing approach that pushes the system at the validity boundaries of the linearity design assumptions (i.e., the design assumption needed by control engineers to use linearised versions of the physical models).
We have provided a qualitative characterisation of the CPS control layer input space and proposed a novel test case parametrisation.
Leveraging our test case parametrisation, we use the  qualitative characterisation to generate test cases that cover the frequency-amplitude plane, propose a novel metric to identify stress test cases, and define metamorphic relations describing the expected CPS behaviour (filtering, linear, non-linear).
We have applied the proposed approach to six case studies and evaluated the output of the testing process.
Our results show that our testing approach effectively generates test cases that falsify the linearity design assumptions and push the SUT at the bounds of its design scope.
Furthermore, the proposed behaviour metrics $\dnl$ and $\dof$, together with the metamorphic relations, help identify stress test cases, and especially those outside of the CPS design scope.

\partparagraph{Future Work}
Future work has to investigate the impact of the use of different input shapes on test case generation.
Furthermore, as discussed in Section~\ref{sec:experimental-discussion}, it may be possible to reduce the number of tests while obtaining similar results.
Hence, improved sampling strategies (e.g., adaptive ones~\cite{Chen:2005}) should be investigated.
Another technical extension of this work should involve developing a method to assess whether a test, which does not necessarily consist of a repeated input sequence, belongs to the design scope of the CPS or not.
More specifically, this would require a study on the generalisability of the proposed metrics and MRs in handling arbitrary signals.

On a higher level, the main contribution of this work is to open a new perspective on the testing of control-based CPSs: stress testing driven by control design assumptions.
In future work, we plan to extend this approach to the testing of other assumptions not considered here, such as  not accounting for the control mode changes and  the interaction between different control loops.
Furthermore, another important research direction is testing the combined falsification of the different assumptions, e.g., when the falsification of the linearised model is combined with the falsification of the mode changes, and investigate how such  combinations affect the design scope of the SUT.

\section*{Acknowledgements}
This project has received funding from the European Union's Horizon 2020 research and innovation programme under grant agreement No 871259 (ADMORPH project) and No. 957254 (COSMOS); from  NSERC of Canada under the Discovery and CRC programs; from the Wallenberg AI, Autonomous Systems and Software Program (WASP) funded by the Knut and Alice Wallenberg Foundation.
This publication reflects only the authors' view and the European Commission is not responsible for any use that may be made of the information it contains.
At the time of doing this work, Claudio Mandrioli and Martina Maggio were members of ELLIIT Strategic Research Area at Lund University.

\bibliographystyle{ACM-Reference-Format}


\begin{thebibliography}{57}


\ifx \showCODEN    \undefined \def \showCODEN     #1{\unskip}     \fi
\ifx \showDOI      \undefined \def \showDOI       #1{#1}\fi
\ifx \showISBNx    \undefined \def \showISBNx     #1{\unskip}     \fi
\ifx \showISBNxiii \undefined \def \showISBNxiii  #1{\unskip}     \fi
\ifx \showISSN     \undefined \def \showISSN      #1{\unskip}     \fi
\ifx \showLCCN     \undefined \def \showLCCN      #1{\unskip}     \fi
\ifx \shownote     \undefined \def \shownote      #1{#1}          \fi
\ifx \showarticletitle \undefined \def \showarticletitle #1{#1}   \fi
\ifx \showURL      \undefined \def \showURL       {\relax}        \fi
\providecommand\bibfield[2]{#2}
\providecommand\bibinfo[2]{#2}
\providecommand\natexlab[1]{#1}
\providecommand\showeprint[2][]{arXiv:#2}

\bibitem[\protect\citeauthoryear{Abbas and Bonakdarpour}{Abbas and
  Bonakdarpour}{2022}]%
        {abbas:2022}
\bibfield{author}{\bibinfo{person}{Houssam Abbas} {and} \bibinfo{person}{Borzoo
  Bonakdarpour}.} \bibinfo{year}{2022}\natexlab{}.
\newblock \showarticletitle{Leveraging System Dynamics In Runtime Verification
  Of Cyber-Physical Systems}. In \bibinfo{booktitle}{\emph{Leveraging
  Applications of Formal Methods, Verification and Validation. Verification
  Principles: 11th International Symposium, ISoLA 2022, Rhodes, Greece, October
  22–30, 2022, Proceedings, Part I}} (Rhodes, Greece).
  \bibinfo{publisher}{Springer-Verlag}, \bibinfo{address}{Berlin, Heidelberg},
  \bibinfo{pages}{264–278}.
\newblock
\showISBNx{978-3-031-19848-9}
\urldef\tempurl%
\url{https://doi.org/10.1007/978-3-031-19849-6_16}
\showDOI{\tempurl}


\bibitem[\protect\citeauthoryear{Aleti and Grunske}{Aleti and Grunske}{2015}]%
        {Aleti:2015}
\bibfield{author}{\bibinfo{person}{Aldeida Aleti} {and} \bibinfo{person}{Lars
  Grunske}.} \bibinfo{year}{2015}\natexlab{}.
\newblock \showarticletitle{Test data generation with a Kalman filter-based
  adaptive genetic algorithm}.
\newblock \bibinfo{journal}{\emph{Journal of Systems and Software}}
  \bibinfo{volume}{103} (\bibinfo{year}{2015}), \bibinfo{pages}{343--352}.
\newblock
\showISSN{0164-1212}
\urldef\tempurl%
\url{https://doi.org/10.1016/j.jss.2014.11.035}
\showDOI{\tempurl}


\bibitem[\protect\citeauthoryear{Astrom and Murray}{Astrom and Murray}{2008}]%
        {Astrom:2008}
\bibfield{author}{\bibinfo{person}{Karl~Johan Astrom} {and}
  \bibinfo{person}{Richard~M. Murray}.} \bibinfo{year}{2008}\natexlab{}.
\newblock \bibinfo{booktitle}{\emph{Feedback Systems: An Introduction for
  Scientists and Engineers}}.
\newblock \bibinfo{publisher}{Princeton University Press},
  \bibinfo{address}{USA}.
\newblock
\showISBNx{0691135762}


\bibitem[\protect\citeauthoryear{Ayerdi, Terragni, Arrieta, Tonella, Sagardui,
  and Arratibel}{Ayerdi et~al\mbox{.}}{2021}]%
        {Ayeredi:2021}
\bibfield{author}{\bibinfo{person}{Jon Ayerdi}, \bibinfo{person}{Valerio
  Terragni}, \bibinfo{person}{Aitor Arrieta}, \bibinfo{person}{Paolo Tonella},
  \bibinfo{person}{Goiuria Sagardui}, {and} \bibinfo{person}{Maite Arratibel}.}
  \bibinfo{year}{2021}\natexlab{}.
\newblock \showarticletitle{Generating Metamorphic Relations for Cyber-Physical
  Systems with Genetic Programming: An Industrial Case Study}. In
  \bibinfo{booktitle}{\emph{Proceedings of the 29th ACM Joint Meeting on
  European Software Engineering Conference and Symposium on the Foundations of
  Software Engineering}} (Athens, Greece)
  \emph{(\bibinfo{series}{ESEC/FSE'2021})}. \bibinfo{publisher}{ACM},
  \bibinfo{address}{New York, NY, USA}, \bibinfo{pages}{1264--1274}.
\newblock
\showISBNx{9781450385626}
\urldef\tempurl%
\url{https://doi.org/10.1145/3468264.3473920}
\showDOI{\tempurl}


\bibitem[\protect\citeauthoryear{Balasubramaniam, Bagheri, Elbaum, and
  Bradley}{Balasubramaniam et~al\mbox{.}}{2020}]%
        {Balasubramaniam:2020}
\bibfield{author}{\bibinfo{person}{Balaji Balasubramaniam},
  \bibinfo{person}{Hamid Bagheri}, \bibinfo{person}{Sebastian Elbaum}, {and}
  \bibinfo{person}{Justin Bradley}.} \bibinfo{year}{2020}\natexlab{}.
\newblock \showarticletitle{Investigating Controller Evolution and Divergence
  through Mining and Mutation}. In \bibinfo{booktitle}{\emph{2020 ACM/IEEE 11th
  International Conference on Cyber-Physical Systems (ICCPS)}}.
  \bibinfo{publisher}{IEEE}, \bibinfo{address}{United States of America},
  \bibinfo{pages}{151--161}.
\newblock
\urldef\tempurl%
\url{https://doi.org/10.1109/ICCPS48487.2020.00022}
\showDOI{\tempurl}


\bibitem[\protect\citeauthoryear{Bittanti}{Bittanti}{2019}]%
        {bittanti:2019}
\bibfield{author}{\bibinfo{person}{Sergio Bittanti}.}
  \bibinfo{year}{2019}\natexlab{}.
\newblock \bibinfo{booktitle}{\emph{Model Identification}}.
\newblock \bibinfo{publisher}{John Wiley and Sons, Ltd},
  \bibinfo{address}{United States of America}, Chapter~4,
  \bibinfo{pages}{81--105}.
\newblock
\showISBNx{9781119546405}
\urldef\tempurl%
\url{https://doi.org/10.1002/9781119546405.ch4}
\showDOI{\tempurl}
\showeprint{https://onlinelibrary.wiley.com/doi/pdf/10.1002/9781119546405.ch4}


\bibitem[\protect\citeauthoryear{Bozhko, von~der Brüggen, and
  Brandenburg}{Bozhko et~al\mbox{.}}{2021}]%
        {Bozhko:2021}
\bibfield{author}{\bibinfo{person}{Sergey Bozhko}, \bibinfo{person}{Georg
  von~der Brüggen}, {and} \bibinfo{person}{Björn~B. Brandenburg}.}
  \bibinfo{year}{2021}\natexlab{}.
\newblock \showarticletitle{Monte Carlo Response-Time Analysis}. In
  \bibinfo{booktitle}{\emph{2021 IEEE Real-Time Systems Symposium (RTSS)}}.
  \bibinfo{publisher}{IEEE}, \bibinfo{address}{United States of America},
  \bibinfo{pages}{342--355}.
\newblock
\urldef\tempurl%
\url{https://doi.org/10.1109/RTSS52674.2021.00039}
\showDOI{\tempurl}


\bibitem[\protect\citeauthoryear{Bradley and Bagheri}{Bradley and
  Bagheri}{2020}]%
        {Bradley:2020}
\bibfield{author}{\bibinfo{person}{Justin~M. Bradley} {and}
  \bibinfo{person}{Hamid Bagheri}.} \bibinfo{year}{2020}\natexlab{}.
\newblock \showarticletitle{Control Software: Research Directions in the
  Intersection of Control Theory and Software Engineering}. In
  \bibinfo{booktitle}{\emph{AIAA Scitech 2020 Forum}}.
  \bibinfo{publisher}{Aerospace Research Center}, \bibinfo{address}{United
  States of America}.
\newblock
\urldef\tempurl%
\url{https://doi.org/10.2514/6.2020-2102}
\showDOI{\tempurl}
\showeprint{https://arc.aiaa.org/doi/pdf/10.2514/6.2020-2102}


\bibitem[\protect\citeauthoryear{Briand, Nejati, Sabetzadeh, and
  Bianculli}{Briand et~al\mbox{.}}{2016}]%
        {Briand:2016}
\bibfield{author}{\bibinfo{person}{Lionel Briand}, \bibinfo{person}{Shiva
  Nejati}, \bibinfo{person}{Mehrdad Sabetzadeh}, {and}
  \bibinfo{person}{Domenico Bianculli}.} \bibinfo{year}{2016}\natexlab{}.
\newblock \showarticletitle{Testing the Untestable: Model Testing of Complex
  Software-Intensive Systems}. In \bibinfo{booktitle}{\emph{Proceedings of the
  38th International Conference on Software Engineering Companion}} (Austin,
  Texas) \emph{(\bibinfo{series}{ICSE '16})}. \bibinfo{publisher}{Association
  for Computing Machinery}, \bibinfo{address}{New York, NY, USA},
  \bibinfo{pages}{789–792}.
\newblock
\showISBNx{9781450342056}
\urldef\tempurl%
\url{https://doi.org/10.1145/2889160.2889212}
\showDOI{\tempurl}


\bibitem[\protect\citeauthoryear{Bringmann and Kr\"{a}mer}{Bringmann and
  Kr\"{a}mer}{2006}]%
        {Bringmann:2006}
\bibfield{author}{\bibinfo{person}{Eckard Bringmann} {and}
  \bibinfo{person}{Andreas Kr\"{a}mer}.} \bibinfo{year}{2006}\natexlab{}.
\newblock \showarticletitle{Systematic Testing of the Continuous Behavior of
  Automotive Systems}. In \bibinfo{booktitle}{\emph{Proceedings of the 2006
  International Workshop on Software Engineering for Automotive Systems}}
  (Shanghai, China) \emph{(\bibinfo{series}{SEAS '06})}.
  \bibinfo{publisher}{Association for Computing Machinery},
  \bibinfo{address}{New York, NY, USA}, \bibinfo{pages}{13–20}.
\newblock
\showISBNx{1595934022}
\urldef\tempurl%
\url{https://doi.org/10.1145/1138474.1138479}
\showDOI{\tempurl}


\bibitem[\protect\citeauthoryear{Bringmann and Krämer}{Bringmann and
  Krämer}{2008}]%
        {Bringmann:2008}
\bibfield{author}{\bibinfo{person}{Eckard Bringmann} {and}
  \bibinfo{person}{Andreas Krämer}.} \bibinfo{year}{2008}\natexlab{}.
\newblock \showarticletitle{Model-Based Testing of Automotive Systems}. In
  \bibinfo{booktitle}{\emph{2008 1st International Conference on Software
  Testing, Verification, and Validation}}. \bibinfo{publisher}{IEEE},
  \bibinfo{address}{United States of America}, \bibinfo{pages}{485--493}.
\newblock
\urldef\tempurl%
\url{https://doi.org/10.1109/ICST.2008.45}
\showDOI{\tempurl}


\bibitem[\protect\citeauthoryear{{Broy}, {Kruger}, {Pretschner}, and
  {Salzmann}}{{Broy} et~al\mbox{.}}{2007}]%
        {Broy:2007}
\bibfield{author}{\bibinfo{person}{M. {Broy}}, \bibinfo{person}{I.~H.
  {Kruger}}, \bibinfo{person}{A. {Pretschner}}, {and} \bibinfo{person}{C.
  {Salzmann}}.} \bibinfo{year}{2007}\natexlab{}.
\newblock \showarticletitle{Engineering Automotive Software}.
\newblock \bibinfo{journal}{\emph{Proc. IEEE}} \bibinfo{volume}{95},
  \bibinfo{number}{2} (\bibinfo{date}{Feb} \bibinfo{year}{2007}),
  \bibinfo{pages}{356--373}.
\newblock
\urldef\tempurl%
\url{https://doi.org/10.1109/JPROC.2006.888386}
\showDOI{\tempurl}


\bibitem[\protect\citeauthoryear{Chakarov, Sankaranarayanan, and
  Fainekos}{Chakarov et~al\mbox{.}}{2012}]%
        {Chakarov:2012}
\bibfield{author}{\bibinfo{person}{Aleksandar Chakarov},
  \bibinfo{person}{Sriram Sankaranarayanan}, {and} \bibinfo{person}{Georgios
  Fainekos}.} \bibinfo{year}{2012}\natexlab{}.
\newblock \showarticletitle{Combining Time and Frequency Domain Specifications
  for Periodic Signals}. In \bibinfo{booktitle}{\emph{Runtime Verification}},
  \bibfield{editor}{\bibinfo{person}{Sarfraz Khurshid} {and}
  \bibinfo{person}{Koushik Sen}} (Eds.). \bibinfo{publisher}{Springer Berlin
  Heidelberg}, \bibinfo{address}{Berlin, Heidelberg},
  \bibinfo{pages}{294--309}.
\newblock
\showISBNx{978-3-642-29860-8}


\bibitem[\protect\citeauthoryear{Chen, Kuo, Liu, Poon, Towey, Tse, and
  Zhou}{Chen et~al\mbox{.}}{2018}]%
        {Chen:2019}
\bibfield{author}{\bibinfo{person}{Tsong~Yueh Chen}, \bibinfo{person}{Fei-Ching
  Kuo}, \bibinfo{person}{Huai Liu}, \bibinfo{person}{Pak-Lok Poon},
  \bibinfo{person}{Dave Towey}, \bibinfo{person}{T.~H. Tse}, {and}
  \bibinfo{person}{Zhi~Quan Zhou}.} \bibinfo{year}{2018}\natexlab{}.
\newblock \showarticletitle{Metamorphic Testing: A Review of Challenges and
  Opportunities}.
\newblock \bibinfo{journal}{\emph{ACM Comput. Surv.}} \bibinfo{volume}{51},
  \bibinfo{number}{1}, Article \bibinfo{articleno}{4} (\bibinfo{date}{jan}
  \bibinfo{year}{2018}), \bibinfo{numpages}{27}~pages.
\newblock
\showISSN{0360-0300}
\urldef\tempurl%
\url{https://doi.org/10.1145/3143561}
\showDOI{\tempurl}


\bibitem[\protect\citeauthoryear{Chen, Leung, and Mak}{Chen
  et~al\mbox{.}}{2005}]%
        {Chen:2005}
\bibfield{author}{\bibinfo{person}{Tsong~Yueh Chen}, \bibinfo{person}{Hing
  Leung}, {and} \bibinfo{person}{Ieng~Kei Mak}.}
  \bibinfo{year}{2005}\natexlab{}.
\newblock \showarticletitle{Adaptive random testing}. In
  \bibinfo{booktitle}{\emph{Proceedings of the 9th Asian Computing Science
  Conference on Advances in Computer Science}}. \bibinfo{publisher}{Springer},
  \bibinfo{address}{Berlin, Heidelberg}, \bibinfo{pages}{320--329}.
\newblock
\urldef\tempurl%
\url{https://doi.org/10.1007/978-3-540-30502-6_23}
\showDOI{\tempurl}


\bibitem[\protect\citeauthoryear{Cooley and Tukey}{Cooley and Tukey}{1965}]%
        {fourier-algo}
\bibfield{author}{\bibinfo{person}{James~W. Cooley} {and}
  \bibinfo{person}{John~W. Tukey}.} \bibinfo{year}{1965}\natexlab{}.
\newblock \showarticletitle{An algorithm for the machine calculation of complex
  Fourier series}.
\newblock \bibinfo{journal}{\emph{Math. Comp.}}  \bibinfo{volume}{19}
  (\bibinfo{year}{1965}), \bibinfo{pages}{297--301}.
\newblock


\bibitem[\protect\citeauthoryear{Dawes and Bianculli}{Dawes and
  Bianculli}{2022}]%
        {Dawes:2022}
\bibfield{author}{\bibinfo{person}{Joshua~Heneage Dawes} {and}
  \bibinfo{person}{Domenico Bianculli}.} \bibinfo{year}{2022}\natexlab{}.
\newblock \showarticletitle{Specifying Source Code and Signal-Based Behaviour
  of Cyber-Physical System Components}. In \bibinfo{booktitle}{\emph{Formal
  Aspects of Component Software: 18th International Conference, FACS 2022,
  Virtual Event, November 10–11, 2022, Proceedings}} (Oslo, Norway).
  \bibinfo{publisher}{Springer-Verlag}, \bibinfo{address}{Berlin, Heidelberg},
  \bibinfo{pages}{20–38}.
\newblock
\showISBNx{978-3-031-20871-3}
\urldef\tempurl%
\url{https://doi.org/10.1007/978-3-031-20872-0_2}
\showDOI{\tempurl}


\bibitem[\protect\citeauthoryear{Desborough and Miller}{Desborough and
  Miller}{2002}]%
        {Desborough:2002}
\bibfield{author}{\bibinfo{person}{Lane Desborough} {and}
  \bibinfo{person}{Randy Miller}.} \bibinfo{year}{2002}\natexlab{}.
\newblock \bibinfo{title}{Increasing Customer Value of Industrial Control
  Performance Monitoring — Honeywell ’ s Experience}.
\newblock
\newblock


\bibitem[\protect\citeauthoryear{Donz{\'e}, Maler, Bartocci, Nickovic, Grosu,
  and Smolka}{Donz{\'e} et~al\mbox{.}}{2012}]%
        {Donze:2012}
\bibfield{author}{\bibinfo{person}{Alexandre Donz{\'e}}, \bibinfo{person}{Oded
  Maler}, \bibinfo{person}{Ezio Bartocci}, \bibinfo{person}{Dejan Nickovic},
  \bibinfo{person}{Radu Grosu}, {and} \bibinfo{person}{Scott Smolka}.}
  \bibinfo{year}{2012}\natexlab{}.
\newblock \showarticletitle{On Temporal Logic and Signal Processing}. In
  \bibinfo{booktitle}{\emph{Automated Technology for Verification and
  Analysis}}, \bibfield{editor}{\bibinfo{person}{Supratik Chakraborty} {and}
  \bibinfo{person}{Madhavan Mukund}} (Eds.). \bibinfo{publisher}{Springer
  Berlin Heidelberg}, \bibinfo{address}{Berlin, Heidelberg},
  \bibinfo{pages}{92--106}.
\newblock
\showISBNx{978-3-642-33386-6}


\bibitem[\protect\citeauthoryear{Dreossi, Dang, Donz{\'e}, Kapinski, Jin, and
  Deshmukh}{Dreossi et~al\mbox{.}}{2015}]%
        {Dreossi:2015}
\bibfield{author}{\bibinfo{person}{Tommaso Dreossi}, \bibinfo{person}{Thao
  Dang}, \bibinfo{person}{Alexandre Donz{\'e}}, \bibinfo{person}{James
  Kapinski}, \bibinfo{person}{Xiaoqing Jin}, {and}
  \bibinfo{person}{Jyotirmoy~V. Deshmukh}.} \bibinfo{year}{2015}\natexlab{}.
\newblock \showarticletitle{Efficient Guiding Strategies for Testing of
  Temporal Properties of Hybrid Systems}. In \bibinfo{booktitle}{\emph{NASA
  Formal Methods}}, \bibfield{editor}{\bibinfo{person}{Klaus Havelund},
  \bibinfo{person}{Gerard Holzmann}, {and} \bibinfo{person}{Rajeev Joshi}}
  (Eds.). \bibinfo{publisher}{Springer International Publishing},
  \bibinfo{address}{Cham}, \bibinfo{pages}{127--142}.
\newblock
\showISBNx{978-3-319-17524-9}


\bibitem[\protect\citeauthoryear{Ghosh, Hobbs, Xu, Duggirala, Anderson,
  Thiagarajan, and Chakraborty}{Ghosh et~al\mbox{.}}{2022}]%
        {Ghosh:2022}
\bibfield{author}{\bibinfo{person}{Bineet Ghosh}, \bibinfo{person}{Clara
  Hobbs}, \bibinfo{person}{Shengjie Xu}, \bibinfo{person}{Parasara~Sridhar
  Duggirala}, \bibinfo{person}{James~H. Anderson}, \bibinfo{person}{P.~S.
  Thiagarajan}, {and} \bibinfo{person}{Samarjit Chakraborty}.}
  \bibinfo{year}{2022}\natexlab{}.
\newblock \showarticletitle{Statistical Hypothesis Testing of Controller
  Implementations Under Timing Uncertainties}. In
  \bibinfo{booktitle}{\emph{2022 IEEE 28th International Conference on Embedded
  and Real-Time Computing Systems and Applications (RTCSA)}}.
  \bibinfo{publisher}{IEEE}, \bibinfo{address}{United States of America},
  \bibinfo{pages}{11--20}.
\newblock
\urldef\tempurl%
\url{https://doi.org/10.1109/RTCSA55878.2022.00008}
\showDOI{\tempurl}


\bibitem[\protect\citeauthoryear{Gille-Maisani and Decaulne}{Gille-Maisani and
  Decaulne}{1959}]%
        {Gille:1959}
\bibfield{author}{\bibinfo{person}{J.C. Gille-Maisani} {and}
  \bibinfo{person}{P. Decaulne}.} \bibinfo{year}{1959}\natexlab{}.
\newblock \bibinfo{booktitle}{\emph{Feedback Control Systems: Analysis,
  Synthesis, and Design}}.
\newblock \bibinfo{publisher}{McGraw-Hill}, \bibinfo{address}{United States of
  America}.
\newblock
\showLCCN{58009856}
\urldef\tempurl%
\url{https://books.google.se/books?id=9WG9zQEACAAJ}
\showURL{%
\tempurl}


\bibitem[\protect\citeauthoryear{Greiff}{Greiff}{2017}]%
        {Greiff:2017}
\bibfield{author}{\bibinfo{person}{Marcus Greiff}.}
  \bibinfo{year}{2017}\natexlab{}.
\newblock \bibinfo{title}{Modelling and Control of the Crazyflie Quadrotor for
  Aggressive and Autonomous Flight by Optical Flow Driven State Estimation}.
\newblock
\newblock
\showISSN{0280-5316}
\newblock
\shownote{Student Paper}.


\bibitem[\protect\citeauthoryear{Greiff, Persson, Sun, Åström, and
  Robertsson}{Greiff et~al\mbox{.}}{2021}]%
        {Greiff:2021}
\bibfield{author}{\bibinfo{person}{Marcus Greiff}, \bibinfo{person}{Patrik
  Persson}, \bibinfo{person}{Zhiyong Sun}, \bibinfo{person}{Karl Åström},
  {and} \bibinfo{person}{Anders Robertsson}.} \bibinfo{year}{2021}\natexlab{}.
\newblock \bibinfo{title}{Quadrotor Control on $SU(2)\times R^3$ with SLAM
  Integration}.
\newblock
\newblock
\urldef\tempurl%
\url{https://doi.org/10.48550/ARXIV.2110.01099}
\showDOI{\tempurl}


\bibitem[\protect\citeauthoryear{He, Wang, Shi, and Liu}{He
  et~al\mbox{.}}{2020}]%
        {He:2020}
\bibfield{author}{\bibinfo{person}{Xiao He}, \bibinfo{person}{Xingwei Wang},
  \bibinfo{person}{Jia Shi}, {and} \bibinfo{person}{Yi Liu}.}
  \bibinfo{year}{2020}\natexlab{}.
\newblock \showarticletitle{Testing High Performance Numerical Simulation
  Programs: Experience, Lessons Learned, and Open Issues}. In
  \bibinfo{booktitle}{\emph{Proceedings of the 29th ACM SIGSOFT International
  Symposium on Software Testing and Analysis}} (Virtual Event, USA)
  \emph{(\bibinfo{series}{ISSTA 2020})}. \bibinfo{publisher}{Association for
  Computing Machinery}, \bibinfo{address}{New York, NY, USA},
  \bibinfo{pages}{502–515}.
\newblock
\showISBNx{9781450380089}
\urldef\tempurl%
\url{https://doi.org/10.1145/3395363.3397382}
\showDOI{\tempurl}


\bibitem[\protect\citeauthoryear{He, Chen, Huang, Wang, Pei, and Yuan}{He
  et~al\mbox{.}}{2019}]%
        {He:2019}
\bibfield{author}{\bibinfo{person}{Zhijian He}, \bibinfo{person}{Yao Chen},
  \bibinfo{person}{Enyan Huang}, \bibinfo{person}{Qixin Wang},
  \bibinfo{person}{Yu Pei}, {and} \bibinfo{person}{Haidong Yuan}.}
  \bibinfo{year}{2019}\natexlab{}.
\newblock \showarticletitle{A System Identification Based Oracle for
  Control-CPS Software Fault Localization}. In \bibinfo{booktitle}{\emph{2019
  IEEE/ACM 41st International Conference on Software Engineering (ICSE)}}.
  \bibinfo{publisher}{IEEE}, \bibinfo{address}{United States of America},
  \bibinfo{pages}{116--127}.
\newblock
\urldef\tempurl%
\url{https://doi.org/10.1109/ICSE.2019.00029}
\showDOI{\tempurl}


\bibitem[\protect\citeauthoryear{Hu and Lin}{Hu and Lin}{2001}]%
        {Hu:2001}
\bibfield{author}{\bibinfo{person}{Tingshu Hu} {and} \bibinfo{person}{Zongli
  Lin}.} \bibinfo{year}{2001}\natexlab{}.
\newblock \bibinfo{booktitle}{\emph{Introduction}}.
\newblock \bibinfo{publisher}{Birkh{\"a}user Boston}, \bibinfo{address}{Boston,
  MA}, \bibinfo{pages}{1--9}.
\newblock
\showISBNx{978-1-4612-0205-9}
\urldef\tempurl%
\url{https://doi.org/10.1007/978-1-4612-0205-9_1}
\showDOI{\tempurl}


\bibitem[\protect\citeauthoryear{Hänsel, Rose, Herber, and Glesner}{Hänsel
  et~al\mbox{.}}{2011}]%
        {Hansel:2011}
\bibfield{author}{\bibinfo{person}{Joachim Hänsel}, \bibinfo{person}{Daniela
  Rose}, \bibinfo{person}{Paula Herber}, {and} \bibinfo{person}{Sabine
  Glesner}.} \bibinfo{year}{2011}\natexlab{}.
\newblock \showarticletitle{An Evolutionary Algorithm for the Generation of
  Timed Test Traces for Embedded Real-Time Systems}. In
  \bibinfo{booktitle}{\emph{2011 Fourth IEEE International Conference on
  Software Testing, Verification and Validation}}. \bibinfo{publisher}{IEEE},
  \bibinfo{address}{United States of America}, \bibinfo{pages}{170--179}.
\newblock
\urldef\tempurl%
\url{https://doi.org/10.1109/ICST.2011.37}
\showDOI{\tempurl}


\bibitem[\protect\citeauthoryear{Khalil}{Khalil}{2002}]%
        {khalil:2002}
\bibfield{author}{\bibinfo{person}{H.K. Khalil}.}
  \bibinfo{year}{2002}\natexlab{}.
\newblock \bibinfo{booktitle}{\emph{Nonlinear Systems}}.
\newblock \bibinfo{publisher}{Prentice Hall}, \bibinfo{address}{United States
  of America}.
\newblock
\showISBNx{9780130673893}
\showLCCN{95045804}
\urldef\tempurl%
\url{https://books.google.se/books?id=t\_d1QgAACAAJ}
\showURL{%
\tempurl}


\bibitem[\protect\citeauthoryear{Lamberg, Beine, Eschmann, Otterbach, Conrad,
  and Fey}{Lamberg et~al\mbox{.}}{2004}]%
        {Lamberg:2004}
\bibfield{author}{\bibinfo{person}{Klaus Lamberg}, \bibinfo{person}{Michael
  Beine}, \bibinfo{person}{Mario Eschmann}, \bibinfo{person}{Rainer Otterbach},
  \bibinfo{person}{Mirko Conrad}, {and} \bibinfo{person}{Ines Fey}.}
  \bibinfo{year}{2004}\natexlab{}.
\newblock \showarticletitle{Model-based Testing of Embedded Automotive Software
  Using Mtest}. In \bibinfo{booktitle}{\emph{SAE 2004 World Congress and
  Exhibition}}. \bibinfo{publisher}{SAE International},
  \bibinfo{address}{United States of America}, \bibinfo{numpages}{11}~pages.
\newblock
\showISSN{0148-7191}
\urldef\tempurl%
\url{https://doi.org/10.4271/2004-01-1593}
\showDOI{\tempurl}


\bibitem[\protect\citeauthoryear{Lee}{Lee}{2015}]%
        {Lee:2015}
\bibfield{author}{\bibinfo{person}{Edward~A. Lee}.}
  \bibinfo{year}{2015}\natexlab{}.
\newblock \showarticletitle{The Past, Present and Future of Cyber-Physical
  Systems: A Focus on Models}.
\newblock \bibinfo{journal}{\emph{Sensors}} \bibinfo{volume}{15},
  \bibinfo{number}{3} (\bibinfo{year}{2015}), \bibinfo{pages}{4837--4869}.
\newblock
\showISSN{1424-8220}
\urldef\tempurl%
\url{https://doi.org/10.3390/s150304837}
\showDOI{\tempurl}


\bibitem[\protect\citeauthoryear{Lu, Nolte, Bate, and Cucu-Grosjean}{Lu
  et~al\mbox{.}}{2012}]%
        {Lu:2012}
\bibfield{author}{\bibinfo{person}{Yue Lu}, \bibinfo{person}{Thomas Nolte},
  \bibinfo{person}{Iain Bate}, {and} \bibinfo{person}{Liliana Cucu-Grosjean}.}
  \bibinfo{year}{2012}\natexlab{}.
\newblock \showarticletitle{A Statistical Response-Time Analysis of Real-Time
  Embedded Systems}. In \bibinfo{booktitle}{\emph{2012 IEEE 33rd Real-Time
  Systems Symposium}}. \bibinfo{publisher}{IEEE}, \bibinfo{address}{United
  States of America}, \bibinfo{pages}{351--362}.
\newblock
\urldef\tempurl%
\url{https://doi.org/10.1109/RTSS.2012.85}
\showDOI{\tempurl}


\bibitem[\protect\citeauthoryear{Magnani, Cattaneo, Chiari, and Agosta}{Magnani
  et~al\mbox{.}}{2021}]%
        {Magnani:2021}
\bibfield{author}{\bibinfo{person}{Gabriele Magnani}, \bibinfo{person}{Daniele
  Cattaneo}, \bibinfo{person}{Michele Chiari}, {and} \bibinfo{person}{Giovanni
  Agosta}.} \bibinfo{year}{2021}\natexlab{}.
\newblock \showarticletitle{{The Impact of Precision Tuning on Embedded Systems
  Performance: A Case Study on Field-Oriented Control}}. In
  \bibinfo{booktitle}{\emph{12th Workshop on Parallel Programming and Run-Time
  Management Techniques for Many-core Architectures and 10th Workshop on Design
  Tools and Architectures for Multicore Embedded Computing Platforms
  (PARMA-DITAM 2021)}} \emph{(\bibinfo{series}{Open Access Series in
  Informatics (OASIcs)}, Vol.~\bibinfo{volume}{88})},
  \bibfield{editor}{\bibinfo{person}{Jo\~{a}o Bispo}, \bibinfo{person}{Stefano
  Cherubin}, {and} \bibinfo{person}{Jos\'{e} Flich}} (Eds.).
  \bibinfo{publisher}{Schloss Dagstuhl -- Leibniz-Zentrum f{\"u}r Informatik},
  \bibinfo{address}{Dagstuhl, Germany}, \bibinfo{pages}{3:1--3:13}.
\newblock
\showISBNx{978-3-95977-181-8}
\showISSN{2190-6807}
\urldef\tempurl%
\url{https://doi.org/10.4230/OASIcs.PARMA-DITAM.2021.3}
\showDOI{\tempurl}


\bibitem[\protect\citeauthoryear{Mandrioli, Max, and Maggio}{Mandrioli
  et~al\mbox{.}}{2023}]%
        {Mandrioli:2023}
\bibfield{author}{\bibinfo{person}{Claudio Mandrioli},
  \bibinfo{person}{Nyberg~Carlsson Max}, {and} \bibinfo{person}{Martina
  Maggio}.} \bibinfo{year}{2023}\natexlab{}.
\newblock \showarticletitle{Testing Abstractions for Cyber-Physical Control
  Systems}.
\newblock \bibinfo{journal}{\emph{ACM Transactions on Software Engineering and
  Methodology}} \bibinfo{volume}{1}, \bibinfo{number}{1}, Article
  \bibinfo{articleno}{1} (\bibinfo{date}{Aug.} \bibinfo{year}{2023}),
  \bibinfo{numpages}{31}~pages.
\newblock
\urldef\tempurl%
\url{https://doi.org/10.1145/3617170}
\showDOI{\tempurl}


\bibitem[\protect\citeauthoryear{Marculescu, Feldt, Torkar, and
  Poulding}{Marculescu et~al\mbox{.}}{2015}]%
        {Marculescu:2015}
\bibfield{author}{\bibinfo{person}{Bogdan Marculescu}, \bibinfo{person}{Robert
  Feldt}, \bibinfo{person}{Richard Torkar}, {and} \bibinfo{person}{Simon
  Poulding}.} \bibinfo{year}{2015}\natexlab{}.
\newblock \showarticletitle{An initial industrial evaluation of interactive
  search-based testing for embedded software}.
\newblock \bibinfo{journal}{\emph{Applied Soft Computing}}
  \bibinfo{volume}{29} (\bibinfo{date}{April} \bibinfo{year}{2015}),
  \bibinfo{pages}{26--39}.
\newblock
Issue 0.
\urldef\tempurl%
\url{https://doi.org/10.1016/j.asoc.2014.12.025}
\showDOI{\tempurl}


\bibitem[\protect\citeauthoryear{Matinnejad, Nejati, Briand, and
  Brcukmann}{Matinnejad et~al\mbox{.}}{2014}]%
        {Matinnejad:2014}
\bibfield{author}{\bibinfo{person}{Reza Matinnejad}, \bibinfo{person}{Shiva
  Nejati}, \bibinfo{person}{Lionel Briand}, {and} \bibinfo{person}{Thomas
  Brcukmann}.} \bibinfo{year}{2014}\natexlab{}.
\newblock \showarticletitle{MiL Testing of Highly Configurable Continuous
  Controllers: Scalable Search Using Surrogate Models}. In
  \bibinfo{booktitle}{\emph{Proceedings of the 29th ACM/IEEE International
  Conference on Automated Software Engineering}} (Vasteras, Sweden)
  \emph{(\bibinfo{series}{ASE '14})}. \bibinfo{publisher}{Association for
  Computing Machinery}, \bibinfo{address}{New York, NY, USA},
  \bibinfo{pages}{163–174}.
\newblock
\showISBNx{9781450330138}
\urldef\tempurl%
\url{https://doi.org/10.1145/2642937.2642978}
\showDOI{\tempurl}


\bibitem[\protect\citeauthoryear{Matinnejad, Nejati, and Briand}{Matinnejad
  et~al\mbox{.}}{2017}]%
        {Matinnejad:2017}
\bibfield{author}{\bibinfo{person}{Reza Matinnejad}, \bibinfo{person}{Shiva
  Nejati}, {and} \bibinfo{person}{Lionel~C. Briand}.}
  \bibinfo{year}{2017}\natexlab{}.
\newblock \showarticletitle{Automated Testing of Hybrid Simulink/Stateflow
  Controllers: Industrial Case Studies}. In
  \bibinfo{booktitle}{\emph{Proceedings of the 2017 11th Joint Meeting on
  Foundations of Software Engineering}} (Paderborn, Germany)
  \emph{(\bibinfo{series}{ESEC/FSE 2017})}. \bibinfo{publisher}{Association for
  Computing Machinery}, \bibinfo{address}{New York, NY, USA},
  \bibinfo{pages}{938–943}.
\newblock
\showISBNx{9781450351058}
\urldef\tempurl%
\url{https://doi.org/10.1145/3106237.3117770}
\showDOI{\tempurl}


\bibitem[\protect\citeauthoryear{Matinnejad, Nejati, Briand, and
  Bruckmann}{Matinnejad et~al\mbox{.}}{2016}]%
        {Reza:2016}
\bibfield{author}{\bibinfo{person}{Reza Matinnejad}, \bibinfo{person}{Shiva
  Nejati}, \bibinfo{person}{Lionel~C. Briand}, {and} \bibinfo{person}{Thomas
  Bruckmann}.} \bibinfo{year}{2016}\natexlab{}.
\newblock \showarticletitle{Automated Test Suite Generation for Time-Continuous
  Simulink Models}. In \bibinfo{booktitle}{\emph{2016 IEEE/ACM 38th
  International Conference on Software Engineering (ICSE)}}.
  \bibinfo{publisher}{ACM}, \bibinfo{address}{United States of America},
  \bibinfo{pages}{595--606}.
\newblock
\urldef\tempurl%
\url{https://doi.org/10.1145/2884781.2884797}
\showDOI{\tempurl}


\bibitem[\protect\citeauthoryear{Menghi, Nejati, Briand, and Parache}{Menghi
  et~al\mbox{.}}{2019a}]%
        {Menghi:2019}
\bibfield{author}{\bibinfo{person}{Claudio Menghi}, \bibinfo{person}{Shiva
  Nejati}, \bibinfo{person}{Lionel~C. Briand}, {and}
  \bibinfo{person}{Yago~Isasi Parache}.} \bibinfo{year}{2019}\natexlab{a}.
\newblock \showarticletitle{Approximation-Refinement Testing of
  Compute-Intensive Cyber-Physical Models: An Approach Based on System
  Identification}.
\newblock \bibinfo{journal}{\emph{CoRR}}  \bibinfo{volume}{abs/1910.02837}
  (\bibinfo{year}{2019}), \bibinfo{numpages}{12}~pages.
\newblock
\showeprint[arxiv]{1910.02837}
\urldef\tempurl%
\url{http://arxiv.org/abs/1910.02837}
\showURL{%
\tempurl}


\bibitem[\protect\citeauthoryear{Menghi, Nejati, Gaaloul, and Briand}{Menghi
  et~al\mbox{.}}{2019b}]%
        {menghi:2019:oracles}
\bibfield{author}{\bibinfo{person}{Claudio Menghi}, \bibinfo{person}{Shiva
  Nejati}, \bibinfo{person}{Khouloud Gaaloul}, {and} \bibinfo{person}{Lionel~C.
  Briand}.} \bibinfo{year}{2019}\natexlab{b}.
\newblock \showarticletitle{Generating automated and online test oracles for
  Simulink models with continuous and uncertain behaviors}. In
  \bibinfo{booktitle}{\emph{Foundations of Software Engineering}}.
  \bibinfo{publisher}{ACM}, \bibinfo{address}{United States of America},
  \bibinfo{pages}{27--38}.
\newblock


\bibitem[\protect\citeauthoryear{Menghi, Viganò, Bianculli, and Briand}{Menghi
  et~al\mbox{.}}{2021}]%
        {Menghi:2021}
\bibfield{author}{\bibinfo{person}{Claudio Menghi}, \bibinfo{person}{Enrico
  Viganò}, \bibinfo{person}{Domenico Bianculli}, {and}
  \bibinfo{person}{Lionel~C. Briand}.} \bibinfo{year}{2021}\natexlab{}.
\newblock \showarticletitle{Trace-Checking CPS Properties: Bridging the
  Cyber-Physical Gap}. In \bibinfo{booktitle}{\emph{2021 IEEE/ACM 43rd
  International Conference on Software Engineering (ICSE)}}.
  \bibinfo{publisher}{ACM}, \bibinfo{address}{United States of America},
  \bibinfo{pages}{847--859}.
\newblock
\urldef\tempurl%
\url{https://doi.org/10.1109/ICSE43902.2021.00082}
\showDOI{\tempurl}


\bibitem[\protect\citeauthoryear{Mueller, Hehn, and D’Andrea}{Mueller
  et~al\mbox{.}}{2016}]%
        {Mueller:2016}
\bibfield{author}{\bibinfo{person}{Mark~W Mueller}, \bibinfo{person}{Markus
  Hehn}, {and} \bibinfo{person}{Raffaello D’Andrea}.}
  \bibinfo{year}{2016}\natexlab{}.
\newblock \showarticletitle{Covariance Correction Step for Kalman Filtering
  with an Attitude}.
\newblock \bibinfo{journal}{\emph{Journal of Guidance, Control, and Dynamics}}
  \bibinfo{volume}{40}, \bibinfo{number}{9} (\bibinfo{year}{2016}),
  \bibinfo{pages}{1--7}.
\newblock


\bibitem[\protect\citeauthoryear{Nejati, Gaaloul, Menghi, Briand, Foster, and
  Wolfe}{Nejati et~al\mbox{.}}{2019}]%
        {Nejati:2019}
\bibfield{author}{\bibinfo{person}{Shiva Nejati}, \bibinfo{person}{Khouloud
  Gaaloul}, \bibinfo{person}{Claudio Menghi}, \bibinfo{person}{Lionel~C.
  Briand}, \bibinfo{person}{Stephen Foster}, {and} \bibinfo{person}{David
  Wolfe}.} \bibinfo{year}{2019}\natexlab{}.
\newblock \showarticletitle{Evaluating Model Testing and Model Checking for
  Finding Requirements Violations in Simulink Models}. In
  \bibinfo{booktitle}{\emph{Proceedings of the 2019 27th ACM Joint Meeting on
  European Software Engineering Conference and Symposium on the Foundations of
  Software Engineering}} (Tallinn, Estonia) \emph{(\bibinfo{series}{ESEC/FSE
  2019})}. \bibinfo{publisher}{Association for Computing Machinery},
  \bibinfo{address}{New York, NY, USA}, \bibinfo{pages}{1015–1025}.
\newblock
\showISBNx{9781450355728}
\urldef\tempurl%
\url{https://doi.org/10.1145/3338906.3340444}
\showDOI{\tempurl}


\bibitem[\protect\citeauthoryear{Peleska}{Peleska}{2002}]%
        {Peleska:2002}
\bibfield{author}{\bibinfo{person}{Jan Peleska}.}
  \bibinfo{year}{2002}\natexlab{}.
\newblock \showarticletitle{Hardware/Software Integration Testing for the new
  Airbus Aircraft Families.}
\newblock
  \bibinfo{journal}{\emph{http://www.informatik.uni-bremen.de/agbs/jp/papers/peleskaTestCom2002.html}}
  \bibinfo{volume}{82}, \bibinfo{number}{14} (\bibinfo{date}{01}
  \bibinfo{year}{2002}).
\newblock
\showISBNx{978-1-4757-6705-6}
\urldef\tempurl%
\url{https://doi.org/10.1007/978-0-387-35497-2_24}
\showDOI{\tempurl}


\bibitem[\protect\citeauthoryear{Plaku, Kavraki, and Vardi}{Plaku
  et~al\mbox{.}}{2009}]%
        {Plaku:2009}
\bibfield{author}{\bibinfo{person}{Erion Plaku}, \bibinfo{person}{Lydia~E.
  Kavraki}, {and} \bibinfo{person}{Moshe~Y. Vardi}.}
  \bibinfo{year}{2009}\natexlab{}.
\newblock \showarticletitle{Falsification of LTL Safety Properties in Hybrid
  Systems}. In \bibinfo{booktitle}{\emph{Tools and Algorithms for the
  Construction and Analysis of Systems}},
  \bibfield{editor}{\bibinfo{person}{Stefan Kowalewski} {and}
  \bibinfo{person}{Anna Philippou}} (Eds.). \bibinfo{publisher}{Springer Berlin
  Heidelberg}, \bibinfo{address}{Berlin, Heidelberg},
  \bibinfo{pages}{368--382}.
\newblock
\showISBNx{978-3-642-00768-2}


\bibitem[\protect\citeauthoryear{Priyadarshi~Tripathy}{Priyadarshi~Tripathy}{2008}]%
        {Naik:2008}
\bibfield{author}{\bibinfo{person}{Kshirasagar~Naik Priyadarshi~Tripathy}.}
  \bibinfo{year}{2008}\natexlab{}.
\newblock \bibinfo{booktitle}{\emph{Acceptance Testing}}.
\newblock \bibinfo{publisher}{John Wiley and Sons, Ltd},
  \bibinfo{address}{United States of America}, Chapter~14,
  \bibinfo{pages}{450--470}.
\newblock
\showISBNx{9780470382844}
\urldef\tempurl%
\url{https://doi.org/10.1002/9780470382844.ch14}
\showDOI{\tempurl}
\showeprint{https://onlinelibrary.wiley.com/doi/pdf/10.1002/9780470382844.ch14}


\bibitem[\protect\citeauthoryear{Samad and Balas}{Samad and Balas}{2003}]%
        {Samad:2003}
\bibfield{author}{\bibinfo{person}{Tariq Samad} {and} \bibinfo{person}{Gary
  Balas}.} \bibinfo{year}{2003}\natexlab{}.
\newblock \bibinfo{booktitle}{\emph{Frontmatter}}.
\newblock \bibinfo{publisher}{John Wiley and Sons, Ltd},
  \bibinfo{address}{United States of America}, \bibinfo{pages}{i--xx}.
\newblock
\urldef\tempurl%
\url{https://doi.org/10.1002/047172288X.fmatter}
\showDOI{\tempurl}


\bibitem[\protect\citeauthoryear{Sanchez-Stern, Panchekha, Lerner, and
  Tatlock}{Sanchez-Stern et~al\mbox{.}}{2018}]%
        {Sanchez:2018}
\bibfield{author}{\bibinfo{person}{Alex Sanchez-Stern}, \bibinfo{person}{Pavel
  Panchekha}, \bibinfo{person}{Sorin Lerner}, {and} \bibinfo{person}{Zachary
  Tatlock}.} \bibinfo{year}{2018}\natexlab{}.
\newblock \showarticletitle{Finding Root Causes of Floating Point Error}.
\newblock \bibinfo{journal}{\emph{SIGPLAN Not.}} \bibinfo{volume}{53},
  \bibinfo{number}{4} (\bibinfo{date}{jun} \bibinfo{year}{2018}),
  \bibinfo{pages}{256–269}.
\newblock
\showISSN{0362-1340}
\urldef\tempurl%
\url{https://doi.org/10.1145/3296979.3192411}
\showDOI{\tempurl}


\bibitem[\protect\citeauthoryear{Silano, Aucone, and Iannelli}{Silano
  et~al\mbox{.}}{2018}]%
        {Silano:2018}
\bibfield{author}{\bibinfo{person}{Giuseppe Silano}, \bibinfo{person}{Emanuele
  Aucone}, {and} \bibinfo{person}{Luigi Iannelli}.}
  \bibinfo{year}{2018}\natexlab{}.
\newblock \showarticletitle{CrazyS: A Software-In-The-Loop Platform for the
  Crazyflie 2.0 Nano-Quadcopter}. In \bibinfo{booktitle}{\emph{2018 26th
  Mediterranean Conference on Control and Automation (MED)}}.
  \bibinfo{publisher}{IEEE}, \bibinfo{address}{United States of America},
  \bibinfo{pages}{1--6}.
\newblock
\urldef\tempurl%
\url{https://doi.org/10.1109/MED.2018.8442759}
\showDOI{\tempurl}


\bibitem[\protect\citeauthoryear{Sinharay}{Sinharay}{2010}]%
        {Sinharay:2010}
\bibfield{author}{\bibinfo{person}{S. Sinharay}.}
  \bibinfo{year}{2010}\natexlab{}.
\newblock \showarticletitle{Continuous Probability Distributions}.
\newblock In \bibinfo{booktitle}{\emph{International Encyclopedia of Education
  (Third Edition)} (\bibinfo{edition}{third edition} ed.)},
  \bibfield{editor}{\bibinfo{person}{Penelope Peterson}, \bibinfo{person}{Eva
  Baker}, {and} \bibinfo{person}{Barry McGaw}} (Eds.).
  \bibinfo{publisher}{Elsevier}, \bibinfo{address}{Oxford},
  \bibinfo{pages}{98--102}.
\newblock
\showISBNx{978-0-08-044894-7}
\urldef\tempurl%
\url{https://doi.org/10.1016/B978-0-08-044894-7.01720-6}
\showDOI{\tempurl}


\bibitem[\protect\citeauthoryear{{Timperley}, {Afzal}, {Katz}, {Hernandez}, and
  {Le Goues}}{{Timperley} et~al\mbox{.}}{2018}]%
        {Timperley:2018}
\bibfield{author}{\bibinfo{person}{C.~S. {Timperley}}, \bibinfo{person}{A.
  {Afzal}}, \bibinfo{person}{D.~S. {Katz}}, \bibinfo{person}{J.~M.
  {Hernandez}}, {and} \bibinfo{person}{C. {Le Goues}}.}
  \bibinfo{year}{2018}\natexlab{}.
\newblock \showarticletitle{Crashing Simulated Planes is Cheap: Can Simulation
  Detect Robotics Bugs Early?}. In \bibinfo{booktitle}{\emph{2018 IEEE 11th
  International Conference on Software Testing, Verification and Validation
  (ICST)}}. \bibinfo{publisher}{IEEE}, \bibinfo{address}{United States of
  America}, \bibinfo{pages}{331--342}.
\newblock
\urldef\tempurl%
\url{https://doi.org/10.1109/ICST.2018.00040}
\showDOI{\tempurl}


\bibitem[\protect\citeauthoryear{Vreman, Cervin, and Maggio}{Vreman
  et~al\mbox{.}}{2021}]%
        {Vreman:2021}
\bibfield{author}{\bibinfo{person}{Nils Vreman}, \bibinfo{person}{Anton
  Cervin}, {and} \bibinfo{person}{Martina Maggio}.}
  \bibinfo{year}{2021}\natexlab{}.
\newblock \showarticletitle{Stability and Performance Analysis of Control
  Systems Subject to Bursts of Deadline Misses}. In
  \bibinfo{booktitle}{\emph{33rd Euromicro Conference on Real-Time Systems
  (ECRTS 2021)}}, Vol.~\bibinfo{volume}{196}. \bibinfo{publisher}{Schloss
  Dagstuhl - Leibniz-Zentrum f{\"u}r Informatik}, \bibinfo{address}{Germany},
  \bibinfo{numpages}{23}~pages.
\newblock
\urldef\tempurl%
\url{https://doi.org/10.4230/LIPIcs.ECRTS.2021.15}
\showDOI{\tempurl}
\newblock
\shownote{33rd Euromicro Conference on Real-Time Systems (ECRTS 2021) ;
  Conference date: 05-07-2021 Through 09-07-2021}.


\bibitem[\protect\citeauthoryear{White}{White}{2001}]%
        {White:2001}
\bibfield{author}{\bibinfo{person}{A.L. White}.}
  \bibinfo{year}{2001}\natexlab{}.
\newblock \showarticletitle{Comments on modified condition/decision coverage
  for software testing [of flight control software]}. In
  \bibinfo{booktitle}{\emph{2001 IEEE Aerospace Conference Proceedings (Cat.
  No.01TH8542)}}, Vol.~\bibinfo{volume}{6}. \bibinfo{publisher}{IEEE},
  \bibinfo{address}{United States of America}, \bibinfo{pages}{2821--2827
  vol.6}.
\newblock
\urldef\tempurl%
\url{https://doi.org/10.1109/AERO.2001.931302}
\showDOI{\tempurl}


\bibitem[\protect\citeauthoryear{Wu, Zeng, Wang, and Yu}{Wu
  et~al\mbox{.}}{2017}]%
        {Wu:2017}
\bibfield{author}{\bibinfo{person}{Meng Wu}, \bibinfo{person}{Haibo Zeng},
  \bibinfo{person}{Chao Wang}, {and} \bibinfo{person}{Huafeng Yu}.}
  \bibinfo{year}{2017}\natexlab{}.
\newblock \showarticletitle{INVITED: Safety guard: Runtime enforcement for
  safety-critical cyber-physical systems}. In \bibinfo{booktitle}{\emph{2017
  54th ACM/EDAC/IEEE Design Automation Conference (DAC)}}.
  \bibinfo{publisher}{ACM}, \bibinfo{address}{United States of America},
  \bibinfo{pages}{1--6}.
\newblock
\urldef\tempurl%
\url{https://doi.org/10.1145/3061639.3072957}
\showDOI{\tempurl}


\bibitem[\protect\citeauthoryear{Yamagata, Liu, Akazaki, Duan, and
  Hao}{Yamagata et~al\mbox{.}}{2021}]%
        {Yamagata:2021}
\bibfield{author}{\bibinfo{person}{Yoriyuki Yamagata}, \bibinfo{person}{Shuang
  Liu}, \bibinfo{person}{Takumi Akazaki}, \bibinfo{person}{Yihai Duan}, {and}
  \bibinfo{person}{Jianye Hao}.} \bibinfo{year}{2021}\natexlab{}.
\newblock \showarticletitle{Falsification of Cyber-Physical Systems Using Deep
  Reinforcement Learning}.
\newblock \bibinfo{journal}{\emph{IEEE Transactions on Software Engineering}}
  \bibinfo{volume}{47}, \bibinfo{number}{12} (\bibinfo{year}{2021}),
  \bibinfo{pages}{2823--2840}.
\newblock
\urldef\tempurl%
\url{https://doi.org/10.1109/TSE.2020.2969178}
\showDOI{\tempurl}


\bibitem[\protect\citeauthoryear{Yi, Chen, Mao, and Ji}{Yi
  et~al\mbox{.}}{2017}]%
        {Yi:2017}
\bibfield{author}{\bibinfo{person}{Xin Yi}, \bibinfo{person}{Liqian Chen},
  \bibinfo{person}{Xiaoguang Mao}, {and} \bibinfo{person}{Tao Ji}.}
  \bibinfo{year}{2017}\natexlab{}.
\newblock \showarticletitle{Efficient Global Search for Inputs Triggering High
  Floating-Point Inaccuracies}. In \bibinfo{booktitle}{\emph{2017 24th
  Asia-Pacific Software Engineering Conference (APSEC)}}.
  \bibinfo{publisher}{IEEE}, \bibinfo{address}{United States of America},
  \bibinfo{pages}{11--20}.
\newblock
\urldef\tempurl%
\url{https://doi.org/10.1109/APSEC.2017.7}
\showDOI{\tempurl}


\bibitem[\protect\citeauthoryear{Zimmer, Hedrick, and Lee}{Zimmer
  et~al\mbox{.}}{2015}]%
        {Zimmer:2015}
\bibfield{author}{\bibinfo{person}{Michael Zimmer}, \bibinfo{person}{J.
  Hedrick}, {and} \bibinfo{person}{Edward~A. Lee}.}
  \bibinfo{year}{2015}\natexlab{}.
\newblock \showarticletitle{Ramifications of software implementation and
  deployment: A case study on yaw moment controller design}.
\newblock \bibinfo{journal}{\emph{2015 American Control Conference (ACC)}}
  \bibinfo{volume}{0} (\bibinfo{year}{2015}), \bibinfo{pages}{2014--2019}.
\newblock


\end{thebibliography}


\begin{thebibliography}{57}



\ifx \showCODEN    \undefined \def \showCODEN     #1{\unskip}     \fi
\ifx \showDOI      \undefined \def \showDOI       #1{#1}\fi
\ifx \showISBNx    \undefined \def \showISBNx     #1{\unskip}     \fi
\ifx \showISBNxiii \undefined \def \showISBNxiii  #1{\unskip}     \fi
\ifx \showISSN     \undefined \def \showISSN      #1{\unskip}     \fi
\ifx \showLCCN     \undefined \def \showLCCN      #1{\unskip}     \fi
\ifx \shownote     \undefined \def \shownote      #1{#1}          \fi
\ifx \showarticletitle \undefined \def \showarticletitle #1{#1}   \fi
\ifx \showURL      \undefined \def \showURL       {\relax}        \fi
\providecommand\bibfield[2]{#2}
\providecommand\bibinfo[2]{#2}
\providecommand\natexlab[1]{#1}
\providecommand\showeprint[2][]{arXiv:#2}

\bibitem[\protect\citeauthoryear{Abbas and Bonakdarpour}{Abbas and
  Bonakdarpour}{2022}]{abbas:2022}
\bibfield{author}{\bibinfo{person}{Houssam Abbas} {and} \bibinfo{person}{Borzoo
  Bonakdarpour}.} \bibinfo{year}{2022}\natexlab{}.
\newblock \showarticletitle{Leveraging System Dynamics In Runtime Verification
  Of Cyber-Physical Systems}. In \bibinfo{booktitle}{\emph{Leveraging
  Applications of Formal Methods, Verification and Validation. Verification
  Principles: 11th International Symposium, ISoLA 2022, Rhodes, Greece, October
  22–30, 2022, Proceedings, Part I}} (Rhodes, Greece).
  \bibinfo{publisher}{Springer-Verlag}, \bibinfo{address}{Berlin, Heidelberg},
  \bibinfo{pages}{264–278}.
\newblock
\showISBNx{978-3-031-19848-9}
\urldef\tempurl \url{https://doi.org/10.1007/978-3-031-19849-6_16}
\showDOI{\tempurl}


\bibitem[\protect\citeauthoryear{Aleti and Grunske}{Aleti and Grunske}{2015}]{Aleti:2015}
\bibfield{author}{\bibinfo{person}{Aldeida Aleti} {and} \bibinfo{person}{Lars
  Grunske}.} \bibinfo{year}{2015}\natexlab{}.
\newblock \showarticletitle{Test data generation with a Kalman filter-based
  adaptive genetic algorithm}.
\newblock \bibinfo{journal}{\emph{Journal of Systems and Software}}
  \bibinfo{volume}{103} (\bibinfo{year}{2015}), \bibinfo{pages}{343--352}.
\newblock
\showISSN{0164-1212}
\urldef\tempurl \url{https://doi.org/10.1016/j.jss.2014.11.035}
\showDOI{\tempurl}


\bibitem[\protect\citeauthoryear{Astrom and Murray}{Astrom and Murray}{2008}]{Astrom:2008}
\bibfield{author}{\bibinfo{person}{Karl~Johan Astrom} {and}
  \bibinfo{person}{Richard~M. Murray}.} \bibinfo{year}{2008}\natexlab{}.
\newblock \bibinfo{booktitle}{\emph{Feedback Systems: An Introduction for
  Scientists and Engineers}}.
\newblock \bibinfo{publisher}{Princeton University Press},
  \bibinfo{address}{USA}.
\newblock
\showISBNx{0691135762}


\bibitem[\protect\citeauthoryear{Ayerdi, Terragni, Arrieta, Tonella, Sagardui,
  and Arratibel}{Ayerdi et~al\mbox{.}}{2021}]{Ayeredi:2021}
\bibfield{author}{\bibinfo{person}{Jon Ayerdi}, \bibinfo{person}{Valerio
  Terragni}, \bibinfo{person}{Aitor Arrieta}, \bibinfo{person}{Paolo Tonella},
  \bibinfo{person}{Goiuria Sagardui}, {and} \bibinfo{person}{Maite Arratibel}.}
  \bibinfo{year}{2021}\natexlab{}.
\newblock \showarticletitle{Generating Metamorphic Relations for Cyber-Physical
  Systems with Genetic Programming: An Industrial Case Study}. In
  \bibinfo{booktitle}{\emph{Proceedings of the 29th ACM Joint Meeting on
  European Software Engineering Conference and Symposium on the Foundations of
  Software Engineering}} (Athens, Greece)
  \emph{(\bibinfo{series}{ESEC/FSE'2021})}. \bibinfo{publisher}{ACM},
  \bibinfo{address}{New York, NY, USA}, \bibinfo{pages}{1264--1274}.
\newblock
\showISBNx{9781450385626}
\urldef\tempurl \url{https://doi.org/10.1145/3468264.3473920}
\showDOI{\tempurl}


\bibitem[\protect\citeauthoryear{Balasubramaniam, Bagheri, Elbaum, and
  Bradley}{Balasubramaniam et~al\mbox{.}}{2020}]{Balasubramaniam:2020}
\bibfield{author}{\bibinfo{person}{Balaji Balasubramaniam},
  \bibinfo{person}{Hamid Bagheri}, \bibinfo{person}{Sebastian Elbaum}, {and}
  \bibinfo{person}{Justin Bradley}.} \bibinfo{year}{2020}\natexlab{}.
\newblock \showarticletitle{Investigating Controller Evolution and Divergence
  through Mining and Mutation}. In \bibinfo{booktitle}{\emph{2020 ACM/IEEE 11th
  International Conference on Cyber-Physical Systems (ICCPS)}}.
  \bibinfo{publisher}{IEEE}, \bibinfo{address}{United States of America},
  \bibinfo{pages}{151--161}.
\newblock
\urldef\tempurl \url{https://doi.org/10.1109/ICCPS48487.2020.00022}
\showDOI{\tempurl}


\bibitem[\protect\citeauthoryear{Bittanti}{Bittanti}{2019}]{bittanti:2019}
\bibfield{author}{\bibinfo{person}{Sergio Bittanti}.}
  \bibinfo{year}{2019}\natexlab{}.
\newblock \bibinfo{booktitle}{\emph{Model Identification}}.
\newblock \bibinfo{publisher}{John Wiley and Sons, Ltd},
  \bibinfo{address}{United States of America}, Chapter~4,
  \bibinfo{pages}{81--105}.
\newblock
\showISBNx{9781119546405}
\urldef\tempurl \url{https://doi.org/10.1002/9781119546405.ch4}
\showDOI{\tempurl}
\showeprint{https://onlinelibrary.wiley.com/doi/pdf/10.1002/9781119546405.ch4}


\bibitem[\protect\citeauthoryear{Bozhko, von~der Brüggen, and
  Brandenburg}{Bozhko et~al\mbox{.}}{2021}]{Bozhko:2021}
\bibfield{author}{\bibinfo{person}{Sergey Bozhko}, \bibinfo{person}{Georg
  von~der Brüggen}, {and} \bibinfo{person}{Björn~B. Brandenburg}.}
  \bibinfo{year}{2021}\natexlab{}.
\newblock \showarticletitle{Monte Carlo Response-Time Analysis}. In
  \bibinfo{booktitle}{\emph{2021 IEEE Real-Time Systems Symposium (RTSS)}}.
  \bibinfo{publisher}{IEEE}, \bibinfo{address}{United States of America},
  \bibinfo{pages}{342--355}.
\newblock
\urldef\tempurl \url{https://doi.org/10.1109/RTSS52674.2021.00039}
\showDOI{\tempurl}


\bibitem[\protect\citeauthoryear{Bradley and Bagheri}{Bradley and
  Bagheri}{2020}]{Bradley:2020}
\bibfield{author}{\bibinfo{person}{Justin~M. Bradley} {and}
  \bibinfo{person}{Hamid Bagheri}.} \bibinfo{year}{2020}\natexlab{}.
\newblock \showarticletitle{Control Software: Research Directions in the
  Intersection of Control Theory and Software Engineering}. In
  \bibinfo{booktitle}{\emph{AIAA Scitech 2020 Forum}}.
  \bibinfo{publisher}{Aerospace Research Center}, \bibinfo{address}{United
  States of America}.
\newblock
\urldef\tempurl \url{https://doi.org/10.2514/6.2020-2102}
\showDOI{\tempurl}
\showeprint{https://arc.aiaa.org/doi/pdf/10.2514/6.2020-2102}


\bibitem[\protect\citeauthoryear{Briand, Nejati, Sabetzadeh, and
  Bianculli}{Briand et~al\mbox{.}}{2016}]{Briand:2016}
\bibfield{author}{\bibinfo{person}{Lionel Briand}, \bibinfo{person}{Shiva
  Nejati}, \bibinfo{person}{Mehrdad Sabetzadeh}, {and}
  \bibinfo{person}{Domenico Bianculli}.} \bibinfo{year}{2016}\natexlab{}.
\newblock \showarticletitle{Testing the Untestable: Model Testing of Complex
  Software-Intensive Systems}. In \bibinfo{booktitle}{\emph{Proceedings of the
  38th International Conference on Software Engineering Companion}} (Austin,
  Texas) \emph{(\bibinfo{series}{ICSE '16})}. \bibinfo{publisher}{Association
  for Computing Machinery}, \bibinfo{address}{New York, NY, USA},
  \bibinfo{pages}{789–792}.
\newblock
\showISBNx{9781450342056}
\urldef\tempurl \url{https://doi.org/10.1145/2889160.2889212}
\showDOI{\tempurl}


\bibitem[\protect\citeauthoryear{Bringmann and Kr\"{a}mer}{Bringmann and
  Kr\"{a}mer}{2006}]{Bringmann:2006}
\bibfield{author}{\bibinfo{person}{Eckard Bringmann} {and}
  \bibinfo{person}{Andreas Kr\"{a}mer}.} \bibinfo{year}{2006}\natexlab{}.
\newblock \showarticletitle{Systematic Testing of the Continuous Behavior of
  Automotive Systems}. In \bibinfo{booktitle}{\emph{Proceedings of the 2006
  International Workshop on Software Engineering for Automotive Systems}}
  (Shanghai, China) \emph{(\bibinfo{series}{SEAS '06})}.
  \bibinfo{publisher}{Association for Computing Machinery},
  \bibinfo{address}{New York, NY, USA}, \bibinfo{pages}{13–20}.
\newblock
\showISBNx{1595934022}
\urldef\tempurl \url{https://doi.org/10.1145/1138474.1138479}
\showDOI{\tempurl}


\bibitem[\protect\citeauthoryear{Bringmann and Krämer}{Bringmann and
  Krämer}{2008}]{Bringmann:2008}
\bibfield{author}{\bibinfo{person}{Eckard Bringmann} {and}
  \bibinfo{person}{Andreas Krämer}.} \bibinfo{year}{2008}\natexlab{}.
\newblock \showarticletitle{Model-Based Testing of Automotive Systems}. In
  \bibinfo{booktitle}{\emph{2008 1st International Conference on Software
  Testing, Verification, and Validation}}. \bibinfo{publisher}{IEEE},
  \bibinfo{address}{United States of America}, \bibinfo{pages}{485--493}.
\newblock
\urldef\tempurl \url{https://doi.org/10.1109/ICST.2008.45}
\showDOI{\tempurl}


\bibitem[\protect\citeauthoryear{{Broy}, {Kruger}, {Pretschner}, and
  {Salzmann}}{{Broy} et~al\mbox{.}}{2007}]{Broy:2007}
\bibfield{author}{\bibinfo{person}{M. {Broy}}, \bibinfo{person}{I.~H.
  {Kruger}}, \bibinfo{person}{A. {Pretschner}}, {and} \bibinfo{person}{C.
  {Salzmann}}.} \bibinfo{year}{2007}\natexlab{}.
\newblock \showarticletitle{Engineering Automotive Software}.
\newblock \bibinfo{journal}{\emph{Proc. IEEE}} \bibinfo{volume}{95},
  \bibinfo{number}{2} (\bibinfo{date}{Feb} \bibinfo{year}{2007}),
  \bibinfo{pages}{356--373}.
\newblock
\urldef\tempurl \url{https://doi.org/10.1109/JPROC.2006.888386}
\showDOI{\tempurl}


\bibitem[\protect\citeauthoryear{Chakarov, Sankaranarayanan, and
  Fainekos}{Chakarov et~al\mbox{.}}{2012}]{Chakarov:2012}
\bibfield{author}{\bibinfo{person}{Aleksandar Chakarov},
  \bibinfo{person}{Sriram Sankaranarayanan}, {and} \bibinfo{person}{Georgios
  Fainekos}.} \bibinfo{year}{2012}\natexlab{}.
\newblock \showarticletitle{Combining Time and Frequency Domain Specifications
  for Periodic Signals}. In \bibinfo{booktitle}{\emph{Runtime Verification}},
  \bibfield{editor}{\bibinfo{person}{Sarfraz Khurshid} {and}
  \bibinfo{person}{Koushik Sen}} (Eds.). \bibinfo{publisher}{Springer Berlin
  Heidelberg}, \bibinfo{address}{Berlin, Heidelberg},
  \bibinfo{pages}{294--309}.
\newblock
\showISBNx{978-3-642-29860-8}


\bibitem[\protect\citeauthoryear{Chen, Kuo, Liu, Poon, Towey, Tse, and
  Zhou}{Chen et~al\mbox{.}}{2018}]{Chen:2019}
\bibfield{author}{\bibinfo{person}{Tsong~Yueh Chen}, \bibinfo{person}{Fei-Ching
  Kuo}, \bibinfo{person}{Huai Liu}, \bibinfo{person}{Pak-Lok Poon},
  \bibinfo{person}{Dave Towey}, \bibinfo{person}{T.~H. Tse}, {and}
  \bibinfo{person}{Zhi~Quan Zhou}.} \bibinfo{year}{2018}\natexlab{}.
\newblock \showarticletitle{Metamorphic Testing: A Review of Challenges and
  Opportunities}.
\newblock \bibinfo{journal}{\emph{ACM Comput. Surv.}} \bibinfo{volume}{51},
  \bibinfo{number}{1}, Article \bibinfo{articleno}{4} (\bibinfo{date}{jan}
  \bibinfo{year}{2018}), \bibinfo{numpages}{27}~pages.
\newblock
\showISSN{0360-0300}
\urldef\tempurl \url{https://doi.org/10.1145/3143561}
\showDOI{\tempurl}


\bibitem[\protect\citeauthoryear{Chen, Leung, and Mak}{Chen
  et~al\mbox{.}}{2005}]{Chen:2005}
\bibfield{author}{\bibinfo{person}{Tsong~Yueh Chen}, \bibinfo{person}{Hing
  Leung}, {and} \bibinfo{person}{Ieng~Kei Mak}.}
  \bibinfo{year}{2005}\natexlab{}.
\newblock \showarticletitle{Adaptive random testing}. In
  \bibinfo{booktitle}{\emph{Proceedings of the 9th Asian Computing Science
  Conference on Advances in Computer Science}}. \bibinfo{publisher}{Springer},
  \bibinfo{address}{Berlin, Heidelberg}, \bibinfo{pages}{320--329}.
\newblock
\urldef\tempurl \url{https://doi.org/10.1007/978-3-540-30502-6_23}
\showDOI{\tempurl}


\bibitem[\protect\citeauthoryear{Cooley and Tukey}{Cooley and Tukey}{1965}]{fourier-algo}
\bibfield{author}{\bibinfo{person}{James~W. Cooley} {and}
  \bibinfo{person}{John~W. Tukey}.} \bibinfo{year}{1965}\natexlab{}.
\newblock \showarticletitle{An algorithm for the machine calculation of complex
  Fourier series}.
\newblock \bibinfo{journal}{\emph{Math. Comp.}}  \bibinfo{volume}{19}
  (\bibinfo{year}{1965}), \bibinfo{pages}{297--301}.
\newblock


\bibitem[\protect\citeauthoryear{Dawes and Bianculli}{Dawes and
  Bianculli}{2022}]{Dawes:2022}
\bibfield{author}{\bibinfo{person}{Joshua~Heneage Dawes} {and}
  \bibinfo{person}{Domenico Bianculli}.} \bibinfo{year}{2022}\natexlab{}.
\newblock \showarticletitle{Specifying Source Code and Signal-Based Behaviour
  of Cyber-Physical System Components}. In \bibinfo{booktitle}{\emph{Formal
  Aspects of Component Software: 18th International Conference, FACS 2022,
  Virtual Event, November 10–11, 2022, Proceedings}} (Oslo, Norway).
  \bibinfo{publisher}{Springer-Verlag}, \bibinfo{address}{Berlin, Heidelberg},
  \bibinfo{pages}{20–38}.
\newblock
\showISBNx{978-3-031-20871-3}
\urldef\tempurl \url{https://doi.org/10.1007/978-3-031-20872-0_2}
\showDOI{\tempurl}


\bibitem[\protect\citeauthoryear{Desborough and Miller}{Desborough and
  Miller}{2002}]{Desborough:2002}
\bibfield{author}{\bibinfo{person}{Lane Desborough} {and}
  \bibinfo{person}{Randy Miller}.} \bibinfo{year}{2002}\natexlab{}.
\newblock \bibinfo{title}{Increasing Customer Value of Industrial Control
  Performance Monitoring — Honeywell ’ s Experience}.
\newblock
\newblock


\bibitem[\protect\citeauthoryear{Donz{\'e}, Maler, Bartocci, Nickovic, Grosu,
  and Smolka}{Donz{\'e} et~al\mbox{.}}{2012}]{Donze:2012}
\bibfield{author}{\bibinfo{person}{Alexandre Donz{\'e}}, \bibinfo{person}{Oded
  Maler}, \bibinfo{person}{Ezio Bartocci}, \bibinfo{person}{Dejan Nickovic},
  \bibinfo{person}{Radu Grosu}, {and} \bibinfo{person}{Scott Smolka}.}
  \bibinfo{year}{2012}\natexlab{}.
\newblock \showarticletitle{On Temporal Logic and Signal Processing}. In
  \bibinfo{booktitle}{\emph{Automated Technology for Verification and
  Analysis}}, \bibfield{editor}{\bibinfo{person}{Supratik Chakraborty} {and}
  \bibinfo{person}{Madhavan Mukund}} (Eds.). \bibinfo{publisher}{Springer
  Berlin Heidelberg}, \bibinfo{address}{Berlin, Heidelberg},
  \bibinfo{pages}{92--106}.
\newblock
\showISBNx{978-3-642-33386-6}


\bibitem[\protect\citeauthoryear{Dreossi, Dang, Donz{\'e}, Kapinski, Jin, and
  Deshmukh}{Dreossi et~al\mbox{.}}{2015}]{Dreossi:2015}
\bibfield{author}{\bibinfo{person}{Tommaso Dreossi}, \bibinfo{person}{Thao
  Dang}, \bibinfo{person}{Alexandre Donz{\'e}}, \bibinfo{person}{James
  Kapinski}, \bibinfo{person}{Xiaoqing Jin}, {and}
  \bibinfo{person}{Jyotirmoy~V. Deshmukh}.} \bibinfo{year}{2015}\natexlab{}.
\newblock \showarticletitle{Efficient Guiding Strategies for Testing of
  Temporal Properties of Hybrid Systems}. In \bibinfo{booktitle}{\emph{NASA
  Formal Methods}}, \bibfield{editor}{\bibinfo{person}{Klaus Havelund},
  \bibinfo{person}{Gerard Holzmann}, {and} \bibinfo{person}{Rajeev Joshi}}
  (Eds.). \bibinfo{publisher}{Springer International Publishing},
  \bibinfo{address}{Cham}, \bibinfo{pages}{127--142}.
\newblock
\showISBNx{978-3-319-17524-9}


\bibitem[\protect\citeauthoryear{Ghosh, Hobbs, Xu, Duggirala, Anderson,
  Thiagarajan, and Chakraborty}{Ghosh et~al\mbox{.}}{2022}]{Ghosh:2022}
\bibfield{author}{\bibinfo{person}{Bineet Ghosh}, \bibinfo{person}{Clara
  Hobbs}, \bibinfo{person}{Shengjie Xu}, \bibinfo{person}{Parasara~Sridhar
  Duggirala}, \bibinfo{person}{James~H. Anderson}, \bibinfo{person}{P.~S.
  Thiagarajan}, {and} \bibinfo{person}{Samarjit Chakraborty}.}
  \bibinfo{year}{2022}\natexlab{}.
\newblock \showarticletitle{Statistical Hypothesis Testing of Controller
  Implementations Under Timing Uncertainties}. In
  \bibinfo{booktitle}{\emph{2022 IEEE 28th International Conference on Embedded
  and Real-Time Computing Systems and Applications (RTCSA)}}.
  \bibinfo{publisher}{IEEE}, \bibinfo{address}{United States of America},
  \bibinfo{pages}{11--20}.
\newblock
\urldef\tempurl \url{https://doi.org/10.1109/RTCSA55878.2022.00008}
\showDOI{\tempurl}


\bibitem[\protect\citeauthoryear{Gille-Maisani and Decaulne}{Gille-Maisani and
  Decaulne}{1959}]{Gille:1959}
\bibfield{author}{\bibinfo{person}{J.C. Gille-Maisani} {and}
  \bibinfo{person}{P. Decaulne}.} \bibinfo{year}{1959}\natexlab{}.
\newblock \bibinfo{booktitle}{\emph{Feedback Control Systems: Analysis,
  Synthesis, and Design}}.
\newblock \bibinfo{publisher}{McGraw-Hill}, \bibinfo{address}{United States of
  America}.
\newblock
\showLCCN{58009856}
\urldef\tempurl \url{https://books.google.se/books?id=9WG9zQEACAAJ}
\showURL{\tempurl}


\bibitem[\protect\citeauthoryear{Greiff}{Greiff}{2017}]{Greiff:2017}
\bibfield{author}{\bibinfo{person}{Marcus Greiff}.}
  \bibinfo{year}{2017}\natexlab{}.
\newblock \bibinfo{title}{Modelling and Control of the Crazyflie Quadrotor for
  Aggressive and Autonomous Flight by Optical Flow Driven State Estimation}.
\newblock
\newblock
\showISSN{0280-5316}
\newblock
\shownote{Student Paper}.


\bibitem[\protect\citeauthoryear{Greiff, Persson, Sun, Åström, and
  Robertsson}{Greiff et~al\mbox{.}}{2021}]{Greiff:2021}
\bibfield{author}{\bibinfo{person}{Marcus Greiff}, \bibinfo{person}{Patrik
  Persson}, \bibinfo{person}{Zhiyong Sun}, \bibinfo{person}{Karl Åström},
  {and} \bibinfo{person}{Anders Robertsson}.} \bibinfo{year}{2021}\natexlab{}.
\newblock \bibinfo{title}{Quadrotor Control on $SU(2)\times R^3$ with SLAM
  Integration}.
\newblock
\newblock
\urldef\tempurl \url{https://doi.org/10.48550/ARXIV.2110.01099}
\showDOI{\tempurl}


\bibitem[\protect\citeauthoryear{He, Wang, Shi, and Liu}{He
  et~al\mbox{.}}{2020}]{He:2020}
\bibfield{author}{\bibinfo{person}{Xiao He}, \bibinfo{person}{Xingwei Wang},
  \bibinfo{person}{Jia Shi}, {and} \bibinfo{person}{Yi Liu}.}
  \bibinfo{year}{2020}\natexlab{}.
\newblock \showarticletitle{Testing High Performance Numerical Simulation
  Programs: Experience, Lessons Learned, and Open Issues}. In
  \bibinfo{booktitle}{\emph{Proceedings of the 29th ACM SIGSOFT International
  Symposium on Software Testing and Analysis}} (Virtual Event, USA)
  \emph{(\bibinfo{series}{ISSTA 2020})}. \bibinfo{publisher}{Association for
  Computing Machinery}, \bibinfo{address}{New York, NY, USA},
  \bibinfo{pages}{502–515}.
\newblock
\showISBNx{9781450380089}
\urldef\tempurl \url{https://doi.org/10.1145/3395363.3397382}
\showDOI{\tempurl}


\bibitem[\protect\citeauthoryear{He, Chen, Huang, Wang, Pei, and Yuan}{He
  et~al\mbox{.}}{2019}]{He:2019}
\bibfield{author}{\bibinfo{person}{Zhijian He}, \bibinfo{person}{Yao Chen},
  \bibinfo{person}{Enyan Huang}, \bibinfo{person}{Qixin Wang},
  \bibinfo{person}{Yu Pei}, {and} \bibinfo{person}{Haidong Yuan}.}
  \bibinfo{year}{2019}\natexlab{}.
\newblock \showarticletitle{A System Identification Based Oracle for
  Control-CPS Software Fault Localization}. In \bibinfo{booktitle}{\emph{2019
  IEEE/ACM 41st International Conference on Software Engineering (ICSE)}}.
  \bibinfo{publisher}{IEEE}, \bibinfo{address}{United States of America},
  \bibinfo{pages}{116--127}.
\newblock
\urldef\tempurl \url{https://doi.org/10.1109/ICSE.2019.00029}
\showDOI{\tempurl}


\bibitem[\protect\citeauthoryear{Hu and Lin}{Hu and Lin}{2001}]{Hu:2001}
\bibfield{author}{\bibinfo{person}{Tingshu Hu} {and} \bibinfo{person}{Zongli
  Lin}.} \bibinfo{year}{2001}\natexlab{}.
\newblock \bibinfo{booktitle}{\emph{Introduction}}.
\newblock \bibinfo{publisher}{Birkh{\"a}user Boston}, \bibinfo{address}{Boston,
  MA}, \bibinfo{pages}{1--9}.
\newblock
\showISBNx{978-1-4612-0205-9}
\urldef\tempurl \url{https://doi.org/10.1007/978-1-4612-0205-9_1}
\showDOI{\tempurl}


\bibitem[\protect\citeauthoryear{Hänsel, Rose, Herber, and Glesner}{Hänsel
  et~al\mbox{.}}{2011}]{Hansel:2011}
\bibfield{author}{\bibinfo{person}{Joachim Hänsel}, \bibinfo{person}{Daniela
  Rose}, \bibinfo{person}{Paula Herber}, {and} \bibinfo{person}{Sabine
  Glesner}.} \bibinfo{year}{2011}\natexlab{}.
\newblock \showarticletitle{An Evolutionary Algorithm for the Generation of
  Timed Test Traces for Embedded Real-Time Systems}. In
  \bibinfo{booktitle}{\emph{2011 Fourth IEEE International Conference on
  Software Testing, Verification and Validation}}. \bibinfo{publisher}{IEEE},
  \bibinfo{address}{United States of America}, \bibinfo{pages}{170--179}.
\newblock
\urldef\tempurl \url{https://doi.org/10.1109/ICST.2011.37}
\showDOI{\tempurl}


\bibitem[\protect\citeauthoryear{Khalil}{Khalil}{2002}]{khalil:2002}
\bibfield{author}{\bibinfo{person}{H.K. Khalil}.}
  \bibinfo{year}{2002}\natexlab{}.
\newblock \bibinfo{booktitle}{\emph{Nonlinear Systems}}.
\newblock \bibinfo{publisher}{Prentice Hall}, \bibinfo{address}{United States
  of America}.
\newblock
\showISBNx{9780130673893}
\showLCCN{95045804}
\urldef\tempurl \url{https://books.google.se/books?id=t\_d1QgAACAAJ}
\showURL{\tempurl}


\bibitem[\protect\citeauthoryear{Lamberg, Beine, Eschmann, Otterbach, Conrad,
  and Fey}{Lamberg et~al\mbox{.}}{2004}]{Lamberg:2004}
\bibfield{author}{\bibinfo{person}{Klaus Lamberg}, \bibinfo{person}{Michael
  Beine}, \bibinfo{person}{Mario Eschmann}, \bibinfo{person}{Rainer Otterbach},
  \bibinfo{person}{Mirko Conrad}, {and} \bibinfo{person}{Ines Fey}.}
  \bibinfo{year}{2004}\natexlab{}.
\newblock \showarticletitle{Model-based Testing of Embedded Automotive Software
  Using Mtest}. In \bibinfo{booktitle}{\emph{SAE 2004 World Congress and
  Exhibition}}. \bibinfo{publisher}{SAE International},
  \bibinfo{address}{United States of America}, \bibinfo{numpages}{11}~pages.
\newblock
\showISSN{0148-7191}
\urldef\tempurl \url{https://doi.org/10.4271/2004-01-1593}
\showDOI{\tempurl}


\bibitem[\protect\citeauthoryear{Lee}{Lee}{2015}]{Lee:2015}
\bibfield{author}{\bibinfo{person}{Edward~A. Lee}.}
  \bibinfo{year}{2015}\natexlab{}.
\newblock \showarticletitle{The Past, Present and Future of Cyber-Physical
  Systems: A Focus on Models}.
\newblock \bibinfo{journal}{\emph{Sensors}} \bibinfo{volume}{15},
  \bibinfo{number}{3} (\bibinfo{year}{2015}), \bibinfo{pages}{4837--4869}.
\newblock
\showISSN{1424-8220}
\urldef\tempurl \url{https://doi.org/10.3390/s150304837}
\showDOI{\tempurl}


\bibitem[\protect\citeauthoryear{Lu, Nolte, Bate, and Cucu-Grosjean}{Lu
  et~al\mbox{.}}{2012}]{Lu:2012}
\bibfield{author}{\bibinfo{person}{Yue Lu}, \bibinfo{person}{Thomas Nolte},
  \bibinfo{person}{Iain Bate}, {and} \bibinfo{person}{Liliana Cucu-Grosjean}.}
  \bibinfo{year}{2012}\natexlab{}.
\newblock \showarticletitle{A Statistical Response-Time Analysis of Real-Time
  Embedded Systems}. In \bibinfo{booktitle}{\emph{2012 IEEE 33rd Real-Time
  Systems Symposium}}. \bibinfo{publisher}{IEEE}, \bibinfo{address}{United
  States of America}, \bibinfo{pages}{351--362}.
\newblock
\urldef\tempurl \url{https://doi.org/10.1109/RTSS.2012.85}
\showDOI{\tempurl}


\bibitem[\protect\citeauthoryear{Magnani, Cattaneo, Chiari, and Agosta}{Magnani
  et~al\mbox{.}}{2021}]{Magnani:2021}
\bibfield{author}{\bibinfo{person}{Gabriele Magnani}, \bibinfo{person}{Daniele
  Cattaneo}, \bibinfo{person}{Michele Chiari}, {and} \bibinfo{person}{Giovanni
  Agosta}.} \bibinfo{year}{2021}\natexlab{}.
\newblock \showarticletitle{{The Impact of Precision Tuning on Embedded Systems
  Performance: A Case Study on Field-Oriented Control}}. In
  \bibinfo{booktitle}{\emph{12th Workshop on Parallel Programming and Run-Time
  Management Techniques for Many-core Architectures and 10th Workshop on Design
  Tools and Architectures for Multicore Embedded Computing Platforms
  (PARMA-DITAM 2021)}} \emph{(\bibinfo{series}{Open Access Series in
  Informatics (OASIcs)}, Vol.~\bibinfo{volume}{88})},
  \bibfield{editor}{\bibinfo{person}{Jo\~{a}o Bispo}, \bibinfo{person}{Stefano
  Cherubin}, {and} \bibinfo{person}{Jos\'{e} Flich}} (Eds.).
  \bibinfo{publisher}{Schloss Dagstuhl -- Leibniz-Zentrum f{\"u}r Informatik},
  \bibinfo{address}{Dagstuhl, Germany}, \bibinfo{pages}{3:1--3:13}.
\newblock
\showISBNx{978-3-95977-181-8}
\showISSN{2190-6807}
\urldef\tempurl \url{https://doi.org/10.4230/OASIcs.PARMA-DITAM.2021.3}
\showDOI{\tempurl}


\bibitem[\protect\citeauthoryear{Mandrioli, Max, and Maggio}{Mandrioli
  et~al\mbox{.}}{2023}]{Mandrioli:2023}
\bibfield{author}{\bibinfo{person}{Claudio Mandrioli},
  \bibinfo{person}{Nyberg~Carlsson Max}, {and} \bibinfo{person}{Martina
  Maggio}.} \bibinfo{year}{2023}\natexlab{}.
\newblock \showarticletitle{Testing Abstractions for Cyber-Physical Control
  Systems}.
\newblock \bibinfo{journal}{\emph{ACM Transactions on Software Engineering and
  Methodology}} \bibinfo{volume}{1}, \bibinfo{number}{1}, Article
  \bibinfo{articleno}{1} (\bibinfo{date}{Aug.} \bibinfo{year}{2023}),
  \bibinfo{numpages}{31}~pages.
\newblock
\urldef\tempurl \url{https://doi.org/10.1145/3617170}
\showDOI{\tempurl}


\bibitem[\protect\citeauthoryear{Marculescu, Feldt, Torkar, and
  Poulding}{Marculescu et~al\mbox{.}}{2015}]{Marculescu:2015}
\bibfield{author}{\bibinfo{person}{Bogdan Marculescu}, \bibinfo{person}{Robert
  Feldt}, \bibinfo{person}{Richard Torkar}, {and} \bibinfo{person}{Simon
  Poulding}.} \bibinfo{year}{2015}\natexlab{}.
\newblock \showarticletitle{An initial industrial evaluation of interactive
  search-based testing for embedded software}.
\newblock \bibinfo{journal}{\emph{Applied Soft Computing}}
  \bibinfo{volume}{29} (\bibinfo{date}{April} \bibinfo{year}{2015}),
  \bibinfo{pages}{26--39}.
\newblock
Issue 0.
\urldef\tempurl \url{https://doi.org/10.1016/j.asoc.2014.12.025}
\showDOI{\tempurl}


\bibitem[\protect\citeauthoryear{Matinnejad, Nejati, Briand, and
  Brcukmann}{Matinnejad et~al\mbox{.}}{2014}]{Matinnejad:2014}
\bibfield{author}{\bibinfo{person}{Reza Matinnejad}, \bibinfo{person}{Shiva
  Nejati}, \bibinfo{person}{Lionel Briand}, {and} \bibinfo{person}{Thomas
  Brcukmann}.} \bibinfo{year}{2014}\natexlab{}.
\newblock \showarticletitle{MiL Testing of Highly Configurable Continuous
  Controllers: Scalable Search Using Surrogate Models}. In
  \bibinfo{booktitle}{\emph{Proceedings of the 29th ACM/IEEE International
  Conference on Automated Software Engineering}} (Vasteras, Sweden)
  \emph{(\bibinfo{series}{ASE '14})}. \bibinfo{publisher}{Association for
  Computing Machinery}, \bibinfo{address}{New York, NY, USA},
  \bibinfo{pages}{163–174}.
\newblock
\showISBNx{9781450330138}
\urldef\tempurl \url{https://doi.org/10.1145/2642937.2642978}
\showDOI{\tempurl}


\bibitem[\protect\citeauthoryear{Matinnejad, Nejati, and Briand}{Matinnejad
  et~al\mbox{.}}{2017}]{Matinnejad:2017}
\bibfield{author}{\bibinfo{person}{Reza Matinnejad}, \bibinfo{person}{Shiva
  Nejati}, {and} \bibinfo{person}{Lionel~C. Briand}.}
  \bibinfo{year}{2017}\natexlab{}.
\newblock \showarticletitle{Automated Testing of Hybrid Simulink/Stateflow
  Controllers: Industrial Case Studies}. In
  \bibinfo{booktitle}{\emph{Proceedings of the 2017 11th Joint Meeting on
  Foundations of Software Engineering}} (Paderborn, Germany)
  \emph{(\bibinfo{series}{ESEC/FSE 2017})}. \bibinfo{publisher}{Association for
  Computing Machinery}, \bibinfo{address}{New York, NY, USA},
  \bibinfo{pages}{938–943}.
\newblock
\showISBNx{9781450351058}
\urldef\tempurl \url{https://doi.org/10.1145/3106237.3117770}
\showDOI{\tempurl}


\bibitem[\protect\citeauthoryear{Matinnejad, Nejati, Briand, and
  Bruckmann}{Matinnejad et~al\mbox{.}}{2016}]{Reza:2016}
\bibfield{author}{\bibinfo{person}{Reza Matinnejad}, \bibinfo{person}{Shiva
  Nejati}, \bibinfo{person}{Lionel~C. Briand}, {and} \bibinfo{person}{Thomas
  Bruckmann}.} \bibinfo{year}{2016}\natexlab{}.
\newblock \showarticletitle{Automated Test Suite Generation for Time-Continuous
  Simulink Models}. In \bibinfo{booktitle}{\emph{2016 IEEE/ACM 38th
  International Conference on Software Engineering (ICSE)}}.
  \bibinfo{publisher}{ACM}, \bibinfo{address}{United States of America},
  \bibinfo{pages}{595--606}.
\newblock
\urldef\tempurl \url{https://doi.org/10.1145/2884781.2884797}
\showDOI{\tempurl}


\bibitem[\protect\citeauthoryear{Menghi, Nejati, Briand, and Parache}{Menghi
  et~al\mbox{.}}{2019a}]{Menghi:2019}
\bibfield{author}{\bibinfo{person}{Claudio Menghi}, \bibinfo{person}{Shiva
  Nejati}, \bibinfo{person}{Lionel~C. Briand}, {and}
  \bibinfo{person}{Yago~Isasi Parache}.} \bibinfo{year}{2019}\natexlab{a}.
\newblock \showarticletitle{Approximation-Refinement Testing of
  Compute-Intensive Cyber-Physical Models: An Approach Based on System
  Identification}.
\newblock \bibinfo{journal}{\emph{CoRR}}  \bibinfo{volume}{abs/1910.02837}
  (\bibinfo{year}{2019}), \bibinfo{numpages}{12}~pages.
\newblock
\showeprint[arxiv]{1910.02837}
\urldef\tempurl \url{http://arxiv.org/abs/1910.02837}
\showURL{\tempurl}


\bibitem[\protect\citeauthoryear{Menghi, Nejati, Gaaloul, and Briand}{Menghi
  et~al\mbox{.}}{2019b}]{menghi:2019:oracles}
\bibfield{author}{\bibinfo{person}{Claudio Menghi}, \bibinfo{person}{Shiva
  Nejati}, \bibinfo{person}{Khouloud Gaaloul}, {and} \bibinfo{person}{Lionel~C.
  Briand}.} \bibinfo{year}{2019}\natexlab{b}.
\newblock \showarticletitle{Generating automated and online test oracles for
  Simulink models with continuous and uncertain behaviors}. In
  \bibinfo{booktitle}{\emph{Foundations of Software Engineering}}.
  \bibinfo{publisher}{ACM}, \bibinfo{address}{United States of America},
  \bibinfo{pages}{27--38}.
\newblock


\bibitem[\protect\citeauthoryear{Menghi, Viganò, Bianculli, and Briand}{Menghi
  et~al\mbox{.}}{2021}]{Menghi:2021}
\bibfield{author}{\bibinfo{person}{Claudio Menghi}, \bibinfo{person}{Enrico
  Viganò}, \bibinfo{person}{Domenico Bianculli}, {and}
  \bibinfo{person}{Lionel~C. Briand}.} \bibinfo{year}{2021}\natexlab{}.
\newblock \showarticletitle{Trace-Checking CPS Properties: Bridging the
  Cyber-Physical Gap}. In \bibinfo{booktitle}{\emph{2021 IEEE/ACM 43rd
  International Conference on Software Engineering (ICSE)}}.
  \bibinfo{publisher}{ACM}, \bibinfo{address}{United States of America},
  \bibinfo{pages}{847--859}.
\newblock
\urldef\tempurl \url{https://doi.org/10.1109/ICSE43902.2021.00082}
\showDOI{\tempurl}


\bibitem[\protect\citeauthoryear{Mueller, Hehn, and D’Andrea}{Mueller
  et~al\mbox{.}}{2016}]{Mueller:2016}
\bibfield{author}{\bibinfo{person}{Mark~W Mueller}, \bibinfo{person}{Markus
  Hehn}, {and} \bibinfo{person}{Raffaello D’Andrea}.}
  \bibinfo{year}{2016}\natexlab{}.
\newblock \showarticletitle{Covariance Correction Step for Kalman Filtering
  with an Attitude}.
\newblock \bibinfo{journal}{\emph{Journal of Guidance, Control, and Dynamics}}
  \bibinfo{volume}{40}, \bibinfo{number}{9} (\bibinfo{year}{2016}),
  \bibinfo{pages}{1--7}.
\newblock


\bibitem[\protect\citeauthoryear{Nejati, Gaaloul, Menghi, Briand, Foster, and
  Wolfe}{Nejati et~al\mbox{.}}{2019}]{Nejati:2019}
\bibfield{author}{\bibinfo{person}{Shiva Nejati}, \bibinfo{person}{Khouloud
  Gaaloul}, \bibinfo{person}{Claudio Menghi}, \bibinfo{person}{Lionel~C.
  Briand}, \bibinfo{person}{Stephen Foster}, {and} \bibinfo{person}{David
  Wolfe}.} \bibinfo{year}{2019}\natexlab{}.
\newblock \showarticletitle{Evaluating Model Testing and Model Checking for
  Finding Requirements Violations in Simulink Models}. In
  \bibinfo{booktitle}{\emph{Proceedings of the 2019 27th ACM Joint Meeting on
  European Software Engineering Conference and Symposium on the Foundations of
  Software Engineering}} (Tallinn, Estonia) \emph{(\bibinfo{series}{ESEC/FSE
  2019})}. \bibinfo{publisher}{Association for Computing Machinery},
  \bibinfo{address}{New York, NY, USA}, \bibinfo{pages}{1015–1025}.
\newblock
\showISBNx{9781450355728}
\urldef\tempurl \url{https://doi.org/10.1145/3338906.3340444}
\showDOI{\tempurl}


\bibitem[\protect\citeauthoryear{Peleska}{Peleska}{2002}]{Peleska:2002}
\bibfield{author}{\bibinfo{person}{Jan Peleska}.}
  \bibinfo{year}{2002}\natexlab{}.
\newblock \showarticletitle{Hardware/Software Integration Testing for the new
  Airbus Aircraft Families.}
\newblock
  \bibinfo{journal}{\emph{http://www.informatik.uni-bremen.de/agbs/jp/papers/peleskaTestCom2002.html}}
  \bibinfo{volume}{82}, \bibinfo{number}{14} (\bibinfo{date}{01}
  \bibinfo{year}{2002}).
\newblock
\showISBNx{978-1-4757-6705-6}
\urldef\tempurl \url{https://doi.org/10.1007/978-0-387-35497-2_24}
\showDOI{\tempurl}


\bibitem[\protect\citeauthoryear{Plaku, Kavraki, and Vardi}{Plaku
  et~al\mbox{.}}{2009}]{Plaku:2009}
\bibfield{author}{\bibinfo{person}{Erion Plaku}, \bibinfo{person}{Lydia~E.
  Kavraki}, {and} \bibinfo{person}{Moshe~Y. Vardi}.}
  \bibinfo{year}{2009}\natexlab{}.
\newblock \showarticletitle{Falsification of LTL Safety Properties in Hybrid
  Systems}. In \bibinfo{booktitle}{\emph{Tools and Algorithms for the
  Construction and Analysis of Systems}},
  \bibfield{editor}{\bibinfo{person}{Stefan Kowalewski} {and}
  \bibinfo{person}{Anna Philippou}} (Eds.). \bibinfo{publisher}{Springer Berlin
  Heidelberg}, \bibinfo{address}{Berlin, Heidelberg},
  \bibinfo{pages}{368--382}.
\newblock
\showISBNx{978-3-642-00768-2}


\bibitem[\protect\citeauthoryear{Priyadarshi~Tripathy}{Priyadarshi~Tripathy}{2008}]{Naik:2008}
\bibfield{author}{\bibinfo{person}{Kshirasagar~Naik Priyadarshi~Tripathy}.}
  \bibinfo{year}{2008}\natexlab{}.
\newblock \bibinfo{booktitle}{\emph{Acceptance Testing}}.
\newblock \bibinfo{publisher}{John Wiley and Sons, Ltd},
  \bibinfo{address}{United States of America}, Chapter~14,
  \bibinfo{pages}{450--470}.
\newblock
\showISBNx{9780470382844}
\urldef\tempurl \url{https://doi.org/10.1002/9780470382844.ch14}
\showDOI{\tempurl}
\showeprint{https://onlinelibrary.wiley.com/doi/pdf/10.1002/9780470382844.ch14}


\bibitem[\protect\citeauthoryear{Samad and Balas}{Samad and Balas}{2003}]{Samad:2003}
\bibfield{author}{\bibinfo{person}{Tariq Samad} {and} \bibinfo{person}{Gary
  Balas}.} \bibinfo{year}{2003}\natexlab{}.
\newblock \bibinfo{booktitle}{\emph{Frontmatter}}.
\newblock \bibinfo{publisher}{John Wiley and Sons, Ltd},
  \bibinfo{address}{United States of America}, \bibinfo{pages}{i--xx}.
\newblock
\urldef\tempurl \url{https://doi.org/10.1002/047172288X.fmatter}
\showDOI{\tempurl}


\bibitem[\protect\citeauthoryear{Sanchez-Stern, Panchekha, Lerner, and
  Tatlock}{Sanchez-Stern et~al\mbox{.}}{2018}]{Sanchez:2018}
\bibfield{author}{\bibinfo{person}{Alex Sanchez-Stern}, \bibinfo{person}{Pavel
  Panchekha}, \bibinfo{person}{Sorin Lerner}, {and} \bibinfo{person}{Zachary
  Tatlock}.} \bibinfo{year}{2018}\natexlab{}.
\newblock \showarticletitle{Finding Root Causes of Floating Point Error}.
\newblock \bibinfo{journal}{\emph{SIGPLAN Not.}} \bibinfo{volume}{53},
  \bibinfo{number}{4} (\bibinfo{date}{jun} \bibinfo{year}{2018}),
  \bibinfo{pages}{256–269}.
\newblock
\showISSN{0362-1340}
\urldef\tempurl \url{https://doi.org/10.1145/3296979.3192411}
\showDOI{\tempurl}


\bibitem[\protect\citeauthoryear{Silano, Aucone, and Iannelli}{Silano
  et~al\mbox{.}}{2018}]{Silano:2018}
\bibfield{author}{\bibinfo{person}{Giuseppe Silano}, \bibinfo{person}{Emanuele
  Aucone}, {and} \bibinfo{person}{Luigi Iannelli}.}
  \bibinfo{year}{2018}\natexlab{}.
\newblock \showarticletitle{CrazyS: A Software-In-The-Loop Platform for the
  Crazyflie 2.0 Nano-Quadcopter}. In \bibinfo{booktitle}{\emph{2018 26th
  Mediterranean Conference on Control and Automation (MED)}}.
  \bibinfo{publisher}{IEEE}, \bibinfo{address}{United States of America},
  \bibinfo{pages}{1--6}.
\newblock
\urldef\tempurl \url{https://doi.org/10.1109/MED.2018.8442759}
\showDOI{\tempurl}


\bibitem[\protect\citeauthoryear{Sinharay}{Sinharay}{2010}]{Sinharay:2010}
\bibfield{author}{\bibinfo{person}{S. Sinharay}.}
  \bibinfo{year}{2010}\natexlab{}.
\newblock \showarticletitle{Continuous Probability Distributions}.
\newblock In \bibinfo{booktitle}{\emph{International Encyclopedia of Education
  (Third Edition)} (\bibinfo{edition}{third edition} ed.)},
  \bibfield{editor}{\bibinfo{person}{Penelope Peterson}, \bibinfo{person}{Eva
  Baker}, {and} \bibinfo{person}{Barry McGaw}} (Eds.).
  \bibinfo{publisher}{Elsevier}, \bibinfo{address}{Oxford},
  \bibinfo{pages}{98--102}.
\newblock
\showISBNx{978-0-08-044894-7}
\urldef\tempurl \url{https://doi.org/10.1016/B978-0-08-044894-7.01720-6}
\showDOI{\tempurl}


\bibitem[\protect\citeauthoryear{{Timperley}, {Afzal}, {Katz}, {Hernandez}, and
  {Le Goues}}{{Timperley} et~al\mbox{.}}{2018}]{Timperley:2018}
\bibfield{author}{\bibinfo{person}{C.~S. {Timperley}}, \bibinfo{person}{A.
  {Afzal}}, \bibinfo{person}{D.~S. {Katz}}, \bibinfo{person}{J.~M.
  {Hernandez}}, {and} \bibinfo{person}{C. {Le Goues}}.}
  \bibinfo{year}{2018}\natexlab{}.
\newblock \showarticletitle{Crashing Simulated Planes is Cheap: Can Simulation
  Detect Robotics Bugs Early?}. In \bibinfo{booktitle}{\emph{2018 IEEE 11th
  International Conference on Software Testing, Verification and Validation
  (ICST)}}. \bibinfo{publisher}{IEEE}, \bibinfo{address}{United States of
  America}, \bibinfo{pages}{331--342}.
\newblock
\urldef\tempurl \url{https://doi.org/10.1109/ICST.2018.00040}
\showDOI{\tempurl}


\bibitem[\protect\citeauthoryear{Vreman, Cervin, and Maggio}{Vreman
  et~al\mbox{.}}{2021}]{Vreman:2021}
\bibfield{author}{\bibinfo{person}{Nils Vreman}, \bibinfo{person}{Anton
  Cervin}, {and} \bibinfo{person}{Martina Maggio}.}
  \bibinfo{year}{2021}\natexlab{}.
\newblock \showarticletitle{Stability and Performance Analysis of Control
  Systems Subject to Bursts of Deadline Misses}. In
  \bibinfo{booktitle}{\emph{33rd Euromicro Conference on Real-Time Systems
  (ECRTS 2021)}}, Vol.~\bibinfo{volume}{196}. \bibinfo{publisher}{Schloss
  Dagstuhl - Leibniz-Zentrum f{\"u}r Informatik}, \bibinfo{address}{Germany},
  \bibinfo{numpages}{23}~pages.
\newblock
\urldef\tempurl \url{https://doi.org/10.4230/LIPIcs.ECRTS.2021.15}
\showDOI{\tempurl}
\newblock
\shownote{33rd Euromicro Conference on Real-Time Systems (ECRTS 2021) ;
  Conference date: 05-07-2021 Through 09-07-2021}.


\bibitem[\protect\citeauthoryear{White}{White}{2001}]{White:2001}
\bibfield{author}{\bibinfo{person}{A.L. White}.}
  \bibinfo{year}{2001}\natexlab{}.
\newblock \showarticletitle{Comments on modified condition/decision coverage
  for software testing [of flight control software]}. In
  \bibinfo{booktitle}{\emph{2001 IEEE Aerospace Conference Proceedings (Cat.
  No.01TH8542)}}, Vol.~\bibinfo{volume}{6}. \bibinfo{publisher}{IEEE},
  \bibinfo{address}{United States of America}, \bibinfo{pages}{2821--2827
  vol.6}.
\newblock
\urldef\tempurl \url{https://doi.org/10.1109/AERO.2001.931302}
\showDOI{\tempurl}


\bibitem[\protect\citeauthoryear{Wu, Zeng, Wang, and Yu}{Wu
  et~al\mbox{.}}{2017}]{Wu:2017}
\bibfield{author}{\bibinfo{person}{Meng Wu}, \bibinfo{person}{Haibo Zeng},
  \bibinfo{person}{Chao Wang}, {and} \bibinfo{person}{Huafeng Yu}.}
  \bibinfo{year}{2017}\natexlab{}.
\newblock \showarticletitle{INVITED: Safety guard: Runtime enforcement for
  safety-critical cyber-physical systems}. In \bibinfo{booktitle}{\emph{2017
  54th ACM/EDAC/IEEE Design Automation Conference (DAC)}}.
  \bibinfo{publisher}{ACM}, \bibinfo{address}{United States of America},
  \bibinfo{pages}{1--6}.
\newblock
\urldef\tempurl \url{https://doi.org/10.1145/3061639.3072957}
\showDOI{\tempurl}


\bibitem[\protect\citeauthoryear{Yamagata, Liu, Akazaki, Duan, and
  Hao}{Yamagata et~al\mbox{.}}{2021}]{Yamagata:2021}
\bibfield{author}{\bibinfo{person}{Yoriyuki Yamagata}, \bibinfo{person}{Shuang
  Liu}, \bibinfo{person}{Takumi Akazaki}, \bibinfo{person}{Yihai Duan}, {and}
  \bibinfo{person}{Jianye Hao}.} \bibinfo{year}{2021}\natexlab{}.
\newblock \showarticletitle{Falsification of Cyber-Physical Systems Using Deep
  Reinforcement Learning}.
\newblock \bibinfo{journal}{\emph{IEEE Transactions on Software Engineering}}
  \bibinfo{volume}{47}, \bibinfo{number}{12} (\bibinfo{year}{2021}),
  \bibinfo{pages}{2823--2840}.
\newblock
\urldef\tempurl \url{https://doi.org/10.1109/TSE.2020.2969178}
\showDOI{\tempurl}


\bibitem[\protect\citeauthoryear{Yi, Chen, Mao, and Ji}{Yi
  et~al\mbox{.}}{2017}]{Yi:2017}
\bibfield{author}{\bibinfo{person}{Xin Yi}, \bibinfo{person}{Liqian Chen},
  \bibinfo{person}{Xiaoguang Mao}, {and} \bibinfo{person}{Tao Ji}.}
  \bibinfo{year}{2017}\natexlab{}.
\newblock \showarticletitle{Efficient Global Search for Inputs Triggering High
  Floating-Point Inaccuracies}. In \bibinfo{booktitle}{\emph{2017 24th
  Asia-Pacific Software Engineering Conference (APSEC)}}.
  \bibinfo{publisher}{IEEE}, \bibinfo{address}{United States of America},
  \bibinfo{pages}{11--20}.
\newblock
\urldef\tempurl \url{https://doi.org/10.1109/APSEC.2017.7}
\showDOI{\tempurl}


\bibitem[\protect\citeauthoryear{Zimmer, Hedrick, and Lee}{Zimmer
  et~al\mbox{.}}{2015}]{Zimmer:2015}
\bibfield{author}{\bibinfo{person}{Michael Zimmer}, \bibinfo{person}{J.
  Hedrick}, {and} \bibinfo{person}{Edward~A. Lee}.}
  \bibinfo{year}{2015}\natexlab{}.
\newblock \showarticletitle{Ramifications of software implementation and
  deployment: A case study on yaw moment controller design}.
\newblock \bibinfo{journal}{\emph{2015 American Control Conference (ACC)}}
  \bibinfo{volume}{0} (\bibinfo{year}{2015}), \bibinfo{pages}{2014--2019}.
\newblock


\end{thebibliography}

\end{document}